\documentclass[11pt,letterpaper]{article}
\pdfoutput=1
\usepackage{jheppub}

\usepackage[utf8]{inputenc}

\usepackage{color}
\usepackage{graphicx}
\usepackage{tabularx}
\usepackage{xspace}

\usepackage{verbatim}
\usepackage{amsmath}
\usepackage{amssymb}
\usepackage[caption=false]{subfig}
\usepackage{url}
\usepackage{bbold}
\usepackage{slashed}
\usepackage{array}

\usepackage{multirow}
\usepackage{threeparttable}
\usepackage{paralist}

\newcommand{\GeV}{\text{GeV}}

\newcommand{\genang}[2]{{\lambda^{#1}_{#2}}}

\newcommand{\df}{\text{d}}
\newcommand{\vev}[1]{\langle #1 \rangle}

\DeclareRobustCommand{\Sec}[1]{Sec.~\ref{#1}}

\DeclareRobustCommand{\Fig}[1]{Fig.~\ref{#1}}

\DeclareRobustCommand{\Eq}[1]{Eq.~(\ref{#1})}

\DeclareRobustCommand{\Ref}[1]{Ref.~\cite{#1}}
\DeclareRobustCommand{\Refs}[1]{Refs.~\cite{#1}}

\newcommand{\be}{\begin{equation}}
\newcommand{\ee}{\end{equation}}

\usepackage{xspace}

\newcommand{\beq}{\begin{eqnarray}}
\newcommand{\eeq}{\end{eqnarray}}

\begin{document}

\title{Systematics of quark/gluon tagging}

\author[a]{Philippe Gras,}
\author[b]{Stefan H\"{o}che,}
\author[c]{Deepak Kar,}
\author[d]{Andrew Larkoski,}
\author[e]{\\ Leif L\"{o}nnblad,}
\author[f,g]{Simon Pl\"atzer,}
\author[h,i]{Andrzej Si\'{o}dmok,}
\author[j]{Peter Skands,}
\author[k,\dagger]{\\ Gregory Soyez,}
\emailAdd{gregory.soyez@cea.fr}
\author[l,\dagger]{and Jesse Thaler}
\emailAdd{jthaler@mit.edu}

\note[$\dagger$]{Coordinators for an initial version of this study appearing in \Ref{Badger:2016bpw}.}
\affiliation[a]{IRFU, CEA, Universit\'e Paris-Saclay, Gif-sur-Yvette, France}
\affiliation[b]{SLAC National Accelerator Laboratory, Menlo Park, CA 94025, USA}
\affiliation[c]{School of Physics, University of the Witwatersrand, Johannesburg, Wits 2050, South Africa}
\affiliation[d]{Physics Department, Reed College, Portland, OR 97202}
\affiliation[e]{Department of Astronomy and Theoretical Physics, Lund University, Sweden}
\affiliation[f]{Institute for Particle Physics Phenomenology, University of Durham, Durham DH1 3LE, UK}
\affiliation[g]{Particle Physics Group, School of Physics and Astronomy, \\ University of Manchester, Manchester M13 9PL, UK}
\affiliation[h]{CERN, TH Department, CH–1211 Geneva, Switzerland}
\affiliation[i]{Institute of Nuclear Physics, Polish Academy of Sciences, \\ ul.\ Radzikowskiego 152, 31-342 Krak\'ow, Poland}
\affiliation[j]{School of Physics and Astronomy, Monash University, VIC-3800, Australia}
\affiliation[k]{IPhT, CEA Saclay, CNRS UMR 3681, F-91191 Gif-sur-Yvette, France}
\affiliation[l]{Center for Theoretical Physics, Massachusetts Institute of Technology, \\ Cambridge, MA 02139, U.S.A.}

\abstract{By measuring the substructure of a jet, one can assign it a ``quark'' or ``gluon'' tag.  In the eikonal (double-logarithmic) limit, quark/gluon discrimination is determined solely by the color factor of the initiating parton ($C_F$ versus $C_A$).  In this paper, we confront the challenges faced when going beyond this leading-order understanding, using both parton-shower generators and first-principles calculations to assess the impact of higher-order perturbative and nonperturbative physics.  Working in the idealized context of electron-positron collisions, where one can define a proxy for quark and gluon jets based on the Lorentz structure of the production vertex, we find a fascinating interplay between perturbative shower effects and nonperturbative hadronization effects.   Turning to proton-proton collisions, we highlight a core set of measurements that would constrain current uncertainties in quark/gluon tagging and improve the overall modeling of jets at the Large Hadron Collider.
}

\preprint{\begin{flushright}
MIT--CTP 4885 \\
CoEPP-MN-17-2 \\
MCNET-17-04
\end{flushright}
}

\maketitle

\section{Overview}
\label{sec:overview}

Jets are robust tools for studying short-distance collisions involving quarks and gluons.  With a suitable jet definition, one can connect jet measurements made on clusters of hadrons to perturbative calculations made on clusters of partons.  More ambitiously, one can try to tag jets with a suitably-defined flavor label, thereby enhancing the fraction of, say, quark-tagged jets over gluon-tagged jets.  This is relevant for searches for physics beyond the standard model, where signals of interest are often dominated by quarks while the corresponding backgrounds are dominated by gluons.  A wide variety of quark/gluon discriminants have been proposed \cite{Nilles:1980ys,Jones:1988ay,Fodor:1989ir,Jones:1990rz,Lonnblad:1990qp,Pumplin:1991kc,Gallicchio:2011xq,Gallicchio:2012ez,Krohn:2012fg,Pandolfi:1480598,Chatrchyan:2012sn,Larkoski:2013eya,Larkoski:2014pca,Bhattacherjee:2015psa,FerreiradeLima:2016gcz,Bhattacherjee:2016bpy,Komiske:2016rsd,Davighi:2017hok}, and there is a growing catalog of quark/gluon studies at the Large Hadron Collider (LHC) \cite{Aad:2014gea,Aad:2014bia,Khachatryan:2014dea,Aad:2015owa,Khachatryan:2015bnx,Aad:2016oit}.

In order to achieve robust quark/gluon tagging, though, one needs theoretical and experimental control over quark/gluon radiation patterns.  At the level of eikonal partons, a hard quark radiates soft gluons proportional to its $C_F = 4/3$ color factor while a hard gluon radiates soft gluons proportional to $C_A = 3$, and quark/gluon tagging performance is simply a function of $C_A/C_F$.  As we will see, quark/gluon discrimination performance is highly sensitive to  perturbative effects beyond the eikonal limit, such as $g \to q \overline{q}$ splittings and color coherence, as well as to nonperturbative effects such as color reconnection and hadronization.   While these effects are modeled (to differing degrees) in parton-shower generators, they are relatively unconstrained by existing collider measurements, especially in the gluon channel.

The goal of this paper is to highlight these uncertainties in quark/gluon tagging, using both parton-shower generators and first-principles calculations.  We start in the idealized context of electron-positron collisions, where one can study final-state quark/gluon radiation patterns in the absence of initial-state complications.  Here, we find modest differences in the predicted distributions for quark/gluon discriminants, which then translate to large differences in the predicted quark/gluon separation power.  Motivated by these uncertainties, we propose a set of LHC measurements that should help improve the modeling of jets in general and quark/gluon tagging in particular.  A summary and outline of this paper follows.

A common misconception about quark/gluon tagging is that it is an intrinsically ill-defined problem.  Of course, quark and gluon partons carry color while jets are composed of color-singlet hadrons, so the labels ``quark'' and ``gluon'' are fundamentally ambiguous. 
But this is philosophically no different from the fact that a ``jet'' is
fundamentally ambiguous and one must therefore always specify a concrete jet
finding procedure.  As discussed in \Sec{sec:def}, one can indeed create a
well-defined quark/gluon tagging procedure based on unambiguous hadron-level measurements.  In this way, even if what one means by ``quark'' or ``gluon'' is based on a naive or ambiguous concept (like Born-level cross sections or eikonal limits), quark/gluon discrimination is still a well-defined technique for enhancing desired signals over unwanted backgrounds.

In order to quantify quark/gluon discrimination power, there is a wide range of possible quark/gluon discriminants and a similarly large range of performance metrics, both discussed in \Sec{sec:preliminaries}.  As a concrete set of discriminants, we consider the generalized angularities $\lambda_\beta^\kappa$ \cite{Larkoski:2014pca} (see also \cite{Berger:2003iw,Almeida:2008yp,Ellis:2010rwa,Larkoski:2014uqa}),
\begin{equation}
\label{eq:genang_intro}
\lambda^{\kappa}_{\beta} = \sum_{i \in \text{jet}} z_i^\kappa \theta_i^\beta,
\end{equation}
with the notation to be explained in \Sec{sec:genang}.  We consider five different $(\kappa, \beta)$ working points, which roughly map onto five variables in common use in the literature:
\begin{equation}
\arraycolsep=5pt
\begin{array}{ccccc}
(0,0) & (2,0) & (1,0.5) & (1,1) & (1,2) \\
\text{multiplicity} &  p_T^D &  \text{LHA} & \text{width} & \text{mass}
\end{array}
\end{equation}
Here, multiplicity is the hadron multiplicity within the
jet, $p_T^D$ was defined in
\Refs{Pandolfi:1480598,Chatrchyan:2012sn}, LHA refers to the
``Les Houches Angularity'' (named after the workshop venue where this study was initiated \cite{Badger:2016bpw}),
width is closely related to jet broadening
\cite{Catani:1992jc,Rakow:1981qn,Ellis:1986ig}, and mass is closely
related to jet thrust \cite{Farhi:1977sg}.  To quantify discrimination
performance, we focus on classifier separation (a default output of the 
TMVA package \cite{2007physics...3039H}):
\begin{equation}
\label{eq:deltadef_intro}
\Delta =  \frac{1}{2} \int \text{d} \lambda \, \frac{\bigl(p_q(\lambda) - p_g(\lambda)\bigr)^2}{p_q(\lambda) + p_g(\lambda)},
\end{equation}
where $p_q$ ($p_g$) is the probability distribution for $\lambda$ in a generated quark jet (gluon jet) sample. This and other potential performance
metrics are discussed in \Sec{sec:classsep}. 

To gain a baseline analytic understanding, we use resummed
calculations in \Sec{sec:analytic} to provide a first-order
approximation for quark/gluon radiation patterns.  For $\kappa = 1$,
the generalized angularities are infrared and collinear (IRC) safe,
and therefore calculable in (resummed) perturbation theory.  At
leading-logarithmic (LL) accuracy, the IRC-safe angularities satisfy a
property called Casimir scaling, and the resulting classifier
separation $\Delta$ is a universal function of $C_A/C_F$, independent
of the value of $\beta$.  At present, the distributions for
generalized angularities are known to next-to-leading-logarithmic
(NLL) accuracy \cite{Larkoski:2013eya,Larkoski:2014pca}.  Here, we
include the resummation of the leading-color nonglobal logarithms
\cite{Dasgupta:2001sh}, though we neglect the resummation of pure
jet-radius logarithms \cite{Dasgupta:2014yra}, and soft
single-logarithmic corrections proportional to powers of the jet
radius.  These NLL calculations are effectively at parton-level, so to
obtain hadron-level distributions, we estimate the impact of
nonperturbative effects using shape functions
\cite{Korchemsky:1999kt,Korchemsky:2000kp}.

To gain a more realistic understanding with a full hadronization model, we use parton-shower generators in \Sec{sec:ee} to predict quark/gluon separation power.  In an idealized setup with $e^+e^-$ collisions, we can use the following processes as proxies for quark and gluon jets:
\begin{align}
\text{``quark jets''}: \quad & e^+e^- \to (\gamma/Z)^* \to u \overline{u}, \\
\text{``gluon jets''}: \quad & e^+e^- \to h^* \to g g,
\end{align}
where $h$ is the Higgs boson.  These processes are physically
distinguishable by the quantum numbers of the associated color-singlet
production operator, giving a way to define truth-level quarks and
gluons labels without reference to the final state.\footnote{Of course, the quantum numbers of the color singlet operator are not measurable event by event.  The idea here is to have a fundamental definition of ``quark'' and ``gluon'' that does not reference QCD partons directly.}  We
compare seven different parton-shower generators both before
hadronization (``parton level'') and after hadronization (``hadron
level''):
\begin{itemize}
\item \textsc{Pythia 8.215} \cite{Sjostrand:2014zea},
\item \textsc{Herwig++ 2.7.1} \cite{Bahr:2008pv,Bellm:2013hwb},\footnote{We use the default angular-ordered shower for these studies.  Subsequent to this paper, \Ref{Reichelt:2017hts} performed a study to improve quark/gluon modeling in \textsc{Herwig 7.1} \cite{Bellm:2015jjp,Bellm:2017bvx}.}
\item \textsc{Sherpa 2.2.1} \cite{Gleisberg:2008ta},
\item \textsc{Vincia 2.001} \cite{Fischer:2016vfv},
\item \textsc{Deductor 1.0.2} \cite{Nagy:2014mqa} (with hadronization performed by \textsc{Pythia 8.212}),\footnote{Note that this
\textsc{Deductor} plus \textsc{Pythia} combination has not yet been tuned to data.}
\item \textsc{Ariadne 5.0.$\beta$} \cite{Flensburg:2011kk},\footnote{This version of \textsc{Ariadne} is not yet public, but available from the author on request.  For $e^+ e^-$ collisions, the physics is the same as in \textsc{Ariadne 4} \cite{Lonnblad:1992tz}.}
\item \textsc{Dire 1.0.0} \cite{Hoche:2015sya} (with cluster hadronization performed by \textsc{Sherpa 2.1.1}).
\end{itemize}
To test other generators, the analysis code used for this study is available as a \textsc{Rivet} routine \cite{Buckley:2010ar}, which can be downloaded from \url{https://github.com/gsoyez/lh2015-qg}.

As we will see, the differences between these generators arise from
physics at the interface between perturbative showering and
nonperturbative fragmentation.  One might think that the largest
differences between generators would appear for IRC-unsafe observables
like multiplicity and $p_T^D$, where nonperturbative hadronization
plays an important role.  Surprisingly, comparably-sized differences
are also seen for the IRC-safe angularities, indicating that these
generators have different behavior even at the level of the
perturbative final-state shower.  In \Sec{sec:ee_scales}, we study
these differences as a function of the collision energy $Q$, the jet
radius $R$, and the strong coupling constant $\alpha_s$, showing that
the generators have somewhat different discrimination trends.  In
\Sec{sec:ee_settings}, we compare the default parton shower
configurations to physically-motivated changes, showing that modest
changes to the shower/hadronization parameters can give rather large
differences in quark/gluon separation power.

At the end of the day, most of the disagreement between generators is due to gluon radiation patterns.  This is not so surprising, since most of these generators have been tuned to reproduce distributions from $e^+ e^-$ colliders, and quark (but less so gluon) radiation patterns are highly constrained by event shape measurements at LEP \cite{Heister:2003aj,Abdallah:2003xz,Achard:2004sv,Abbiendi:2004qz}.  In \Sec{sec:pp}, we suggest a possible analysis strategy at the LHC to specifically constrain gluon radiation patterns.  At a hadron collider, the distinction between quark jets and gluon jets is rather subtle, since radiation patterns depend on color connections between the measured final-state jets and the unmeasured initial-state partons.  That said, we find that one can already learn a lot from hadron-level measurements, without trying to isolate ``pure'' quark or gluon samples.  In particular, we advocate measuring the generalized angularities on quark/gluon enriched samples:
\begin{align}
\text{``quark enriched''}: \quad & pp \to Z + \text{jet}, \\
\text{``gluon enriched''}: \quad & pp \to \text{dijets},
\end{align}
where ``enriched'' means that the Born-level process contributing to these channels is dominated by the corresponding jet flavor.  By making judicious kinematic cuts, we could further flavor-enrich these samples \cite{Gallicchio:2011xc}, though we will not pursue that in this paper for simplicity.

We present our final recommendations and conclusions in \Sec{sec:conclude}.  The main take home message from this study is that, contrary to the standard lore, the $e^+e^-$ measurements currently used for tuning are insufficient to constrain uncertainties in the final state shower.  There are alternative $e^+e^-$ measurements, however, that can play an important role in constraining gluon radiation patterns.  Ultimately, gluon-enriched measurements at the LHC will be crucial to achieve robust quark/gluon
discrimination.

\section{What is a quark/gluon jet?}
\label{sec:def}

\begin{figure}
\centering
\subfloat{
\includegraphics[width=0.8\columnwidth]{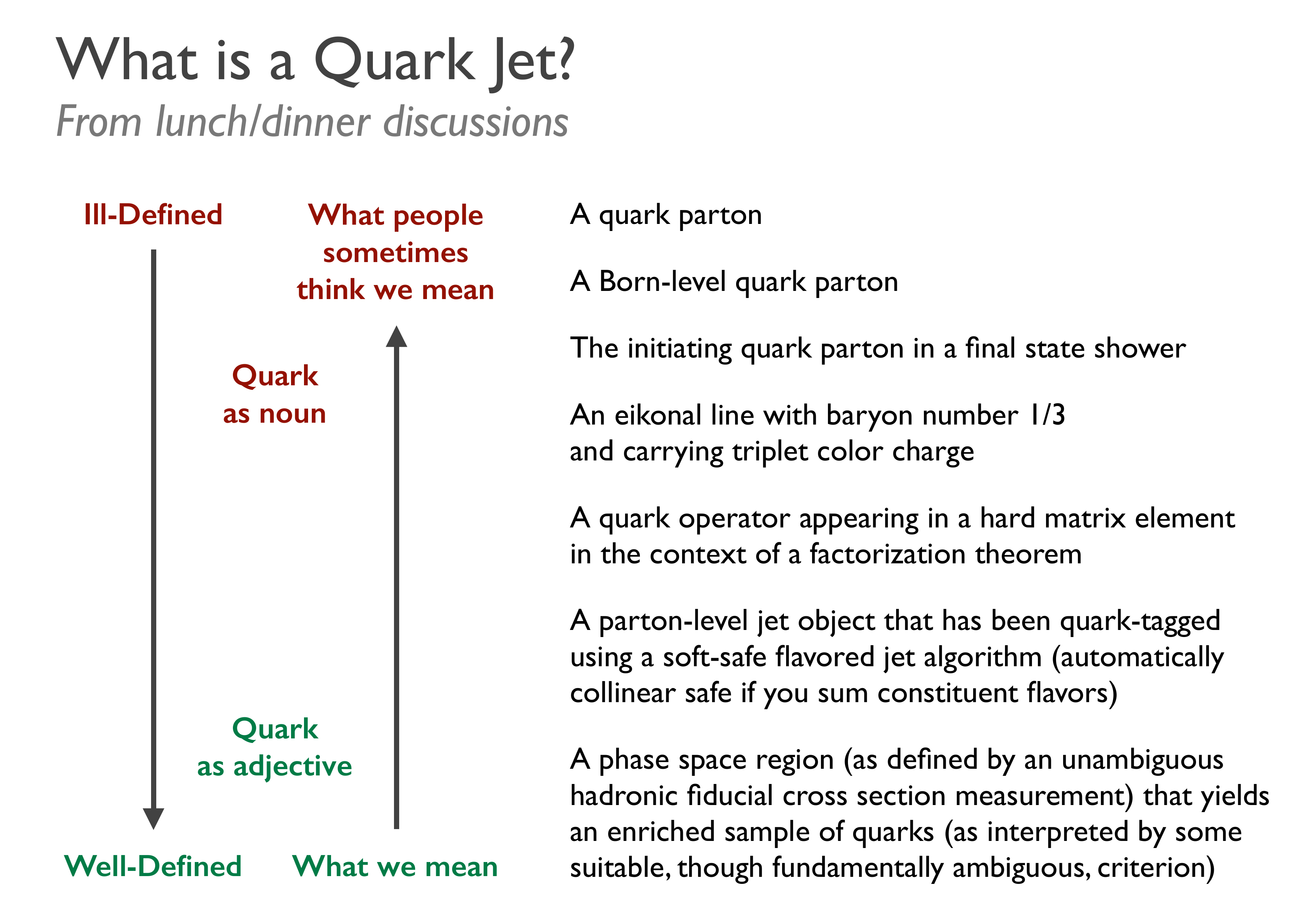}
}
\caption{Original slide from the June 10, 2015 summary report of the quark/gluon Les Houches subgroup \cite{Badger:2016bpw}.}
\label{fig:summary_slide}
\end{figure}

As part of the 2015 Les Houches workshop on ``Physics at TeV Colliders'' \cite{Badger:2016bpw}, an attempt was made to
define exactly what is meant by a ``quark jet'' or ``gluon jet'' (see \Fig{fig:summary_slide}).
Here are some suggested options for defining a quark jet, in
(approximate) order from most ill-defined to most well-defined.
Related statement can be made for gluon jets.

~\\
\noindent \textbf{A quark jet is...}
\begin{itemize}
\item \textbf{A quark parton.}  This definition (incorrectly) assumes that there is a one-to-one map between a jet and its initiating parton.  Because it neglects the important role of additional radiation in determining the structure of a jet, we immediately dismiss this definition.
\item \textbf{A Born-level quark parton.}  This definition at least acknowledges the importance of radiative corrections to jet production, but it leaves open the question of how exactly to define the underlying Born-level process from an observed final state.  (For one answer valid at the parton level, see flavored jet algorithms below.)
\item \textbf{An initiating quark parton in a final state parton
    shower.}  We suspect that this is the definition most LHC
  experimentalists have in mind.  This definition assumes that the parton-shower history is meaningful, though,
  which may not be the case beyond the strongly-ordered or
  LL approximations.  Because the parton shower is
  semi-classical, this definition neglects the impact of genuinely
  quantum radiative corrections as well as nonperturbative
  hadronization.
\item \textbf{A maximum-$p_T$ quark parton within a jet in a final state parton shower.}  This definition uses the hardest parton within the active jet area encountered at any stage of the shower evolution, including the initial hard scattering process.  This ``max-$p_T$'' prescription is a variant on the initiating parton prescription above (see further discussion in \Ref{Buckley:2015gua}).  It differs from the initiating parton by a calculable amount in a LL shower \cite{Dasgupta:2014yra} and is based on the same (naive) assumption that the parton-shower history is meaningful. 
\item \textbf{An eikonal line with baryon number 1/3 and carrying triplet color charge.}  This is another semi-classical definition that attempts to use a well-defined limit of QCD to define quarks in terms of light-like Wilson lines.  Philosophically, this is similar to the parton-shower picture, with a similar concern about how to extrapolate this definition away from the strict eikonal limit.
\item \textbf{A parton-level jet object that has been quark-tagged using an IRC-safe flavored jet algorithm.}  This is the strategy adopted in \Ref{Banfi:2006hf}.  While this definition neglects the impact of hadronization, it does allow for the calculation of quark jet cross sections at all perturbative orders, including quantum corrections.
\end{itemize}
The unifying theme in the above definitions is that they try to identify a quark as an object unto itself, without reference to the specific final state of interest.  However, it is well-known that a ``quark'' in one process may not look like a ``quark'' in other process, due to color correlations with the rest of the event, especially the initial state in $pp$ collisions.  The next definition attempts to deal with the process dependence in defining quarks. 
\begin{itemize}
\item \textbf{A quark operator appearing in a hard matrix element in the context of a factorization theorem.}  This is similar to the attitude taken in \Ref{Gallicchio:2011xc}.  In the context of a well-defined cross section measurement, one can (sometimes) go to a limit of phase space where the hard production of short-distance quarks and gluons factorizes from the subsequent long-distance fragmentation.  This yields a nice (gauge-covariant) operator definition of a quark jet, which can be made precise for observables based on jet grooming \cite{Frye:2016okc,Frye:2016aiz}.  That said, even if a factorization theorem does exist for the measurement of interest, this definition is potentially ambiguous beyond leading power.
\end{itemize}
The definition we adopt for this study is inspired by the idea that one should think about quark/gluon tagging in the context of a specific measurement, regardless of whether the observable in question has a rigorous factorization theorem.
\begin{itemize}
\item \textbf{A phase space region (as defined by an unambiguous
    hadronic fiducial cross section measurement) that yields an
    enriched sample of quarks (as interpreted by some suitable, though
    fundamentally ambiguous, criterion).}  Here, the goal is to
  \emph{tag} a phase space region as being quark-like, rather than try
  to determine a truth definition of a quark.  This definition has the
  advantage of being explicitly tied to hadronic final states and to
  the discriminant variables of interest. \emph{The main
  challenge with this definition is how to determine the criterion
  that corresponds to successful quark enrichment.}  For that, we
  have to rely to some degree on the other less well-defined notions
  of what a quark jet is.
\end{itemize}

To better understand this last definition, consider a quark/gluon discriminant $\lambda$.  Since $\lambda$ can be measured on any jet, one can unambiguously determine the cross section $\text{d} \sigma / \text{d} \lambda$ for any jet sample of interest.  But measuring $\lambda$ does not directly lead to the probability that the jet is a quark jet, nor to the probability distribution  $p_q(\lambda)$ for $\lambda$ within a quark jet sample.  Rather, the process of measuring $\lambda$ must be followed by a separate process of interpreting how the value of $\lambda$ should be used as part of an analysis.

For example, the user could choose that small $\lambda$ jets should be tagged as ``quark-like'' while large $\lambda$ jets should be tagged as ``gluon-like''. Alternatively, the user might combine $\lambda$ with other discriminant variables as part of a more sophisticated classification scheme.  The key point is that one first measures hadron-level discriminant variables on a final state of interest, and only later does one interpret exactly what those discriminants accomplish (which could be different depending on the physics goals of a specific analysis).  Typically, one might use a Born-level or eikonal analysis to define which regions of phase space should be associated with ``quarks'' or ``gluons'', but even if these phase space regions are based on naive or ambiguous logic, $\lambda$ itself is a well-defined discriminant variable.

In \Sec{sec:ee}, we will consider the generalized
angularities $\lambda_{\beta}^\kappa$ as our discriminant variables
and we will assess the degree to which the measured values of
$\lambda_{\beta}^\kappa$ agree with a quark/gluon interpretation based
on Born-level production modes.  This is clearly an idealization,
though one that makes some sense in the context of $e^+e^-$
collisions, since  truth-level ``quark'' and ``gluon'' labels can be
defined by the Lorentz structure of the production vertex.  In
\Sec{sec:pp}, we will recommend that the LHC
experiments perform measurements of $\lambda_\beta^\kappa$ in
well-defined hadron-level final states, without necessarily attempting
to determine separate $p_q(\lambda_\beta^\kappa)$ and
$p_g(\lambda_\beta^\kappa)$ distributions.  Eventually, one would want
to use these hadron-level measurements to infer something about
parton-level quark/gluon radiation patterns.  Even without that
interpretation step, though, direct measurements of $\text{d} \sigma /
\text{d} \lambda_\beta^\kappa$ would provide valuable information for
parton-shower tuning.  This in turn would help $\lambda_\beta^\kappa$ become a more robust and powerful discriminant in searches for new physics beyond the standard model. 

\section{Quantifying tagging performance}
\label{sec:preliminaries}

\subsection{Generalized angularities}
\label{sec:genang}

\begin{figure}
\centering
\subfloat{
\includegraphics[scale = 0.7]{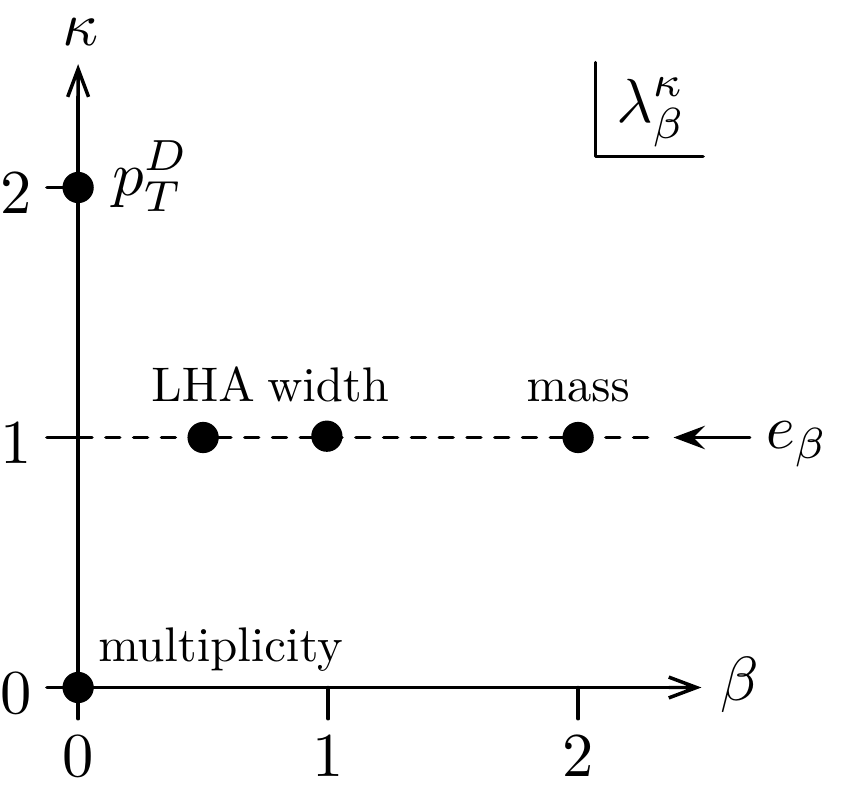}
}
\caption{Two-parameter family of generalized angularities, adapted from \Ref{Larkoski:2014pca}.  The dots correspond to the five benchmark angularities used in this study, with ``LHA'' referring to the Les Houches Angularity.  The horizontal line at $\kappa = 1$ corresponds to the IRC-safe angularities, $e_\beta = \lambda^{1}_{\beta}$.}
\label{fig:lambda_space}
\end{figure}

A wide variety of quark/gluon discriminants have been proposed (see \Ref{Gallicchio:2012ez} for an extensive catalog), but here we limit ourselves to a two-parameter family of generalized angularities \cite{Larkoski:2014pca}, shown in \Fig{fig:lambda_space}.  These are defined as (repeating \Eq{eq:genang_intro} for convenience)
\begin{equation}
\label{eq:genang}
\lambda^{\kappa}_{\beta} = \sum_{i \in \text{jet}} z_i^\kappa \theta_i^\beta,
\end{equation}
where $i$ runs over the jet constituents, $z_i \in [0,1]$ is a momentum fraction, and $\theta_i \in [0,1]$ is a (normalized) angle to the jet axis.  The parameters $\kappa \ge 0$ and $\beta \ge 0$ determine the momentum and angle weighting, respectively.  For $\kappa = 1$, the generalized angularities are IRC safe and hence calculable in perturbation theory \cite{Larkoski:2014uqa} (see also \cite{Ellis:2010rwa,Larkoski:2013paa,Larkoski:2014tva,Procura:2014cba,Hornig:2016ahz}), and we will sometimes use the shorthand
\be
e_\beta \equiv \genang{1}{\beta}. 
\ee
For general $\kappa \not= 1$, there are quasi-perturbative techniques based on generalized fragmentation functions \cite{Larkoski:2014pca} (see also \cite{Krohn:2012fg,Waalewijn:2012sv,Chang:2013rca,Chang:2013iba}).  In our parton-shower studies, we determine $\lambda^{\kappa}_{\beta}$ using all constituents of a jet, though one could also consider using charged-particle-only angularities to improve robustness to pileup (at the expense of losing some particle-level information).

For our $e^+ e^-$ study, we cluster jets with \textsc{FastJet 3.2.1} \cite{Cacciari:2005hq,Cacciari:2011ma} using the $ee$-variant of the
anti-$k_t$ algorithm \cite{Cacciari:2008gp}, with $|\vec{p}|$-ordered
winner-take-all recombination
\cite{Larkoski:2014uqa,Bertolini:2013iqa,Salam:WTAUnpublished} to
determine the jet axis $\hat{n}$.  Unlike standard $E$-scheme
recombination \cite{Blazey:2000qt}, the winner-take-all scheme yields
a jet axis $\hat{n}$ that does not necessarily align with the jet
three-momentum $\vec{p}$; this turns out to be a desirable feature
for avoiding soft recoil effects
\cite{Larkoski:2013eya,Larkoski:2014uqa,Catani:1992jc,Dokshitzer:1998kz,Banfi:2004yd}.  We define
\begin{equation}
z_i \equiv \frac{E_i}{E_{J}}, \qquad \theta_i \equiv \frac{\Omega_{i \hat{n}}}{R},
\end{equation}
where $E_J$ is the jet energy, $E_i$ is the particle energy, $\Omega_{i \hat{n}}$ is the
opening angle to the jet axis, and $R$ is the jet radius (taken to be
$R = 0.6$ by default, unless explicitly mentioned otherwise).

For our $pp$ study, we use the standard $pp$ version of anti-$k_t$ with  $E$-scheme recombination, defining
\begin{equation}\label{eq:def-z-theta-pp}
z_i \equiv \frac{p_{Ti}}{\sum_{j \in \text{jet}} p_{Tj}}, \qquad \theta_i \equiv \frac{R_{i \hat{n}}}{R},
\end{equation}
where $p_{Ti}$ is the particle transverse momentum and $R_{i \hat{n}}$
is the rapidity-azimuth distance to the jet axis.  To define a
recoil-free axis, we recluster the jet using the Cambridge/Aachen
(C/A) algorithm \cite{Dokshitzer:1997in,Wobisch:1998wt} with
$p_T$-ordered winner-take-all recombination. Note that with this
choice of recombination scheme, the $p_T$ of the recoil-free axis is effectively
the scalar sum $\sum_{j \in \text{jet}} p_{Tj}$ used in
\Eq{eq:def-z-theta-pp}. In addition to directly measuring the
angularities, we also want to test the impact of jet grooming (see
e.g.~\cite{Butterworth:2008iy,Ellis:2009su,Ellis:2009me,Krohn:2009th}).
As one grooming example, we use the modified mass drop tagger (mMDT)
with $\mu = 1$ \cite{Butterworth:2008iy,Dasgupta:2013ihk}
(equivalently, soft drop declustering with $\beta = 0$
\cite{Larkoski:2014wba}).  This grooming procedure starts from the
C/A-reclustered jet, which yields an angular-ordered clustering tree.
This tree is then declustered, removing the softer branch until \be
\label{eq:mMDTcriteria}
\frac{\min[p_{T1}, p_{T2}]}{p_{T1} + p_{T2}} > z_{\rm cut},
\ee
where 1 and 2 label the two branches of a splitting.  By applying the mMDT procedure, we can test how quark/gluon discrimination performance and robustness is affected by removing soft radiation from a jet.  For concreteness, we always set $z_{\rm cut} = 0.1$ to match the studies in \Refs{Larkoski:2014wba,Larkoski:2014bia,Larkoski:2015lea,CMS-PAS-HIN-16-006}.

By adjusting $\kappa$ and $\beta$ in the angularities, one can probe different aspects of the jet fragmentation.  We consider five benchmark values for $(\kappa, \beta)$ indicated by the black dots in \Fig{fig:lambda_space}:
\begin{equation}
\label{eq:benchmarkang}
\begin{aligned}
(0,0) &= \text{hadron multiplicity},\\
(2,0) &\Rightarrow p_T^D \text{  \cite{Pandolfi:1480598,Chatrchyan:2012sn} (specifically $\lambda^{2}_{0} = (p_T^D)^2$)},\\
(1,0.5) & = \text{Les Houches Angularity (LHA)},\\
(1,1) &= \text{width or broadening \cite{Catani:1992jc,Rakow:1981qn,Ellis:1986ig}},\\
(1,2) & \Rightarrow \text{mass or thrust \cite{Farhi:1977sg}
  (specifically $\lambda^{1}_{2} \simeq m_{\rm jet}^2 / E_{\rm
    jet}^2$)}.
\end{aligned}
\end{equation}
Except for the LHA, these angularities (or their close cousins) have
already been used in quark/gluon discrimination studies.  The LHA has
been included to have an IRC safe angularity that weights energies
more heavily than angles, similar in spirit to the $\beta = 0.2$ value
advocated in \Ref{Larkoski:2013eya} for energy correlation
functions. Most of the results in this paper are shown in terms of the
LHA; results for the other four benchmark values are available in the
source files for the \texttt{arXiv} preprint \cite{ArXivSource}, where each figure in this paper
corresponds to multipage file.

For the IRC-safe case of $\kappa = 1$, there is an alternative version
of the angularities based on energy correlation functions \cite{Larkoski:2013eya} (see also \cite{Banfi:2004yd,Jankowiak:2011qa,Moult:2016cvt}),
\begin{equation}
\text{ecf}_\beta = \sum_{i < j \in \text{jet}} z_i z_j \theta_{ij}^\beta \simeq \lambda^{1}_{\beta},
\end{equation}
where equality holds in the extreme eikonal limit.\footnote{This equality also relies on using a recoil-free axis choice $\hat{n}$ to define $\theta_i$.  Amusingly, $\lim_{\beta \to 0} \text{ecf}_\beta = (1 - \lambda^{2}_{0})/2$ (i.e.~$\kappa = 2$, $\beta = 0$), such that the $\beta \to 0$ limit of the IRC-safe energy correlation functions corresponds to the IRC-unsafe $p_T^D$.}  For the $e^+ e^-$ case, the pairwise angle $\theta_{ij}$ is typically normalized to the jet radius as $\theta_{ij} \equiv \Omega_{ij}/R$.   To avoid a proliferation of curves, we will not show any results for $\text{ecf}_\beta$.  We will also neglect quark/gluon discriminants that take into account azimuthal asymmetries within the jet, though observables like the covariance tensor \cite{Gallicchio:2012ez} and 2-subjettiness \cite{Thaler:2010tr,Thaler:2011gf,Salam:2016yht} can improve quark/gluon discrimination.  See \Ref{FerreiradeLima:2016gcz} for a related study of quark/gluon systematics for shower deconstruction \cite{Soper:2011cr,Soper:2012pb,Soper:2014rya} and energy correlation functions \cite{Larkoski:2013eya}.

\subsection{Classifier separation}
\label{sec:classsep}

Since we will be testing many parton-shower variants, we need a way to
quantify quark/gluon separation power in a robust way that can easily
be summarized by a single number.  For that purpose we use classifier
separation  (repeating and reorganizing \Eq{eq:deltadef_intro} for convenience),
\begin{equation}
\label{eq:deltadef}
\Delta
 =  \frac{1}{2} \int \text{d} \lambda \,
    \frac{\bigl(p_q(\lambda) - p_g(\lambda)\bigr)^2}{p_q(\lambda) + p_g(\lambda)}
 =  1- 2\int \text{d} \lambda \, \frac{p_q(\lambda) \, p_g(\lambda)}{p_q(\lambda) + p_g(\lambda)},
\end{equation}
where $p_q$ ($p_g$) is the probability distribution for the quark jet (gluon jet) sample as a function of the classifier $\lambda$.  Here, $\Delta = 0$ corresponds to no discrimination power and $\Delta = 1$ corresponds to perfect discrimination power.

\begin{figure}
\centering
\subfloat[]{
\label{fig:roc_curve}
\includegraphics[scale = 0.8]{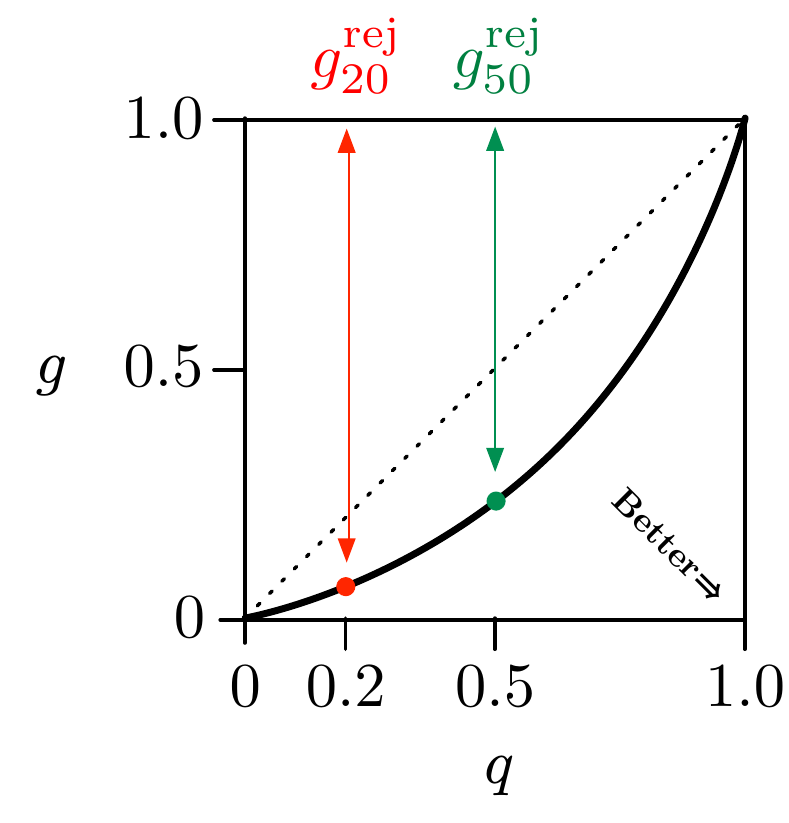}
}
$\qquad$
$\qquad$
\subfloat[]{
\label{fig:truth_overlap}
\includegraphics[scale = 0.8]{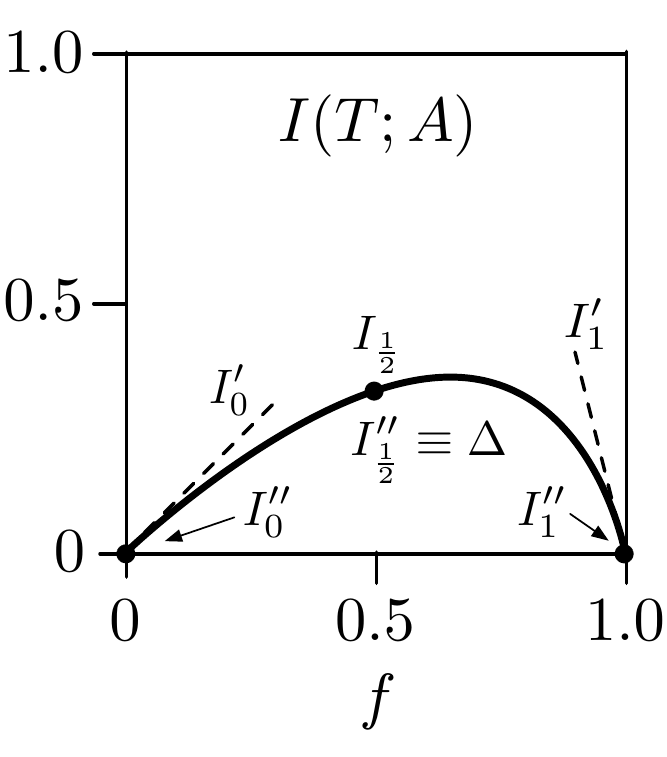}
}
\caption{Alternative metrics for discrimination power.  (a) A ROC curve, showing the gluon rejection rate at fixed 20\% and 50\% quark efficiency.  While not shown in this paper, some of this ROC information is available from the \texttt{arXiv} preprint source files \cite{ArXivSource}.  (b) Mutual information $I(T;A)$ as a function of the quark fraction $f$, showing the relationship to classifier separation $\Delta \equiv I_{\frac{1}{2}}'' $ and other information-theoretic quantities.  The mutual information at $f = 1/2$ (i.e.~$I_{\frac{1}{2}}$) is also available from the \texttt{arXiv} preprint source files \cite{ArXivSource}.}
\end{figure}

A more common way to talk about discrimination power is in terms of receiver operating characteristic (ROC) curves, shown in \Fig{fig:roc_curve}.  At a point ($q$,$g$) on the ROC curve, where $q,g \in [0,1]$, one can define a selection that yields $q$ efficiency for quarks and $g$ mistag rate for gluons, or equivalently, a $(1-g)$ efficiency for gluons for a $(1-q)$ mistag rate for quarks.  There are various ways to turn the ROC curve into a single number, and in the \texttt{arXiv} preprint source files \cite{ArXivSource}, every figure for $\Delta$ is part of a multipage file that also has results for
\begin{align}
\{ g^{\rm  rej}_{20} ,   g^{\rm  rej}_{50}\} &: \quad \text{Gluon rejection rate at \{20\%, 50\%\} quark efficiency;}\\
\{ q^{\rm  rej}_{20} ,   q^{\rm  rej}_{50}\} &: \quad \text{Quark rejection rate at \{20\%, 50\%\} gluon efficiency;}\\
s^{\rm rej} &: \quad \text{Symmetric rejection rate at $s^{\rm rej}$ efficiency.}
\end{align}
Since we are more interested in understanding the relative performance
between parton showers rather than the absolute performance, we will not show full ROC curves in this paper, though they can be easily derived from the $p_q$ and $p_g$ distributions.  If one observable has an everywhere better ROC curve than another (i.e.~it is Pareto optimal), then it will also have a larger $\Delta$ value.  The converse is not true, however, since depending on the desired working point, a ``bad'' discriminant as measured by $\Delta$ might still be ``good'' by another metric.  In that sense, $\Delta$ contains less information than the full ROC curve.

An alternative way to quantify discrimination power is through mutual information, which counts the number of ``bits'' of information gained from measuring a discriminant variable (see \Ref{Larkoski:2014pca}).  Given a sample with quark fraction $f \in [0,1]$ and gluon fraction $(1-f)$, the mutual information with the truth (a.k.a.\ the truth overlap) is
\begin{equation}
I(T; \Lambda) = \int \text{d} \lambda \left(f \, p_q(\lambda) \log_2 \frac{p_q(\lambda)}{p_{\rm tot}(\lambda)} + (1-f) \, p_g(\lambda) \log_2 \frac{p_g(\lambda)}{p_{\rm tot}(\lambda)}   \right),
\end{equation}
where $T = \{q,g\}$ is the set of truth labels, $\Lambda = \{\lambda\}$ is the (continuous) set of discriminant values, and 
\begin{equation}
p_{\rm tot}(\lambda) = f \, p_q(\lambda) + (1-f) \, p_g(\lambda).
\end{equation}
The choice $f = \frac{1}{2}$ was used in \Ref{Larkoski:2014pca} and is also available in the \texttt{arXiv} preprint source files \cite{ArXivSource},
\be
I(T;A)\big|_{f = \frac{1}{2}} \equiv I_{\frac{1}{2}},
\ee 
though other $f$ choices are plausible.

Though we will not use mutual information in this study, it is amusing to note that the second derivative of $I(T;\Lambda)$ with respect to $f$ is related to classifier separation as
\begin{equation}
\label{eq:altdeltadef}
- \frac{\log 2}{4} \frac{\partial^2 I(T;\Lambda)}{\partial f^2} \Big|_{f = \frac{1}{2}}  \equiv I''_\frac{1}{2} = \Delta.
\end{equation}
More broadly, the dependence of $I(T;A)$ on $f$ can be related to other concepts in statistics, as visualized in \Fig{fig:truth_overlap}.  At $f = 0$ and $f = 1$, the mutual information itself is zero, but the derivatives are:
\begin{align}
\frac{\partial I}{\partial f} \Big|_{f = 0} &\equiv I'_0 = \int \df \lambda \, p_q(\lambda)  \log_2 \frac{p_q(\lambda)}{p_g(\lambda)}, \quad &- \log 2 \,  \frac{\partial^2 I}{\partial f^2} \Big|_{f = 0} &\equiv I''_0 = \int \df \lambda \,  \frac{p_q(\lambda)^2}{p_g(\lambda)},\\
- \frac{\partial I}{\partial f} \Big|_{f = 1} &\equiv I'_1 = \int \df \lambda \, p_g(\lambda) \log_2 \frac{p_g(\lambda)}{p_q(\lambda)}, \quad &- \log 2 \, \frac{\partial^2 I}{\partial f^2} \Big|_{f = 1} &\equiv I''_1 = \int \df \lambda \, \frac{p_g(\lambda)^2}{p_q(\lambda)}.
\end{align}
The first derivative is sometimes called relative entropy and the second derivative is sometimes called discrimination significance.  Unlike classifier separation, these later  metrics do not treat quark and gluon distributions symmetrically.

One advantage of $\Delta$ over $I(T;\Lambda)$ is that the integrand in \Eq{eq:deltadef} is easier to interpret, since it tracks the fractional difference between the signal and background at a given value of $\lambda$.\footnote{Another advantage of $\Delta$ over $I(T; \Lambda)$ arises when trying to assign statistical uncertainties to finite Monte Carlo samples.  Since $\Delta$ is defined as a simple integral, one can use standard error propagation to assign uncertainties to $\Delta$.  By contrast, because of the logarithms in the $I(T; \Lambda)$ integrand, one has to be careful about a potential binning bias \cite{Larkoski:2014pca}.}  Specifically, by plotting 
\begin{equation}
\label{eq:deltaintegrand}
\frac{\text{d} \Delta}{\text{d} \lambda} = \frac{1}{2} \frac{\bigl(p_q(\lambda) - p_g(\lambda) \bigr)^2}{p_q(\lambda) + p_g(\lambda)},
\end{equation}
one can easily identify which regions of phase space contribute the most to quark/gluon discrimination.  One can then ask whether or not the regions exhibiting the most separation power are under sufficient theoretical control, including both the size of perturbative uncertainties and the impact of nonperturbative corrections.  

\section{Analytic quark/gluon predictions}
\label{sec:analytic}

For the IRC-safe angularities with $\kappa = 1$ (namely $e_{0.5}$, $e_1$, and $e_2$ from \Eq{eq:benchmarkang}), we can use analytic calculations to get a baseline expectation for the degree of quark/gluon separation.  At LL accuracy, a jet effectively consists of a single soft gluon emission from a hard quark/gluon, with a suitable Sudakov form factor coming from vetoing additional radiation.  At this order, the strong coupling constant is fixed and only the leading splitting function is used.  In particular, the IRC-safe angularities at LL order satisfy a property called Casimir scaling (reviewed below), such that the discrimination power is independent of $\beta$.

At NLL order, the jet is described by multiple gluon emissions from a hard quark/gluon, including the effects of $\alpha_s$ running and subleading terms in the splitting function, but neglecting matrix element corrections and energy-momentum conservation.  For the IRC-safe angularities, the NLL-accurate distributions were calculated in \Refs{Larkoski:2013eya,Larkoski:2014pca} (see also \cite{Berger:2003iw,Banfi:2004yd,Hornig:2009vb,Ellis:2010rwa,Larkoski:2014uqa}).  For the generalized angularities, generalized fragmentation functions \cite{Krohn:2012fg,Waalewijn:2012sv,Chang:2013rca,Chang:2013iba,Larkoski:2014pca} were used to extend the NLL calculation beyond the IRC-safe regime, though we will not use that technique in the present paper.  To date, the impact of soft nonperturbative physics has not been included in the distributions for the IRC-safe angularities, but we do so below.  

\subsection{Casimir scaling at LL}

As shown in \Refs{Larkoski:2013eya,Larkoski:2014pca}, the IRC-safe angularities satisfy Casimir scaling at LL accuracy, which implies that the quark/gluon discrimination performance only depends on the color factor ratio $C_A/C_F$.  To see this, let us first introduce the notation for the cumulative distribution $\Sigma(\lambda)$, which is defined by
\be
\Sigma(\lambda) = \int_0^\lambda \df \lambda' \, p(\lambda'), \qquad p(\lambda) = \frac{\df \Sigma}{\df \lambda}.
\ee
If an observable satisfies Casimir scaling, then the quark and gluon cumulative distributions can be written as
\be
\label{eq:casimirform}
\Sigma_q(\lambda) = e^{-C_F \, r(\lambda)}, \qquad \Sigma_g(\lambda) = e^{-C_A \, r(\lambda)},
\ee
where $r(\lambda)$ is a monotonically decreasing function of $\lambda$.  Here, the only difference between the quark and gluon distributions is in the color factors $C_F = 4/3$ versus $C_A = 3$.

At LL accuracy, the distributions for the IRC-safe angularities take
precisely this form
\cite{Larkoski:2013eya,Larkoski:2014pca},\footnote{Strictly speaking,
  \Eq{eq:LLform} is only valid in the fixed-coupling
  approximation. Running-coupling corrections already arise at
   LL accuracy, replacing $\alpha_s\log^2e_\beta$ by an all-orders
  series $g_1(\alpha_s\log e_\beta)\log e_\beta$. This does not
  affect the property of Casimir scaling.}
\be
\label{eq:LLform}
\Sigma_i^{\rm LL}(e_\beta) = \exp \left[ - \frac{\alpha_s C_i}{\pi \beta} \log^2 e_\beta  \right],
\ee
where $i$ labels the jet flavor.  This LL result can be understood from the fact that quarks and gluons have the same leading splitting function up to an overall multiplicative color factor
\be
\label{eq:splitting_function_Casimir}
P_i(z) \simeq \frac{2 C_i}{z},
\ee
and therefore the Sudakov form factor (which is what appears in \Eq{eq:LLform}) differs only by the color factor in the exponent.

Observables that satisfy Casimir scaling have universal ROC curves \cite{Larkoski:2013eya} and universal truth overlaps \cite{Larkoski:2014pca}, which are independent of the precise functional form of $r(\lambda)$ and only depends on the ratio $C_A/C_F$.  We can derive the same universality for classifier separation.  Using
\be
p_i(\lambda) = - C_i \, r'(\lambda) \, e^{-C_i \, r(\lambda)}
\ee
and the change of variables $u \equiv e^{-C_F \, r(\lambda)}$, we have
\begin{align}
\Delta &=  -\frac{1}{2} \int \df \lambda \, r'(\lambda) \frac{\bigl(C_F \, e^{-C_F \, r(\lambda)} - C_A \, e^{-C_A \, r(\lambda)}\bigr)^2}{C_F\,  e^{-C_F\,  r(\lambda)}+ C_A \, e^{-C_A \, r(\lambda)}} , \\
& = \frac{1}{2} \int_0^1 \frac{\df u}{u} \frac{\bigl(u - (C_A / C_F) u^{C_A/C_F}\bigr)^2}{u + (C_A/C_F) u^{C_A/C_F}} , \\
& = 2 \left({}_2F_1\left[1,\frac{C_F}{C_A - C_F}; \frac{C_A}{C_A - C_F}; - \frac{C_A}{C_F}\right] - \frac{1}{2}  \right),
\end{align}
where ${}_2F_1[a,b;c;z]$ is the hypergeometric function.\footnote{An alternative way to derive this result is to take $I(T;A)$ from \Ref{Larkoski:2014pca} and use \Eq{eq:altdeltadef} to extract $\Delta$.}  For the case of QCD with $C_A/C_F = 9/4$,
\be
\label{eq:LLbenchmark}
\Delta_{\rm QCD} \simeq  0.1286.
\ee
We mark this benchmark value with an arrow on the subsequent plots for reference.\footnote{In large $N_c$ QCD where $C_A/C_F \to 2$, $\Delta \to \ln 3 - 1 \simeq 0.0986$, so quark/gluon separation is expected to be more challenging as $N_c \to \infty$.} 

Going beyond LL accuracy, Casimir scaling is typically violated, and $\Delta$ depends on the precise observable in question.  Thus, all of the differences we see in our subsequent studies are effects that are truly higher-order or nonperturbative.

\subsection{NLL resummation}

Going to NLL accuracy is straightforward for global logarithms, using the formalism of \Ref{Banfi:2004yd}.  Nonglobal logarithms have not been included in previous angularity calculations, but we include them here using their numerical extraction in the large $N_c$ limit \cite{Dasgupta:2001sh}.  We always assume that the jet radius $R$ is order 1 so we can ignore $\log R$ resummation \cite{Dasgupta:2014yra}.

The cumulative distribution for an IRC safe angularity $e_\beta$ takes the form \cite{Banfi:2004yd,Dasgupta:2001sh}
\be
\label{eq:NLLcaesar}
\Sigma(e_\beta) =  \frac{e^{-\gamma_E R'(e_\beta)}}{\Gamma\left(1+R'(e_\beta)\right)} \, e^{-R(e_\beta)} \, e^{-f_{\rm NGL}(e_\beta)},
\ee
where $R(e_\beta)$ is known as the radiator function, $f_{\rm NGL}(e_\beta)$ encodes nonglobal logarithms, $\gamma_E$ is the Euler-Mascheroni constant, $\Gamma$ is the gamma function, and primes indicate logarithmic derivatives:
\be
\label{eq:R_der}
R'(e_\beta) = -\frac{\partial}{\partial \log e_\beta} R(e_\beta).  \ee
The $e^{-R(e_\beta)}$ factor in \Eq{eq:NLLcaesar} is just the Sudakov
form factor, which exhibits Casimir scaling at LL accuracy, and the
prefactor containing $R'(e_\beta)$ captures the effect of multiple
emissions on the $e_\beta$ distribution.  The
$e^{-f_{\rm NGL}(e_\beta)}$ factor comes from Eq.~(18) of
\Ref{Dasgupta:2001sh}, where it is called $\mathcal{S}$.  Note that
$f_{\rm NGL}$ is proportional to $C_i \,C_A$, so it effectively
preserves Casimir scaling; for this reason, the inclusion of nonglobal
logarithms is not expected to have a large impact on quark/gluon
separation power.

For the IRC-safe angularities, the radiator function is \cite{Larkoski:2013eya,Larkoski:2014pca}
\begin{equation}
\label{eq:radiatorC1}
R(e_\beta) = C_i \int_0^{1} \frac{\df \theta}{\theta}\int_0^1 \df  z\, p_i(z) \frac{\alpha_s(k_t)}{\pi} \Theta\left(z \theta^\beta -e_\beta \right),
\end{equation}
and the strong coupling is evaluated with two-loop running at the $k_t$ emission scale
\be
\label{eq:ktscale}
k_t = z \, \theta \,  R \, E_J.
\ee
The reduced splitting functions (i.e.~splitting functions summed over all allowed $1 \to 2$ processes) are 
\be
\label{eq:reducedsplitting}
p_q(z) = \frac{1 + (1-z)^2}{z}, \qquad p_g(z) = 2 \frac{1-z}{z} + z(1-z) + \frac{T_R \, n_f}{C_A} (z^2 + (1-z)^2),
\ee
where $T_R = 1/2$ and we take the number of light quark flavors to be $n_f = 5$.  Following \Ref{Larkoski:2014pca}, both $R(e_\beta)$ and $R'(e_\beta)$ are truncated to only keep terms that are formally of NLL accuracy.

\subsection{Nonperturbative shape function}
\label{subsec:shapefuncdef}

The quark/gluon studies in \Refs{Larkoski:2013eya,Larkoski:2014pca} used solely the distributions as calculated in \Eq{eq:NLLcaesar} (with $f_{\rm NGL} = 0$).  As we will see, nonperturbative hadronization has a big effect in our parton-shower studies, so we would like to include the corresponding effect in our analytic results.

The IRC-safe angularities are additive observables, meaning that at leading power in the small $e_\beta$ limit, one can decompose them into separate contributions from perturbative and nonperturbative modes:
\be
e_\beta \simeq e_\beta^{\text{(pert)}} + e_\beta^{\text{(NP)}}.
\ee
One can then convolve the perturbative distribution for $\hat{e}_\beta \equiv e_\beta^{\text{(pert)}}$ with a nonperturbative shape function $F$ that describes the distribution of $\epsilon \equiv e_\beta^{\text{(NP)}}$ \cite{Korchemsky:1999kt, Korchemsky:2000kp} (see also \cite{Manohar:1994kq, Dokshitzer:1995zt, Salam:2001bd, Lee:2006nr, Mateu:2012nk}),
\be
\label{eq:naiveconvolution}
\frac{\df \sigma}{\df e_\beta} = \int \df \hat{e}_\beta \, \df \epsilon \,  \hat{\sigma}(\hat{e}_\beta) \, F(\epsilon) \, \delta\left(e_\beta - \hat{e}_\beta - \epsilon\right),
\ee
where $\hat{\sigma} \equiv \df \hat{\sigma}/\df \hat{e}_\beta$ refers to the perturbative result.  The shape function prescription gives sensible results in the small $e_\beta$ limit, but it breaks down at large values of $e_\beta$, since the convolution in \Eq{eq:naiveconvolution} can yield $e_\beta$ values that extend beyond the physical range.  To address this, we need to smoothly turn off the nonperturbative shift as $\hat{e}_\beta$ approaches the physical endpoint $e_\beta^{\rm max}$.  There is no unique way to do this, but we find sensible results using
\be
\label{eq:physicalconvolution}
\frac{\df \sigma}{\df e_\beta} = \int \df \hat{e}_\beta \, \df \epsilon \,   \hat{\sigma}(\hat{e}_\beta) \, F(\epsilon) \, \delta \left(e_\beta - e_\beta^{\rm max} \frac{\hat{e}_\beta + \epsilon}{e_\beta^{\rm max} + \epsilon} \right),
\ee
which ensures that the cross section normalization is not modified even when the impact of the shape function is suppressed.  For simplicity, we take $e_\beta^{\rm max} = 1$ for all of our distributions, though in practice the perturbative distributions do not extend out that far.

The shape function $F$ has to be extracted from data, but we can use simple parametrizations that account for some aspects of its known behavior:
\be
\label{eq:Foptions}
F(\epsilon) = \frac{\pi \epsilon}{2 \epsilon_0^2} \exp\left[-\frac{\pi}{4} \frac{\epsilon^2}{\epsilon_0^2} \right], \qquad F_{\rm alt}(\epsilon) = \frac{4\epsilon}{\epsilon_0^2} \exp \left[-\frac{2 \epsilon}{\epsilon_0} \right].
\ee
Both of these functions go linearly to zero at $\epsilon \to 0$, fall exponentially as $\epsilon \to \infty$, are normalized by $\int \df \epsilon \,  F(\epsilon) = 1$, and have an expectation value
\be
\vev{\epsilon} = \int \df \epsilon \, \epsilon \, F(\epsilon) = \epsilon_0.
\ee
The parameter $\epsilon_0$ can therefore be interpreted as the average shift of the perturbative distribution from nonperturbative effects.   The second form in \Eq{eq:Foptions} was used in \Ref{Stewart:2014nna}, but we take the first form as our default since it has a less pronounced high-side tail.

Following \Ref{Larkoski:2013paa} (see also \cite{Manohar:1994kq,Banfi:2004yd}), one can estimate $\epsilon_0$ as a function of $\beta$.  Nonperturbative modes have $k_t \simeq \Lambda_{\rm QCD}$, so \Eq{eq:ktscale} implies the relationship
\be
z \, \theta \simeq \frac{\Lambda_{\rm QCD}}{R \, E_J}.
\ee
Appealing to local parton-hadron duality \cite{Azimov:1984np}, we can estimate the average contribution to $z \, \theta^\beta$ from nonperturbative soft gluon emissions as
\begin{align}
\epsilon_0 &= \frac{\Omega_0}{R \, E_J} \int_0^1 \frac{\df z}{z} \int_0^1 \frac{ \df \theta}{ \theta} \, z \, \theta^\beta \, \delta\left(z \theta - \frac{\Xi_0}{R \, E_J} \right) \nonumber \\
&= \frac{1}{\beta - 1} \frac{\Omega_0}{R \, E_J} \left( 1 - \left(\frac{\Xi_0}{R \, E_J} \right)^{\beta -1} \right),  \label{eq:epsilon0ansatz}
\end{align}
where we are using the soft (and collinear) gluon matrix element to determine the phase space integration, and $\Omega_0$ and $\Xi_0$ are nonperturbative parameters that are both of order $\Lambda_{\rm QCD}$.  This estimate of $\epsilon_0$ can also be understood by considering two types of nonperturbative modes:
\begin{align}
\text{NP Soft}: & \quad z \simeq \frac{\Lambda_{\rm QCD}}{R \, E_J}, \quad \theta \simeq 1, \\
\text{NP Collinear}: & \quad z \simeq 1, \quad \theta \simeq \frac{\Lambda_{\rm QCD}}{R \, E_J}.
\end{align}
These contribute to the angularities as
\be
e_\beta^{\text{(NP)}} \simeq \underbrace{\frac{\Lambda_{\rm QCD}}{R \, E_J}}_{\text{soft}} + \underbrace{\left( \frac{\Lambda_{\rm QCD}}{R \, E_J} \right)^\beta}_{\text{collinear}}.
\ee
We see that soft modes dominate for $\beta > 1$, collinear modes dominate for $\beta < 1$, and soft and collinear modes are equally important for $\beta = 1$.  This behavior is indeed encoded in \Eq{eq:epsilon0ansatz}, which smoothly interpolates between these regimes, yielding a logarithmic structure of $\frac{\Omega_0}{R \, E_J} \log \frac{R \, E_J}{\Xi_0}$ for $\beta = 1$ exactly.

Comparing quarks and gluons, we expect the overall size of the nonperturbative shift $\Omega_0$ should scale proportional to the Casimir factors, as in \Eq{eq:splitting_function_Casimir}.  The scaling of $\Xi_0$ is less clear, since it controls nonperturbative collinear radiation, which is less well studied than nonperturbative soft radiation.  For our baseline distributions, we assume that $\Xi_0$ also obeys Casimir scaling:
\be
\label{eq:Xiscaling}
\frac{\Omega_0^g}{\Omega_0^q} = \frac{\Xi_0^g}{\Xi_0^q} = \frac{C_A}{C_F}.
\ee
By tying $\Omega_0$ and $\Xi_0$ together, this has the effect of reducing the phase space for nonperturbative emissions from gluons, which is particularly important for $\beta < 1$.  Ideally, one would want a more rigorous justification for the assumptions in \Eq{eq:Xiscaling} (as well as the convolution structure in \Eq{eq:physicalconvolution}), though that is beyond the scope of the present work.

\begin{figure}
\centering
\subfloat[]{
\label{fig:NPshift_05}
\includegraphics[width = 0.32\columnwidth]{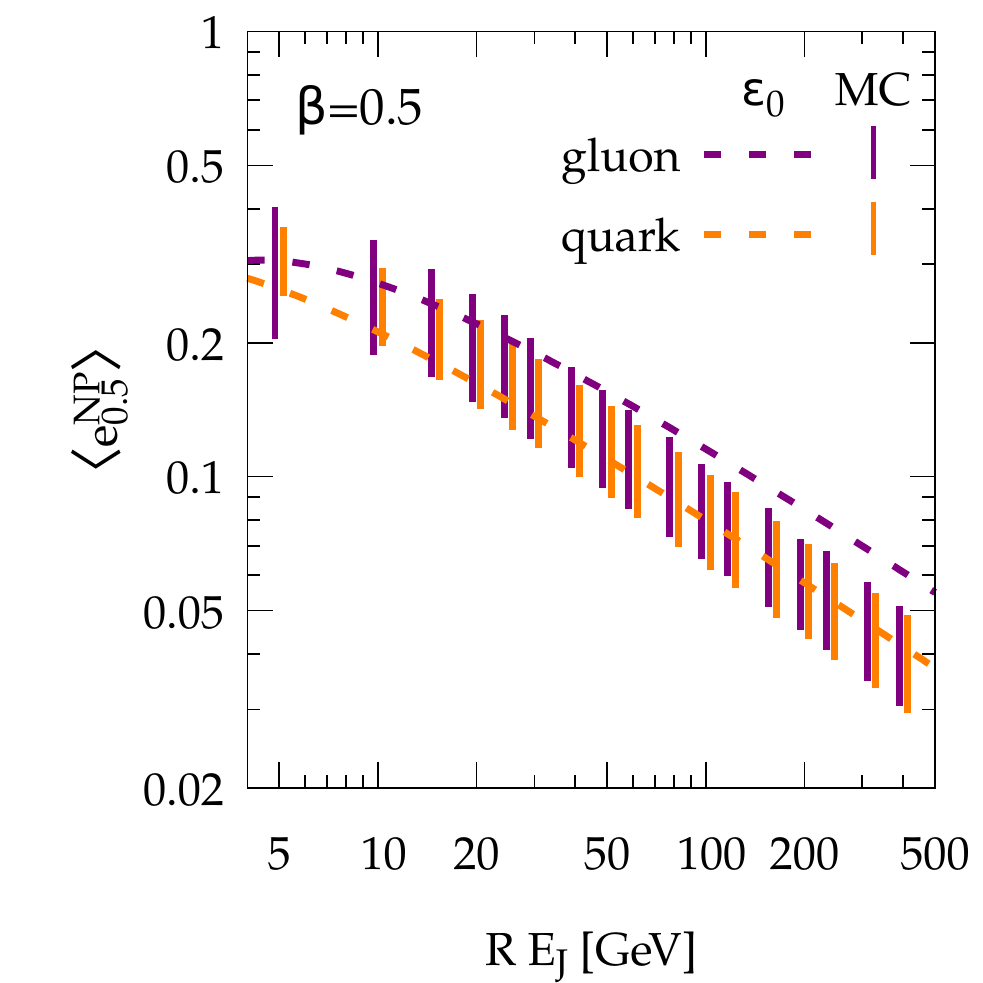}
}
\subfloat[]{
\label{fig:NPshift_10}
\includegraphics[width = 0.32\columnwidth]{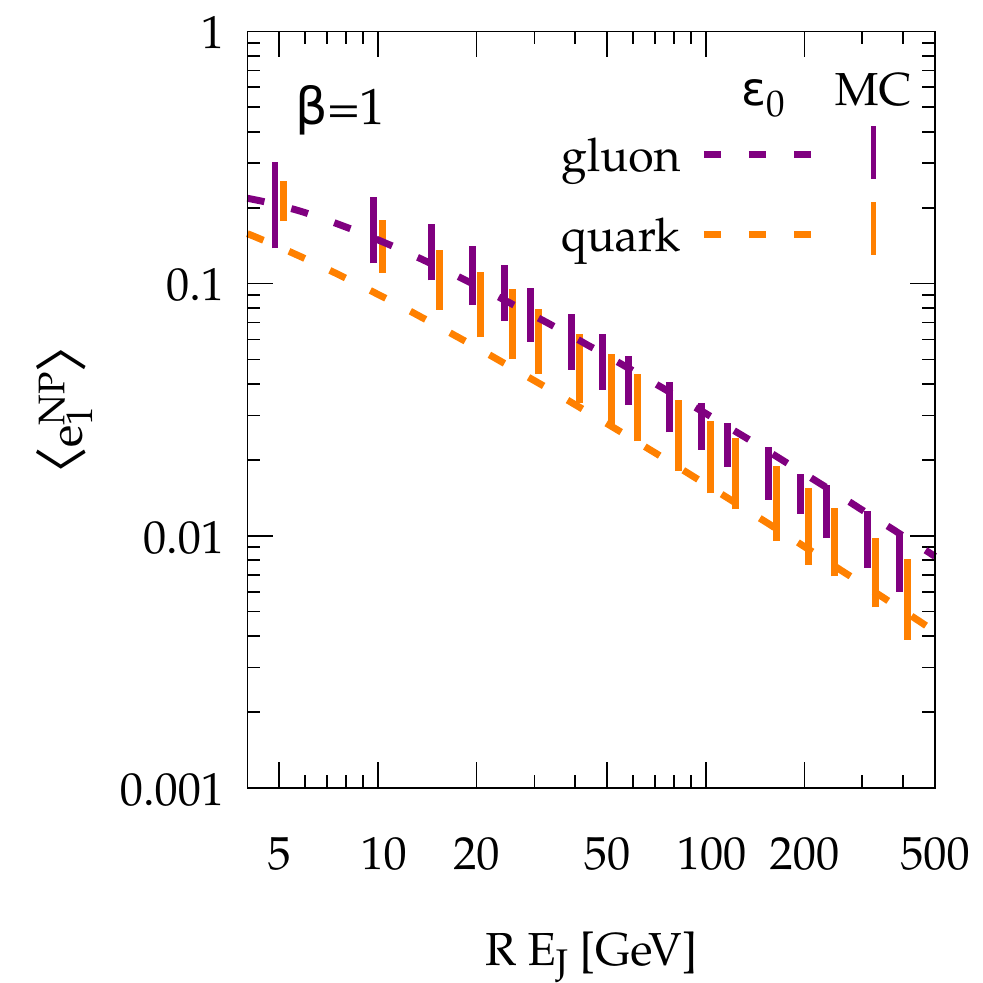}
}
\subfloat[]{
\label{fig:NPshift_20}
\includegraphics[width = 0.32\columnwidth]{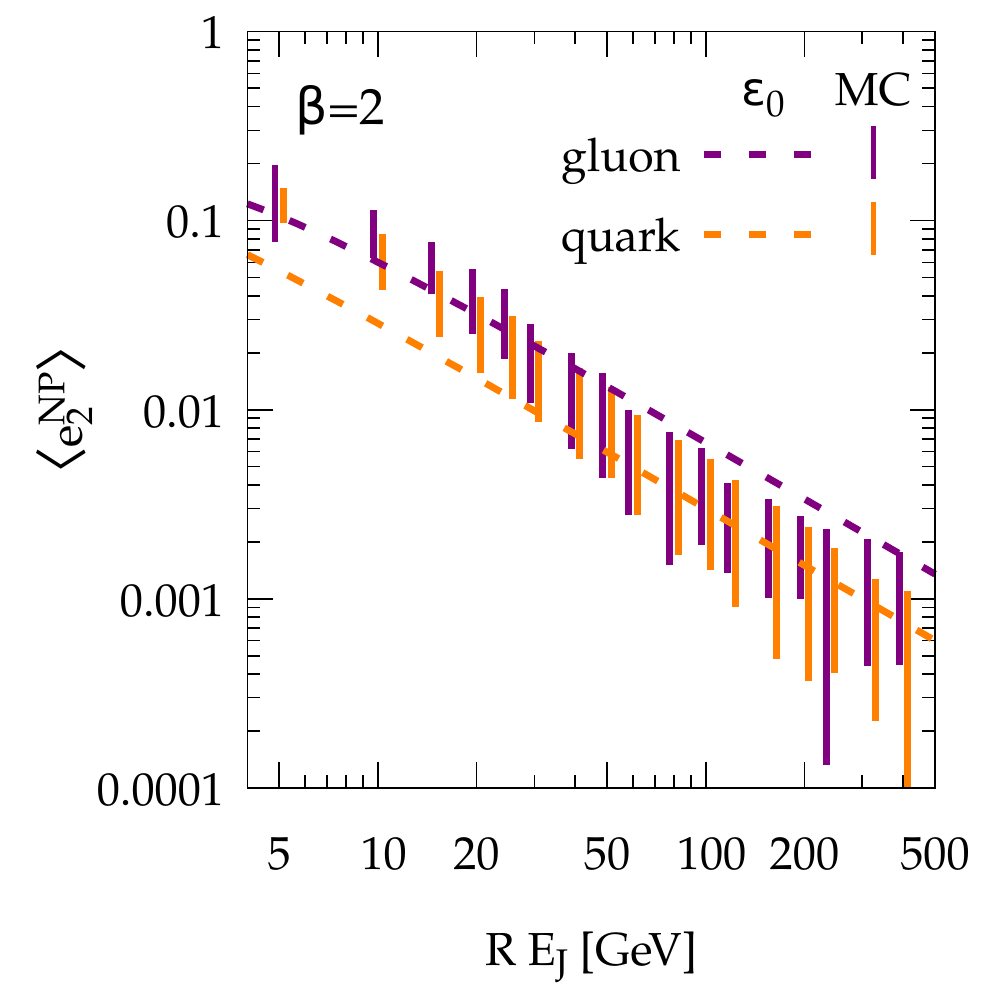}
}
\caption{Average nonperturbative shifts to the IRC-safe angularities
  as a function of $R \, E_J$ for (a) $\beta = 0.5$, (b) $\beta = 1$,
  and (c) $\beta = 2$.  The vertical bands correspond to the range of
  shifts seen by turning hadronization off and on in the different
  parton-shower generators.  The dashed line corresponds to the best
  fit to $\Omega_0$ and $\Xi_0$, assuming the functional form for
  $\epsilon_0$ in \Eq{eq:epsilon0ansatz} and the assumption of Casimir
  scaling in \Eq{eq:Xiscaling}.  While the $R \, E_J$ trend agrees,
  the hadronization corrections (as implemented in the parton showers)
  do not appear to exhibit Casimir scaling.}
\label{fig:NPshift}
\end{figure}

To test whether \Eq{eq:epsilon0ansatz} is a plausible estimate for nonperturbative corrections, we take the parton-shower generators studied in the next section and study how the average value of $e_\beta$ shifts as hadronization is turned off and on, and use that to estimate $\epsilon_0$.  We emphasize that a hadronization model used with a parton shower is not the same as a shape function in an analytic calculation, so one has to be careful drawing conclusions about the size of $\epsilon_0$ from a study like this.  In particular, effects that are captured by $F(\epsilon)$ in an analytic calculation could either be part of the perturbative showering or nonperturbative hadronization in a generator.  That said, we expect that the scaling of the $e_\beta$ shift as a function of $E_J$, $R$, and $\beta$ should be roughly the same.

In \Fig{fig:NPshift}, we show the size of the $e_\beta$ shift as a function of $R \, E_J$ for the three benchmark $\beta$ values, where the band indicates the minimum and maximum shifts seen among \textsc{Pythia}, \textsc{Herwig}, \textsc{Sherpa}, \textsc{Vincia}, \textsc{Deductor}, \textsc{Ariadne}, and \textsc{Dire}.\footnote{We verified that the overall conclusions do not change when considering the separate scaling of $R$ and $E_J$.}  We then compare to the expected shift from \Eq{eq:epsilon0ansatz} with 
\be
\label{eq:OmegaXiValues}
\Omega^i_0 = C_i \times 0.23~\GeV, \qquad \Xi^i_0 = C_i \times
0.37~\GeV, \ee where these values are obtained by doing a
(logarithmic) fit to all of the parton-shower shift values.  While the
parton shower trends with $R \, E_J$ and $\beta$ roughly agree with
\Eq{eq:epsilon0ansatz}, there is no evidence for the Casimir scaling
hypothesis in \Eq{eq:Xiscaling}.\footnote{This conclusion is not
  simply an artifact of the fitting procedure, as none of the
  individual generators show evidence for Casimir scaling in the
  nonperturbative shift either.}  This is most likely because parton
showers already achieve some degree of Casimir scaling through
multiple (perturbative) soft gluon emissions from the shower. Despite
this caveat, we use the extracted values from \Eq{eq:OmegaXiValues}
for our baseline distributions below.

We also consider two alternative scaling behaviors for $\Omega_0$ and $\Xi_0$.  The first alternative is motivated by the observation that, as far as the perturbative soft gluon matrix element is concerned, the Casimir factor affects the rate of soft gluon emissions but not the associated kinematics.  Thus, one might expect the overall $\Omega_0$ factor in \Eq{eq:epsilon0ansatz} to respect Casimir scaling, but not the $\Xi_0$ factor inside the delta function.  We therefore test a variant with
\be
\label{eq:OmegaXiValuesHalfAlt}
\text{\textsc{No $C_i$ in $\Xi_0$}}:  \qquad \Omega^i_0 = C_i \times 0.22~\GeV, \qquad \Xi^i_0 = 0.70~\GeV,
\ee
where again these values are estimated by fitting to the parton shower $e_\beta$ shifts.  As shown in \Fig{fig:summary_hadron_analytic} below, \Eq{eq:OmegaXiValuesHalfAlt} leads to a dramatic increase in the predicted quark/gluon separation power for $\beta < 1$.  The second alternative is motivated by the absence of any evidence of Casimir scaling in the parton showers from \Fig{fig:NPshift}.  We therefore try taking both nonperturbative parameters to be independent of $C_i$, with
\be
\label{eq:OmegaXiValuesAlt}
\text{\textsc{No $C_i$ in $\epsilon_0$}}:  \qquad \Omega^i_0 = 0.44~\GeV, \qquad \Xi^i_0 = 0.70~\GeV,
\ee
which leads to a corresponding decrease in separation power, since the nonperturbative shape function now has the same behavior for quarks and gluons.

\section{Idealized quark/gluon discrimination}
\label{sec:ee}

We now turn to parton-shower studies of quark/gluon discrimination, starting with the idealized case of $e^+ e^-$ collisions.  While far less complicated than quark/gluon tagging in the LHC environment, this $e^+ e^-$ case study demonstrates the importance of final-state evolution for quark/gluon discrimination, independent from initial-state complications arising in $pp$ collisions.  A \textsc{Rivet} routine \cite{Buckley:2010ar} for this analysis can be downloaded from \url{https://github.com/gsoyez/lh2015-qg} under \verb|MC_LHQG_EE.cc|.

To define the truth-level jet flavor, we use a simple definition:  a quark jet is a jet produced by a parton-shower event generator in $e^+ e^- \to (\gamma/Z)^* \to u \bar{u}$ hard scattering, while a gluon jet is a jet produced in $e^+ e^- \to h^* \to gg$.  Of course, an $e^+e^- \to u \bar u$ event can become a $e^+e^- \to u \bar u g$ event after one step of shower evolution, just as $e^+e^- \to g g$ can become $e^+e^- \to g u \bar u$.  This illustrates the inescapable ambiguity in defining jet flavor.\footnote{In an $e^+e^-$ context, our definition at least respects the Lorentz structure of the production vertex, so in that sense it is a fundamental definition that does not reference (ambiguous) quark or gluon partons directly.}  To partially mitigate the effect of wide-angle emissions, we restrict our analysis to jets that satisfy
\begin{equation}
\label{eq:Erestrict}
\frac{E_{\rm jet}}{Q/2} > 0.8,
\end{equation}
where $Q$ is the center-of-mass collision energy, allowing for up to
two jets studied per event.  Note that this condition acts as a
restriction on out-of-jet radiation, which already suppresses to some
extent non-global effects \cite{Dasgupta:2001sh}.\footnote{Note that
  we have not included the effect of \Eq{eq:Erestrict} in our analytic calculation,
  which in principle affects the functional form of $f_{\rm NGL}$ for non-global logarithms.}
There is also the ambiguity of which parton shower to use, so we
investigate quark/gluon radiation patterns in several event
generators: \textsc{Pythia 8.215} \cite{Sjostrand:2014zea},
\textsc{Herwig 2.7.1} \cite{Bahr:2008pv,Bellm:2013hwb}, \textsc{Sherpa
  2.2.1} \cite{Gleisberg:2008ta}, \textsc{Vincia 2.001}
\cite{Fischer:2016vfv}, \textsc{Deductor 1.0.2} \cite{Nagy:2014mqa}
(with hadronization performed by \textsc{Pythia}), \textsc{Ariadne
  5.0.$\beta$} \cite{Flensburg:2011kk}, and \textsc{Dire 1.0.0}
\cite{Hoche:2015sya} (with cluster hadronization performed by
\textsc{Sherpa}).

\subsection{Baseline analysis}

\begin{figure}
\centering
\subfloat[]{
\includegraphics[width = 0.45\columnwidth]{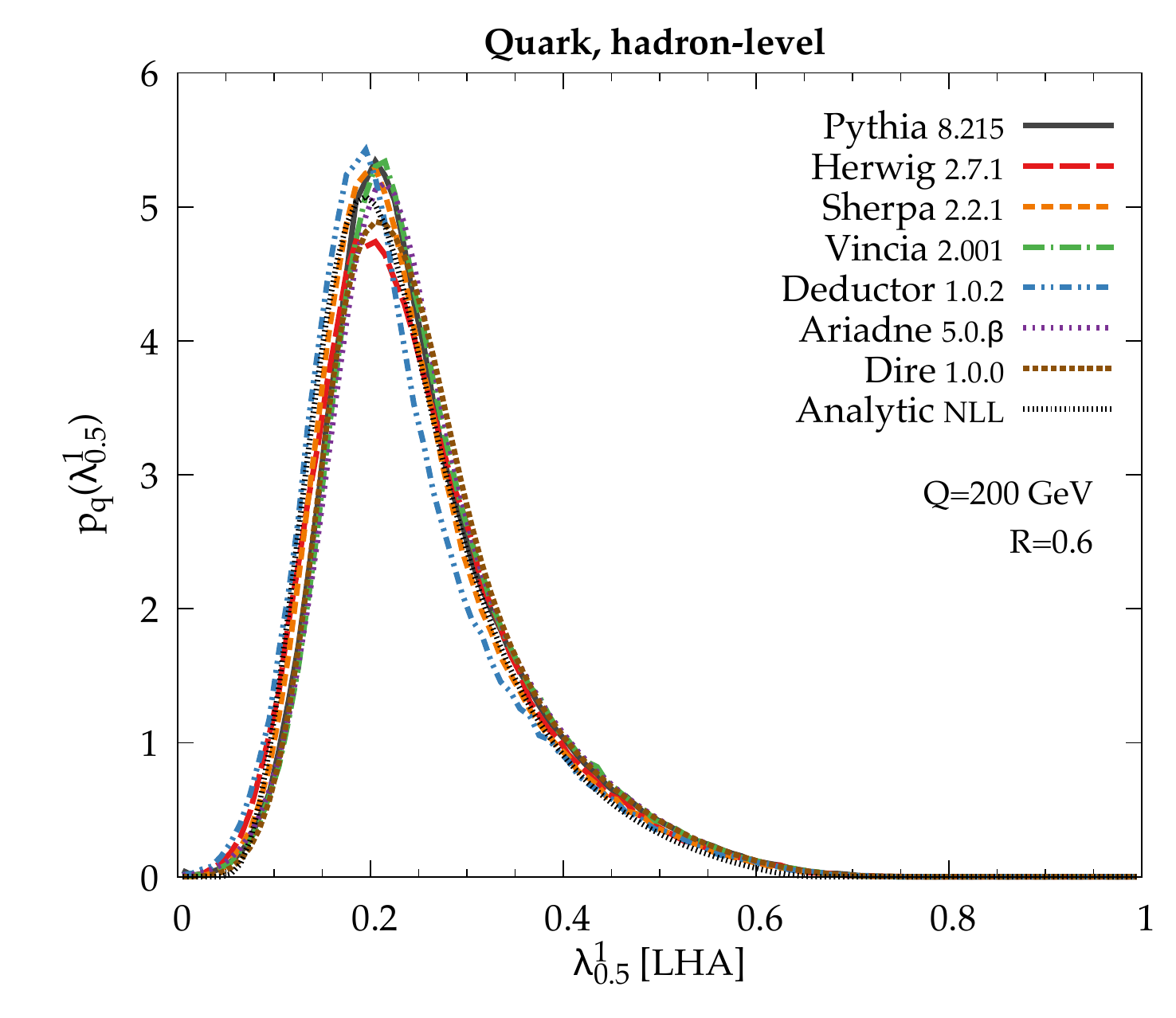}
\label{fig:LHA_hadron_quark}
}
$\qquad$
\subfloat[]{
\includegraphics[width = 0.45\columnwidth]{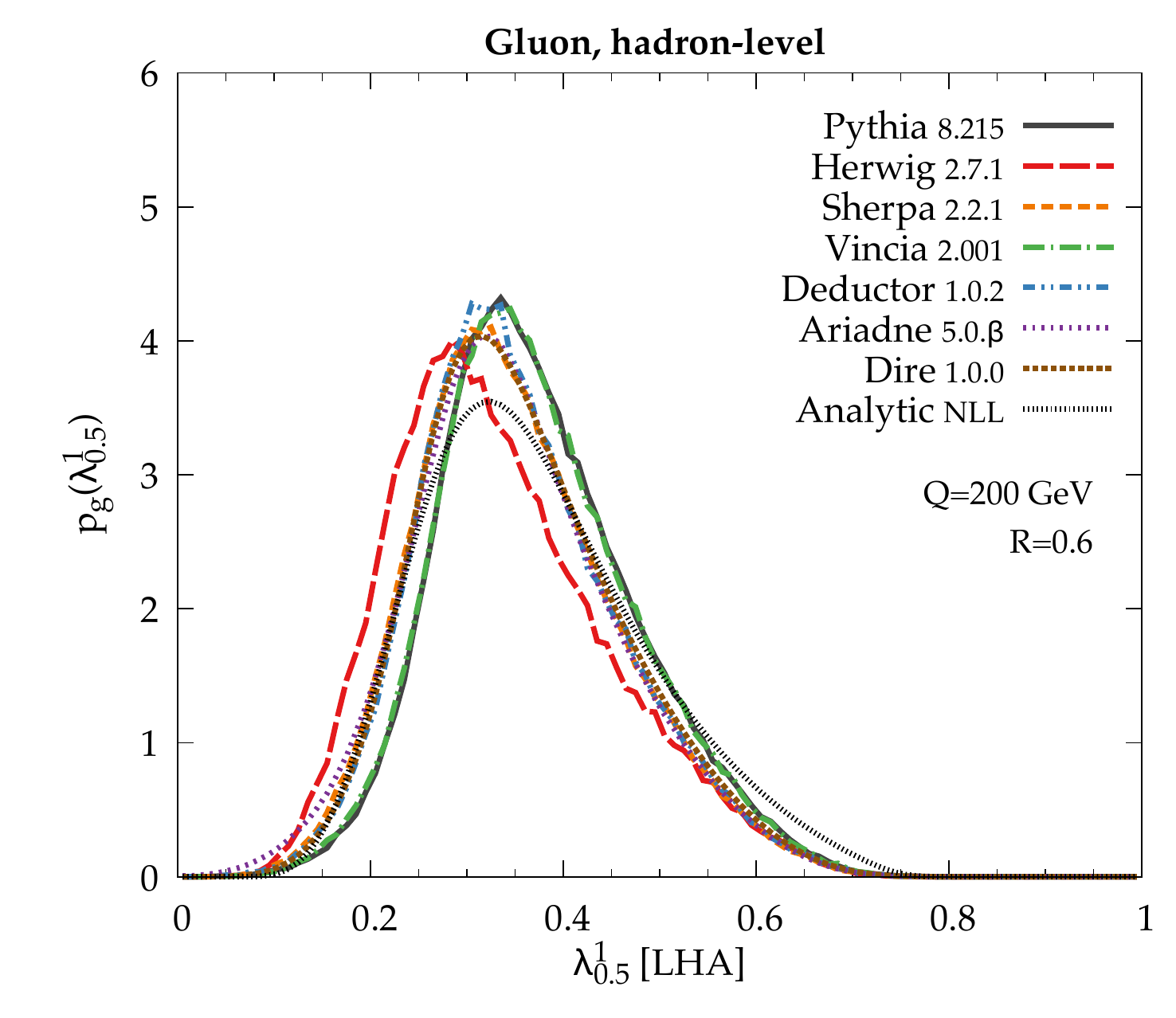}
\label{fig:LHA_hadron_gluon}
}

\subfloat[]{
\includegraphics[width = 0.65\columnwidth]{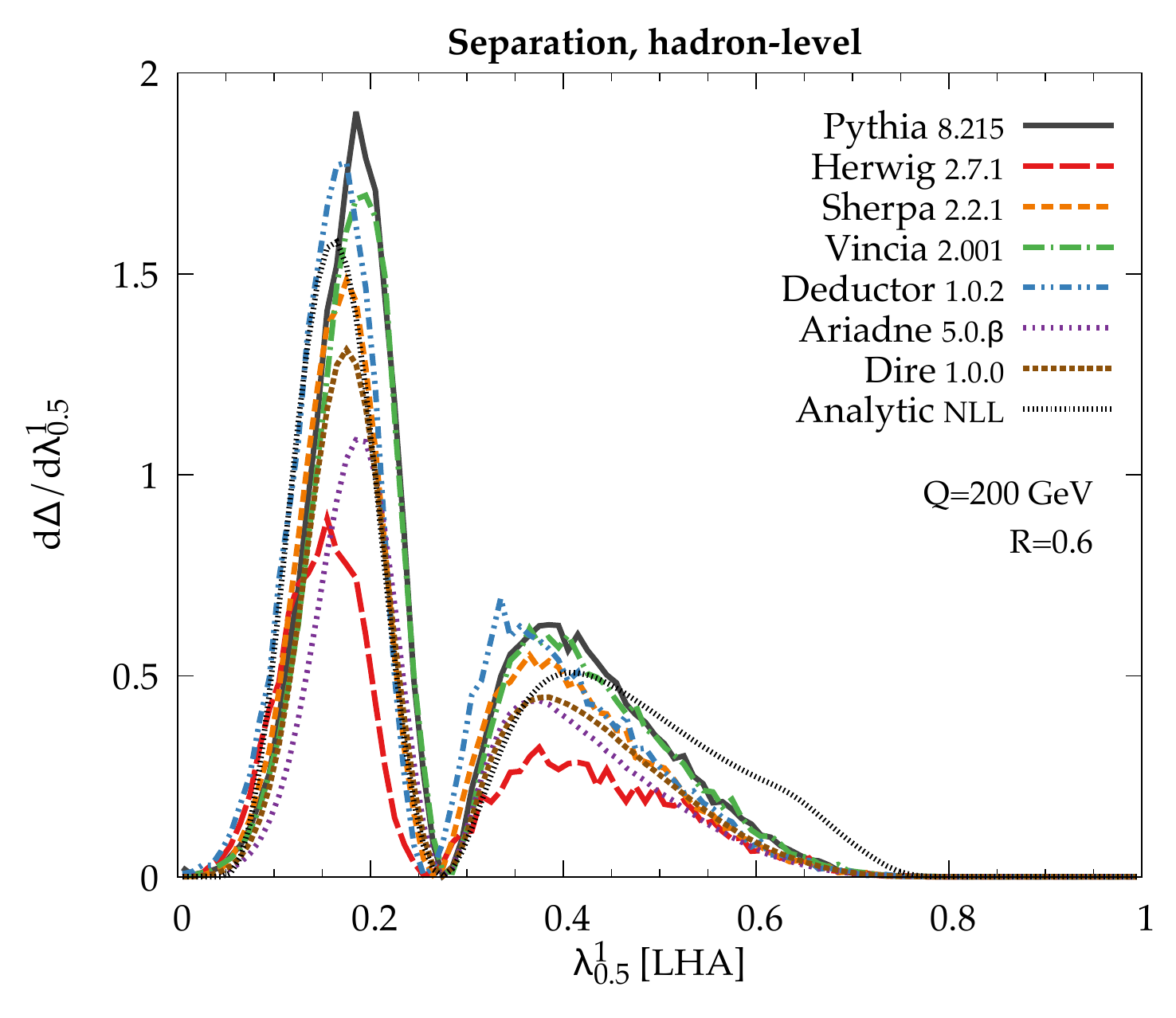}
\label{fig:LHA_hadron_separation}
}
\caption{Hadron-level distributions of the LHA for (a) the $e^+ e^- \to u \bar{u}$ (``quark jet'') sample, (b) the $e^+ e^- \to gg$ (``gluon jet'') sample, and (c) the classifier separation integrand in \Eq{eq:deltaintegrand}.  Seven parton-shower generators---\textsc{Pythia 8.215}, \textsc{Herwig 2.7.1}, \textsc{Sherpa 2.2.1}, \textsc{Vincia 2.001}, \textsc{Deductor 1.0.2}, \textsc{Ariadne 5.0.$\beta$}, and \textsc{Dire 1.0.0}---are run at their baseline settings with center-of-mass energy $Q = 200~\GeV$ and jet radius $R= 0.6$.  We also show the analytic NLL results from \Sec{sec:analytic}.}
\label{fig:LHA_hadron}
\end{figure}

In \Fig{fig:LHA_hadron}, we show hadron-level distributions of the LHA (i.e.~$e_{0.5} = \lambda_{0.5}^1$) in the quark sample ($p_q$) and gluon sample ($p_g$), comparing the baseline settings of seven different parton-shower generators with a center-of-mass collision energy of $Q = 200~\GeV$ and jet radius $R = 0.6$. In the quark sample in \Fig{fig:LHA_hadron_quark}, there is relatively little variation between the generators, which is not surprising since most of these programs have been tuned to match LEP data (though LEP never measured the LHA itself).  Turning to the gluon sample in \Fig{fig:LHA_hadron_gluon}, we see somewhat larger variations between the generators; this is expected since there is no data to directly constrain $e^+ e^- \to gg$ (though there are indirect tests from LEP; see \Sec{sec:conclude}).    It is satisfying that for both the quark and gluon samples, the analytic NLL results from \Sec{sec:analytic} peak at roughly the same locations as the parton showers.  In the \texttt{arXiv} preprint source files \cite{ArXivSource}, one can see comparable levels of agreement for the two other IRC-safe angularities ($e_1$ and $e_2$).

In \Fig{fig:LHA_hadron_separation}, we plot the integrand of classifier separation, $\text{d} \Delta / \text{d} \lambda$ from \Eq{eq:deltaintegrand}. This shows where in the LHA phase space the actual discrimination power lies, with large values of the integrand
corresponding to places where the quark and gluon distributions are
most dissimilar.  Now we see considerable differences between the
generators, reproducing the well-known fact that \textsc{Pythia} is
more optimistic about quark/gluon separation compared to
\textsc{Herwig} \cite{Aad:2014gea}.  The predicted discrimination power from the other five generators and the NLL calculation are intermediate between these
extremes.

\begin{figure}
\centering
\subfloat[]{
\includegraphics[width = 0.45\columnwidth]{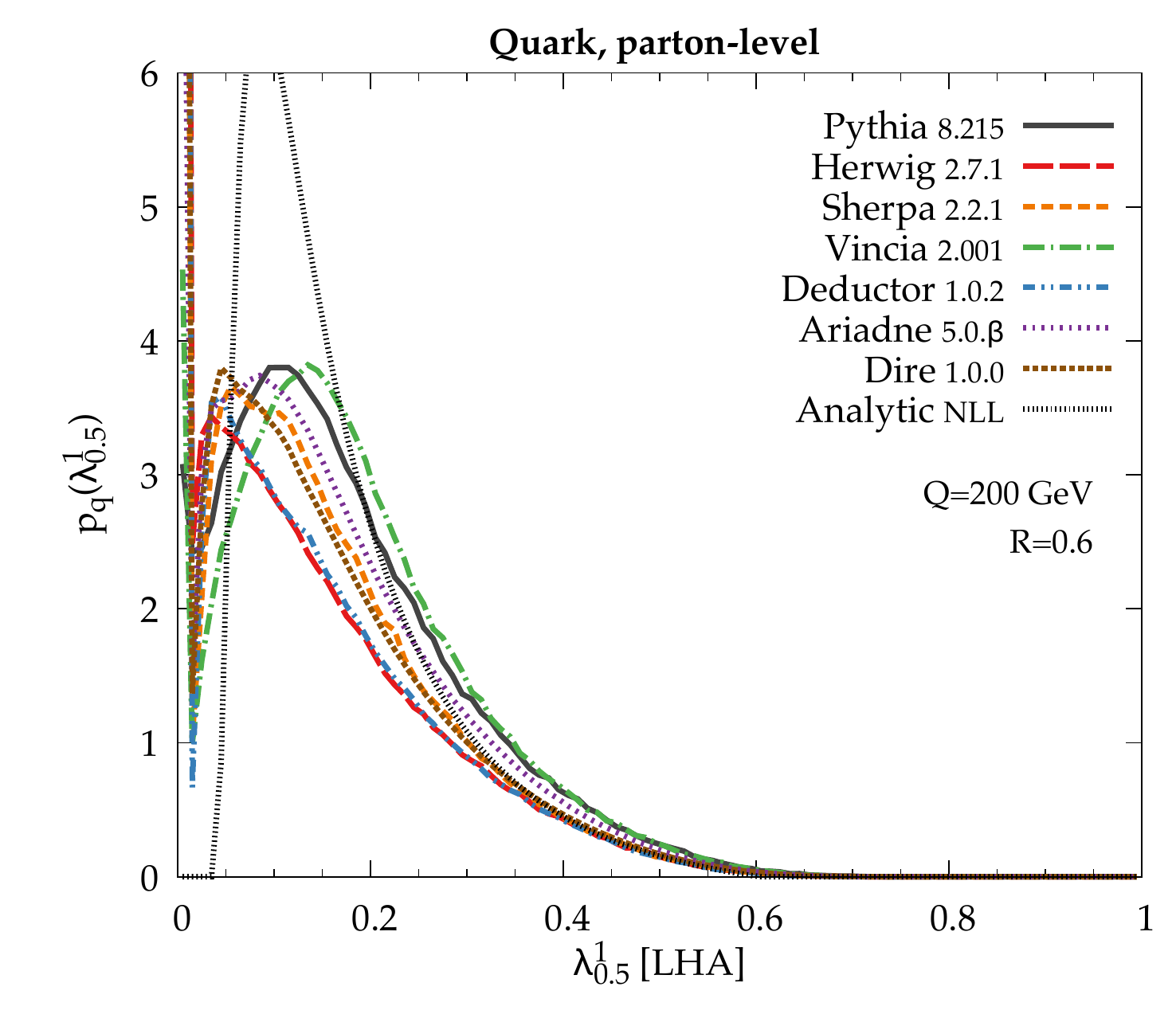}
\label{fig:LHA_parton_quark}
}
$\qquad$
\subfloat[]{
\includegraphics[width = 0.45\columnwidth]{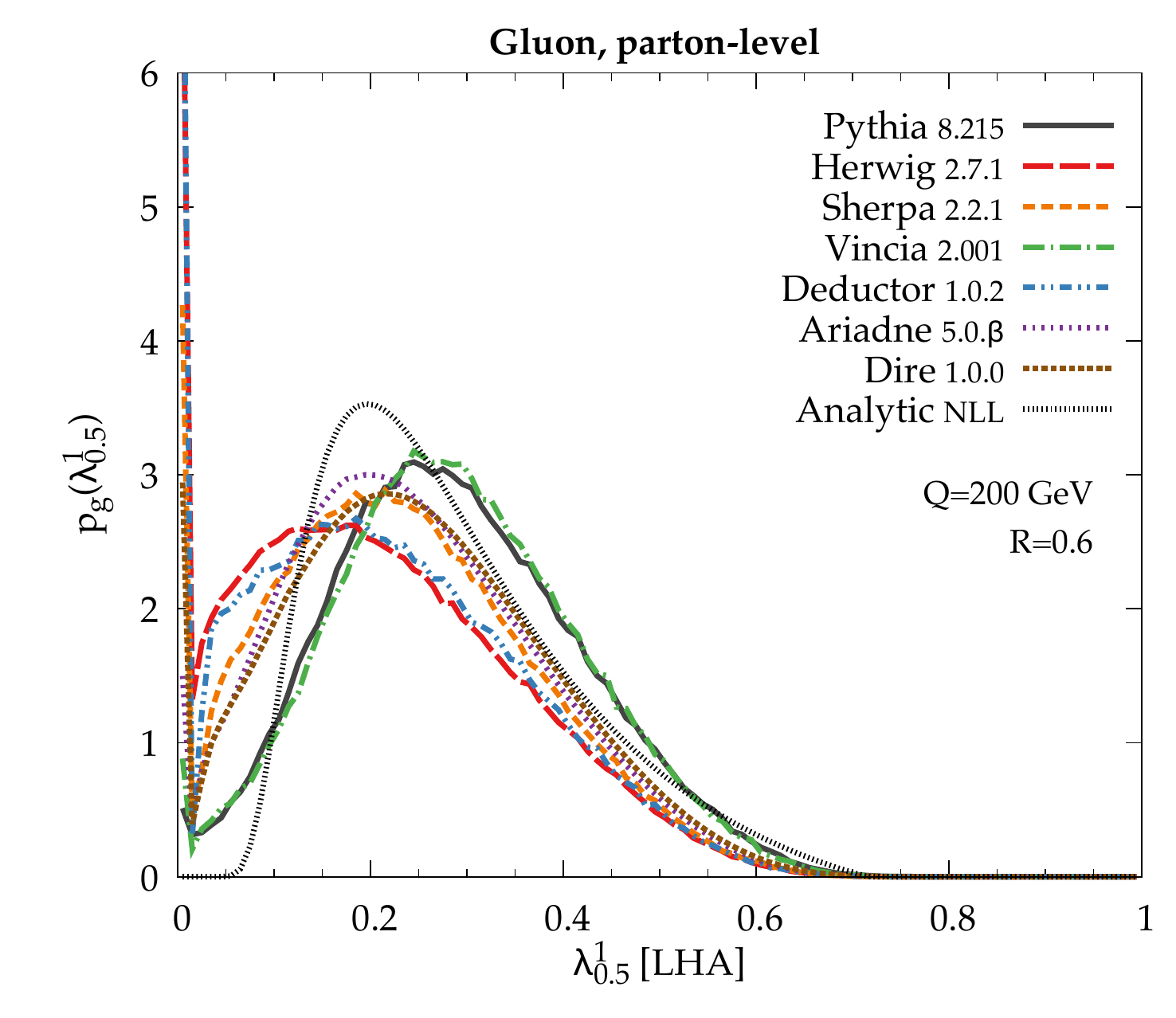}
\label{fig:LHA_parton_gluon}
}

\subfloat[]{
\includegraphics[width = 0.65\columnwidth]{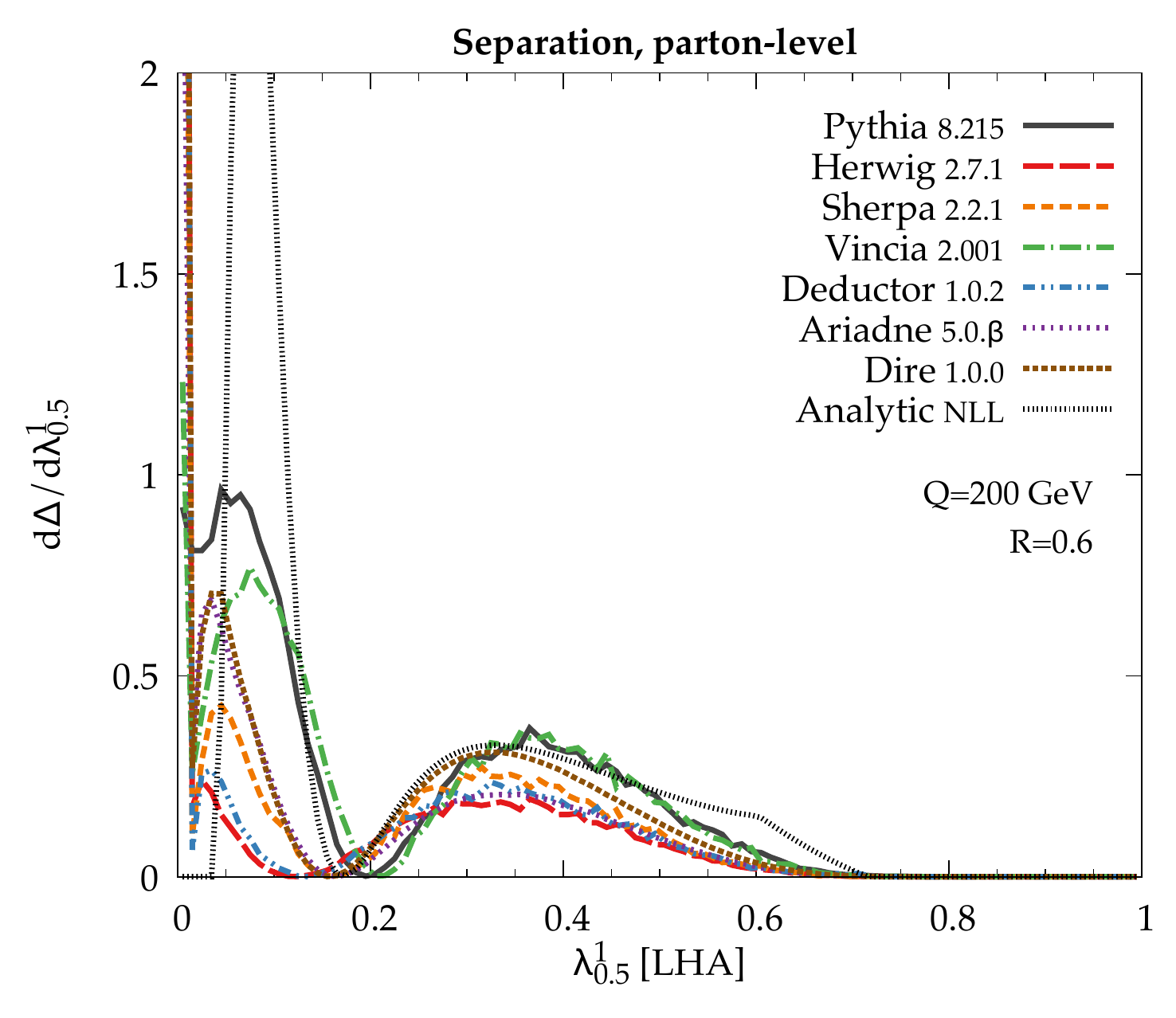}
\label{fig:LHA_parton_separation}
}
\caption{Same as \Fig{fig:LHA_hadron}, but at the parton level.  Note that \textsc{Herwig}, \textsc{Sherpa}, and \textsc{Deductor} all have cross section spikes at $\lambda_{0.5}^1 = 0$ that extend above the plotted range.}
\label{fig:LHA_parton}
\end{figure}

One might expect that the differences between generators are due
simply to their having different hadronization models.  It seems,
however, that the differences already appear at the parton level prior
to hadronization. We should say at the outset that it is nearly impossible to do a true apples-to-apples comparison of parton-level results, since these generators are interfaced to different hadronization models, and only the hadron-level comparison is physically meaningful.  In particular, the crossover between the perturbative and nonperturbative regions is ambiguous and each of these showers has a different effective shower cutoff scale, resulting in different amounts of radiation being generated in the showering versus hadronization steps.\footnote{In general, generators based on string hadronization tend to use a lower shower cutoff scale ($\sim 0.5$ GeV) compared to those based on cluster hadronization ($\sim 1$ GeV).}  Similarly, for the parton-level NLL results, small values of the angularities are artificially suppressed by the $\alpha_s \to \infty$ Landau pole, 
which enhances the Sudakov exponent.\footnote{An alternative approach would be to freeze $\alpha_s$ 
at scales below $\Lambda_{\rm QCD}$ or extend it into the nonperturbative
region as suggested in~\Refs{Dokshitzer:1995qm,Guffanti:2000ep,Gieseke:2007ad}.  Either way, this region of phase space is dominated by the nonperturbative shape function, which is absent from the ``parton-level'' distributions.}

With that caveat in mind, we show parton-level results in \Fig{fig:LHA_parton}.  One immediately notices that three of the
generators---\textsc{Herwig}, \textsc{Sherpa}, and
\textsc{Deductor}---yield a large population of events where the
perturbative shower generates no emissions, even in the gluon sample.  This gives
$\lambda_{0.5}^1 = 0$ such that non-zero values of the LHA are
generated only by the hadronization model.  By contrast,
\textsc{Pythia} and \textsc{Vincia} give overall larger values of the
LHA from the perturbative shower alone, with \textsc{Ariadne} and \textsc{Dire} yielding intermediate results.  As mentioned above, some of this difference can be explained simply by the different shower cutoff scales used in each generator, but it probably also reflects a difference in how semi-perturbative gluon splittings are treated.  Since \Fig{fig:LHA_hadron_quark} shows that all generators give similar distributions for quark jets after hadronization, we
conclude that understanding quark/gluon discrimination is a challenge
at the interface between perturbative showering and nonperturbative
hadronization.

\begin{figure}
\centering
\subfloat[]{
\label{fig:summary_hadron_all}
\includegraphics[width = 0.45\columnwidth]{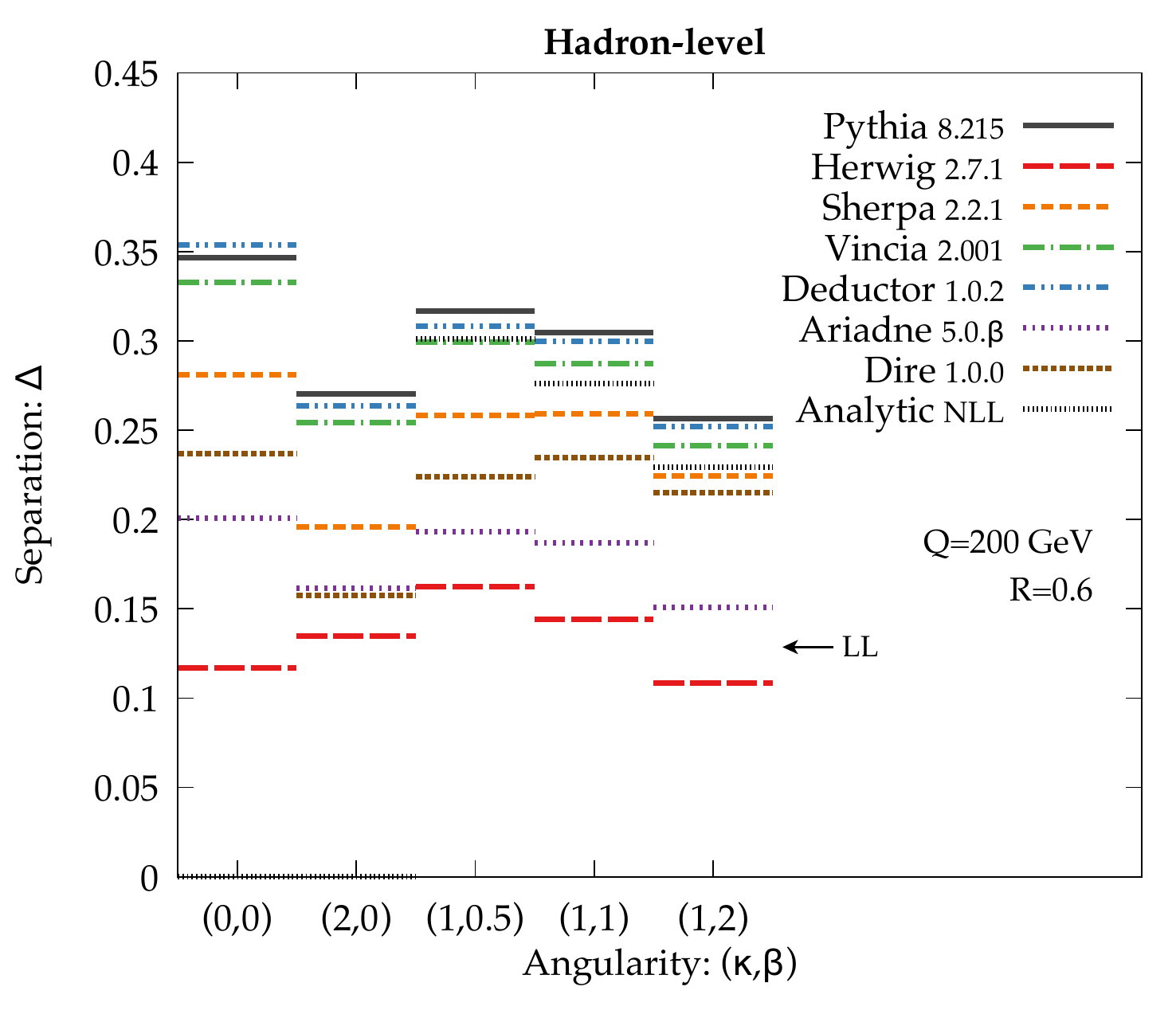}
}
$\qquad$
\subfloat[]{
\label{fig:summary_parton_all}
\includegraphics[width = 0.45\columnwidth]{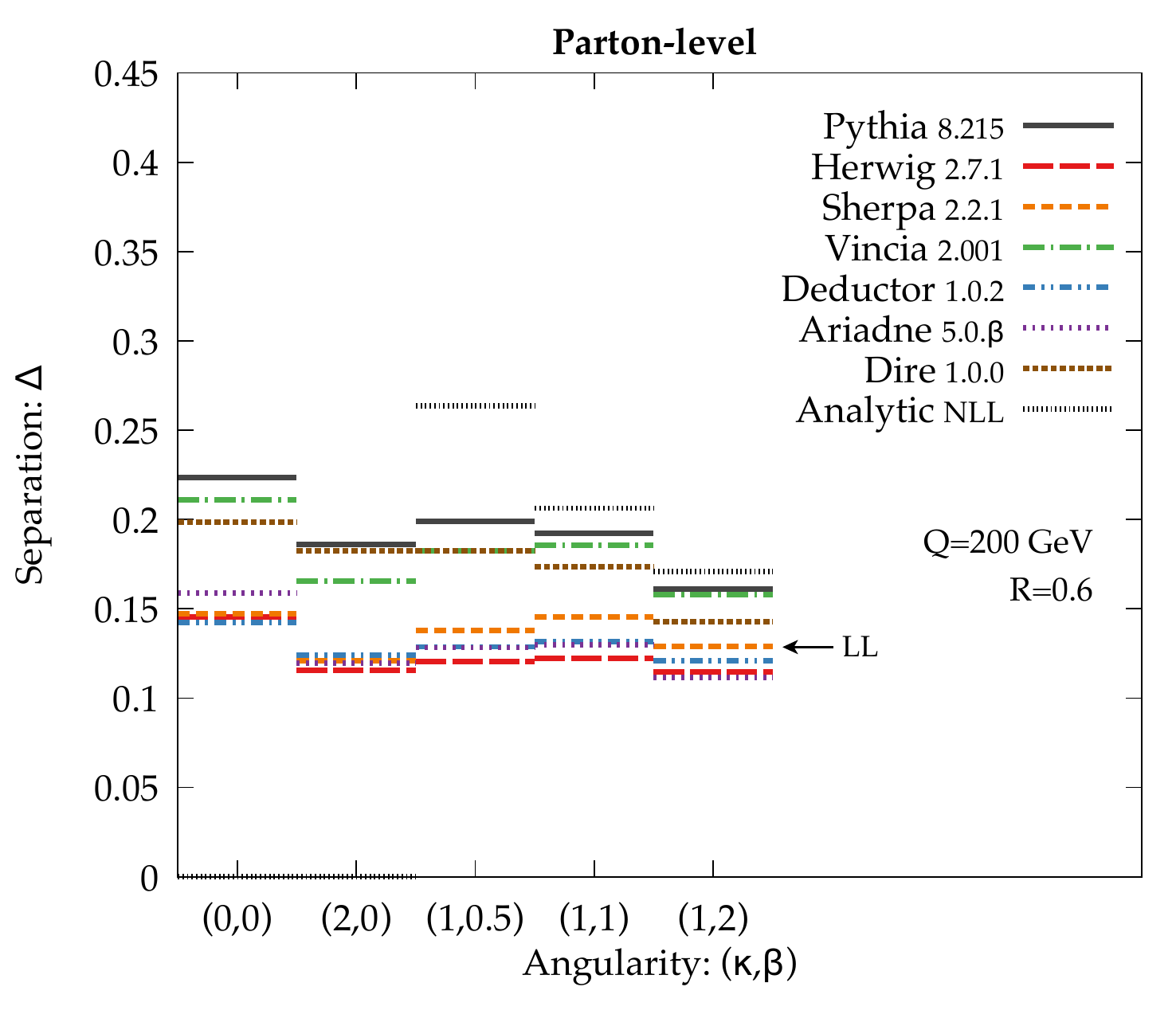}
}
\caption{Classifier separation $\Delta$ for the five benchmark angularities in \Eq{eq:benchmarkang}, determined from the various generators at (a) hadron level and (b) parton level.  The first two columns correspond to IRC-unsafe distributions (multiplicity and $p_T^D$), while the last three columns are the IRC-safe angularities.  The LHA (i.e.~$\kappa = 1$, $\beta = 1/2$) is shown in the middle column.  Results in terms of ROC values appear in the \texttt{arXiv} preprint source files \cite{ArXivSource}, for this and subsequent plots.  The label ``LL''  indicates the value from \Eq{eq:LLbenchmark} predicted by Casimir scaling.}
\label{fig:summary_all}
\end{figure}

To summarize the overall discrimination power, we integrate
\Eq{eq:deltaintegrand} to obtain the value of
classifier separation $\Delta$ for the LHA.  This is shown in
\Fig{fig:summary_all}, which also includes the four
other benchmark angularities from
\Eq{eq:benchmarkang}.  There is a rather large
spread in predicted discrimination power between the generators,
especially at hadron level in
\Fig{fig:summary_hadron_all}.  While such differences
might be expected for IRC-unsafe angularities (multiplicity and
$p_T^D$) which depend on nonperturbative modeling, these differences
persist even for the IRC-safe angularities at parton level (see
\Fig{fig:summary_parton_all}).\footnote{It is interesting that four of the generators---\textsc{Herwig}, \textsc{Sherpa}, \textsc{Deductor}, and \textsc{Ariadne}---have a comparatively narrow spread in predicted discrimination power at parton level, though this spread increases dramatically at hadron level.}  This suggests a more
fundamental difference between the generators that is already present
in the perturbative shower.

For the IRC-safe angularities with $\kappa = 1$, there is a generic
trend seen by all of the hadron-level generators that discrimination
power decreases as $\beta$ increases.  This trend agrees with the
study performed in \Ref{Larkoski:2013eya} and our NLL calculation
here, but disagrees with the ATLAS study in \Ref{Aad:2014gea}, which
found flat (or even increasing) discrimination power with increasing
$\beta$.  Understanding this $\beta$ trend will therefore be crucial
for understanding quark/gluon radiation patterns.

\subsection{Parameter dependence}
\label{sec:ee_scales}

\begin{figure}
\centering
\subfloat[]{
\includegraphics[width = 0.45\columnwidth]{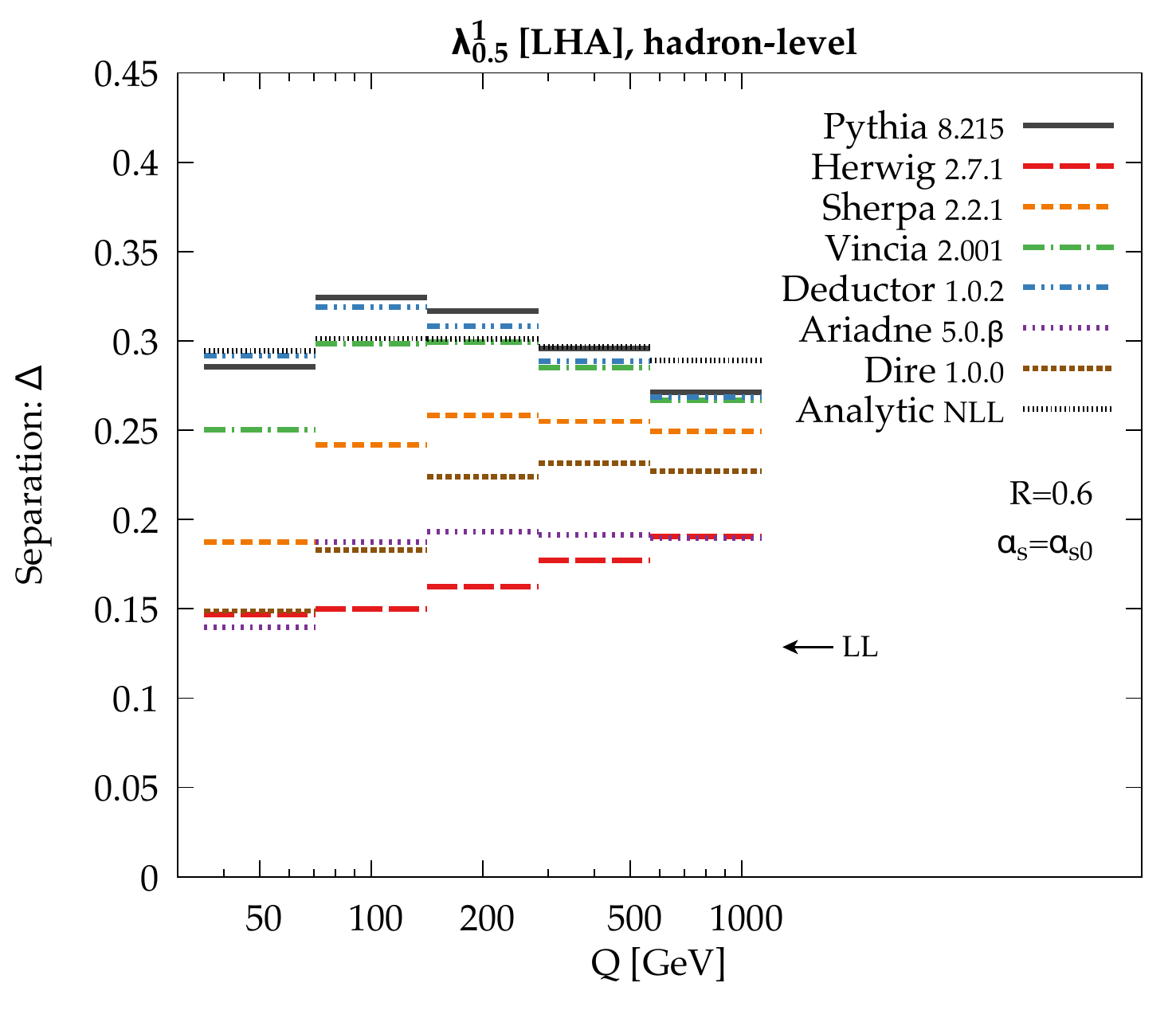}
\label{fig:sweep_Q_hadron}
}
$\qquad$
\subfloat[]{
\includegraphics[width = 0.45\columnwidth]{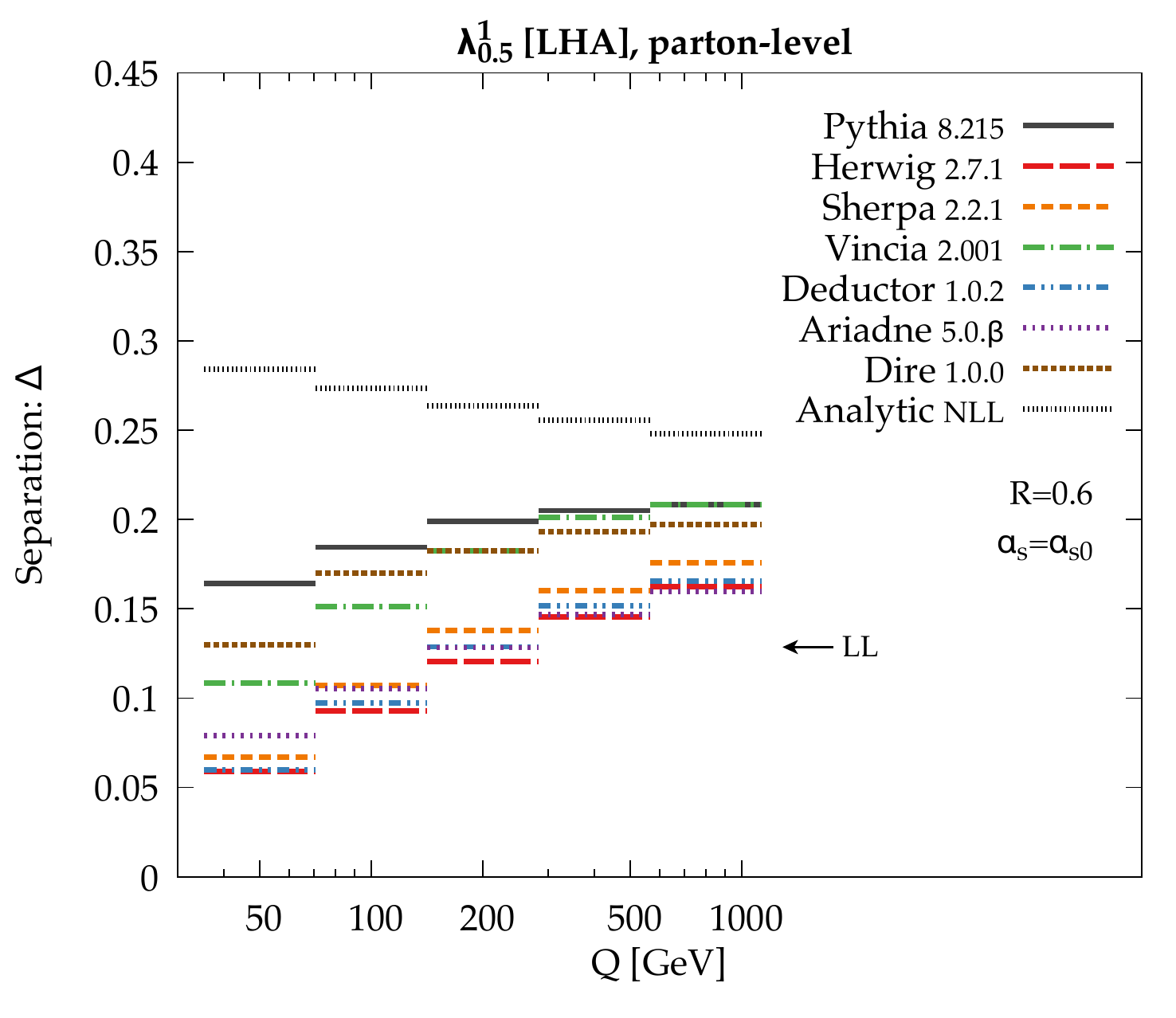}
\label{fig:sweep_Q_parton}
}

\subfloat[]{
\includegraphics[width = 0.45\columnwidth]{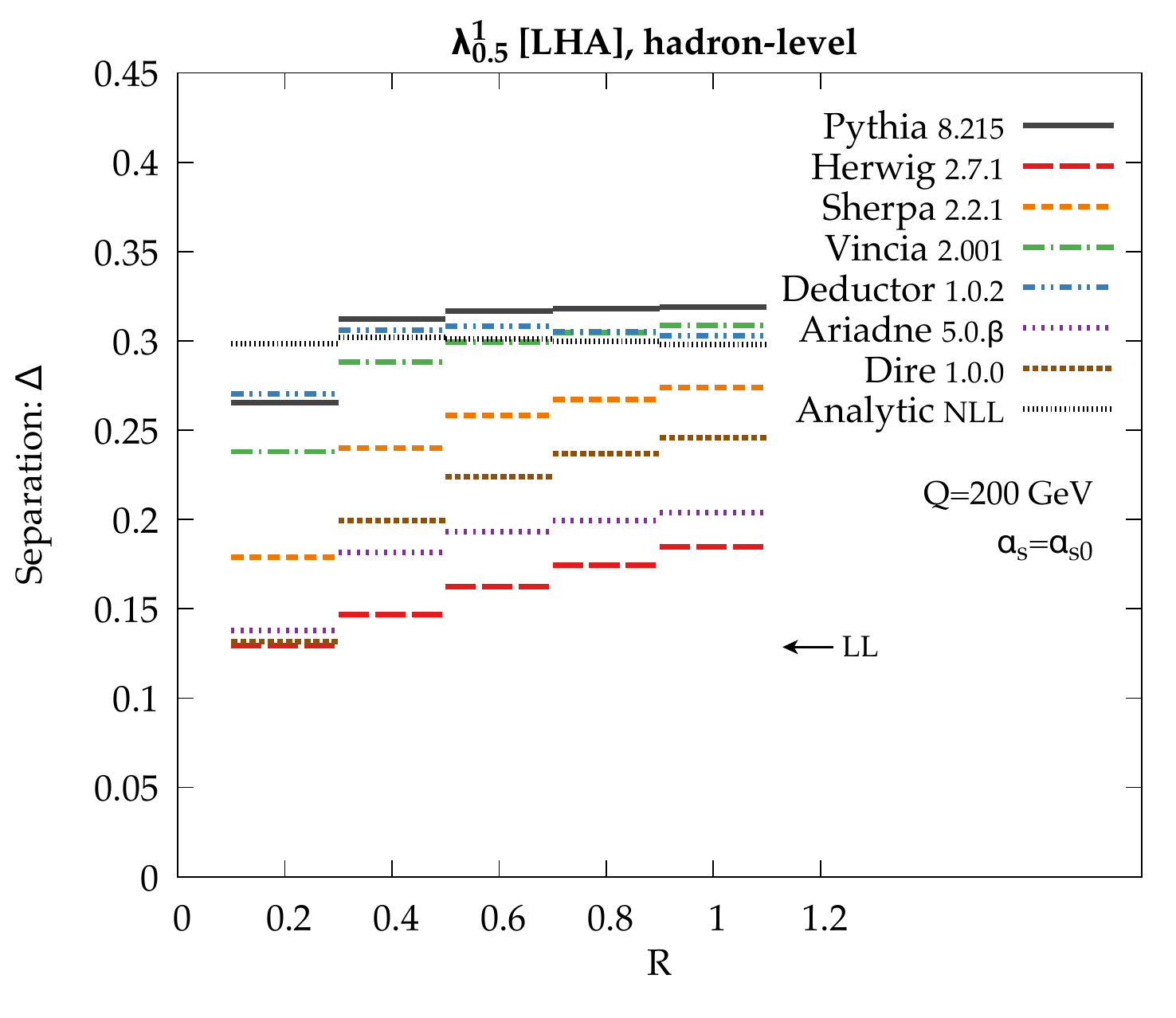}
\label{fig:sweep_R_hadron}
}
$\qquad$
\subfloat[]{
\includegraphics[width = 0.45\columnwidth]{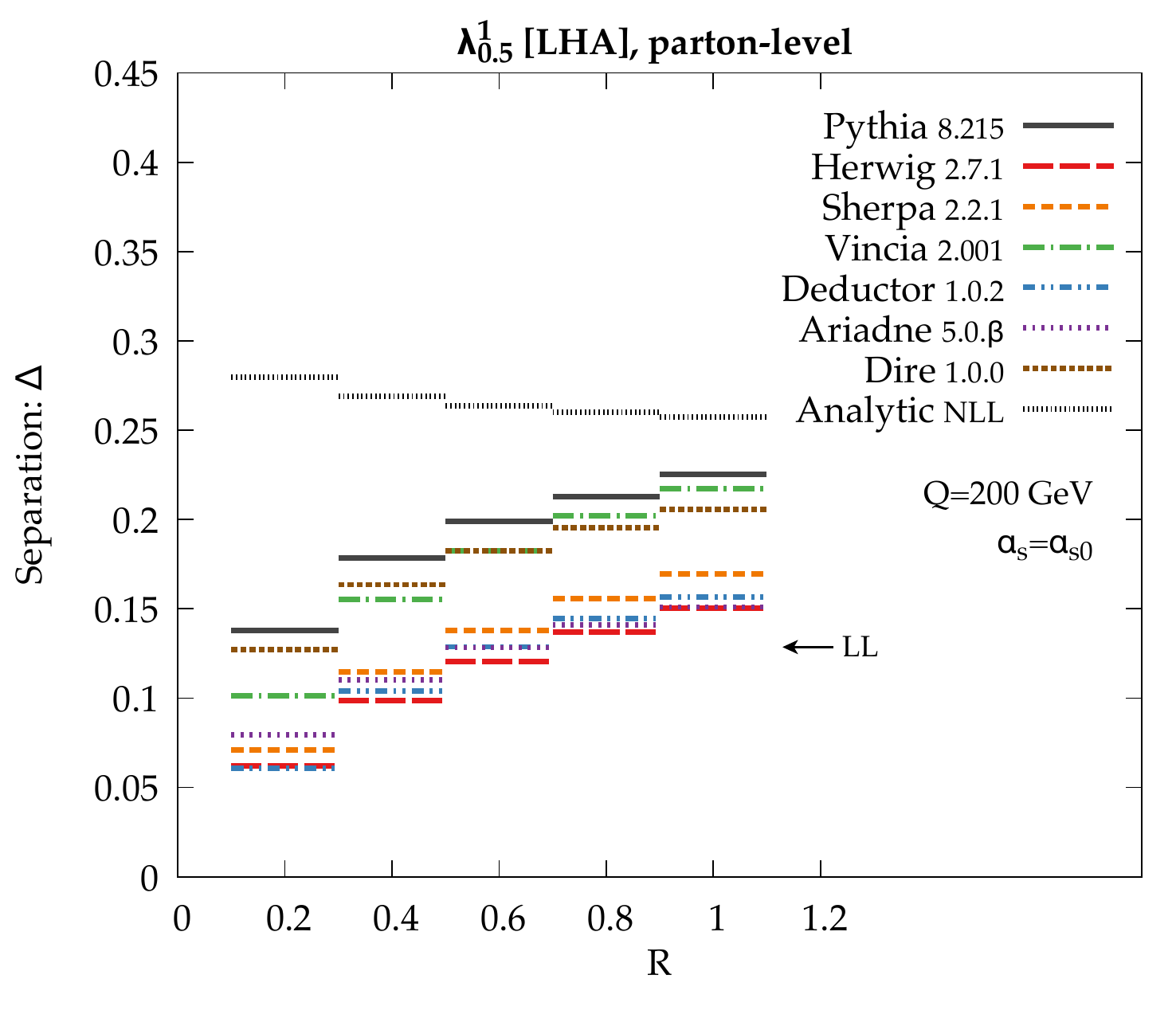}
\label{fig:sweep_R_parton}
}

\subfloat[]{
\includegraphics[width = 0.45\columnwidth]{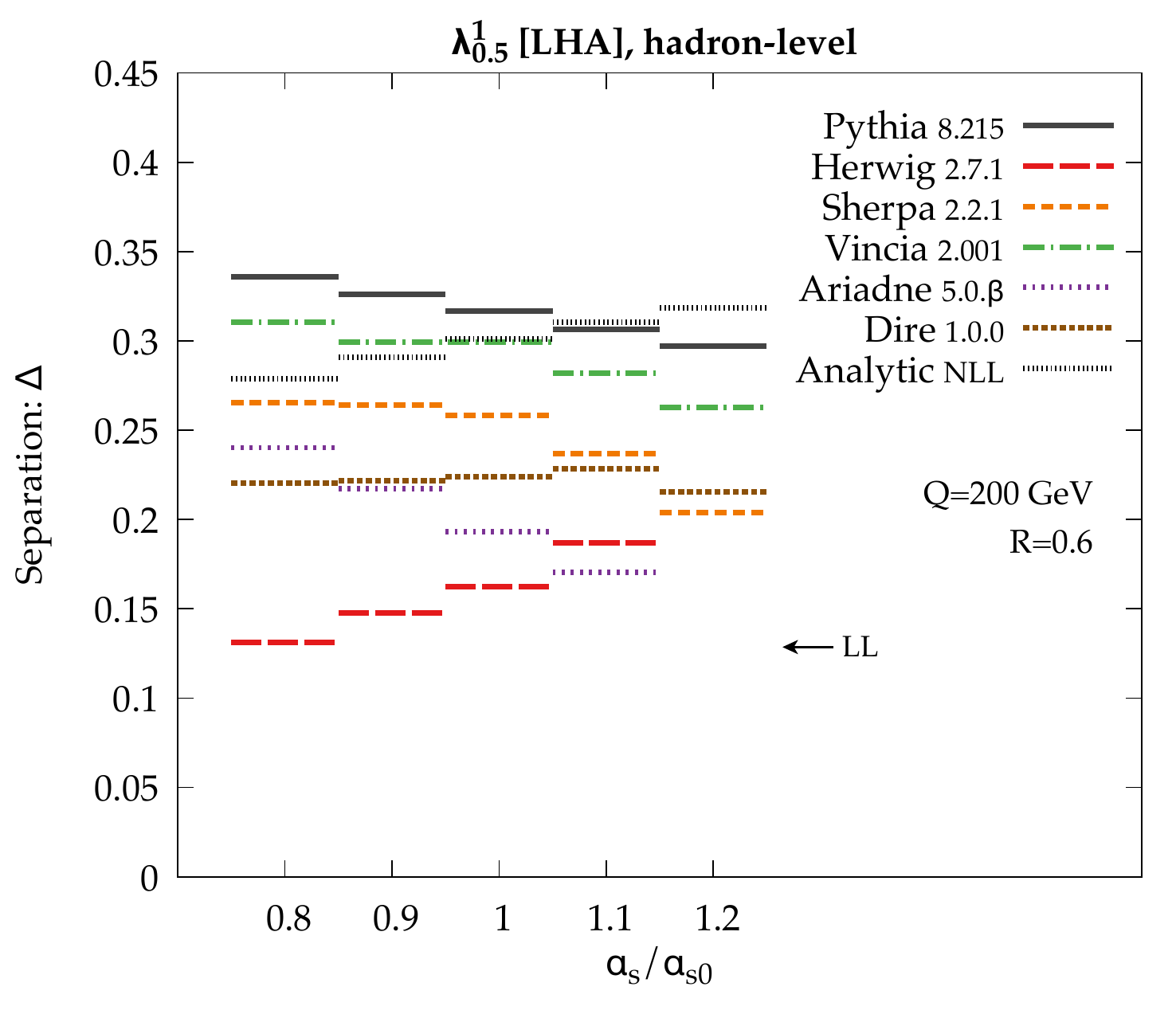}
\label{fig:sweep_as_hadron}
}
$\qquad$
\subfloat[]{
\includegraphics[width = 0.45\columnwidth]{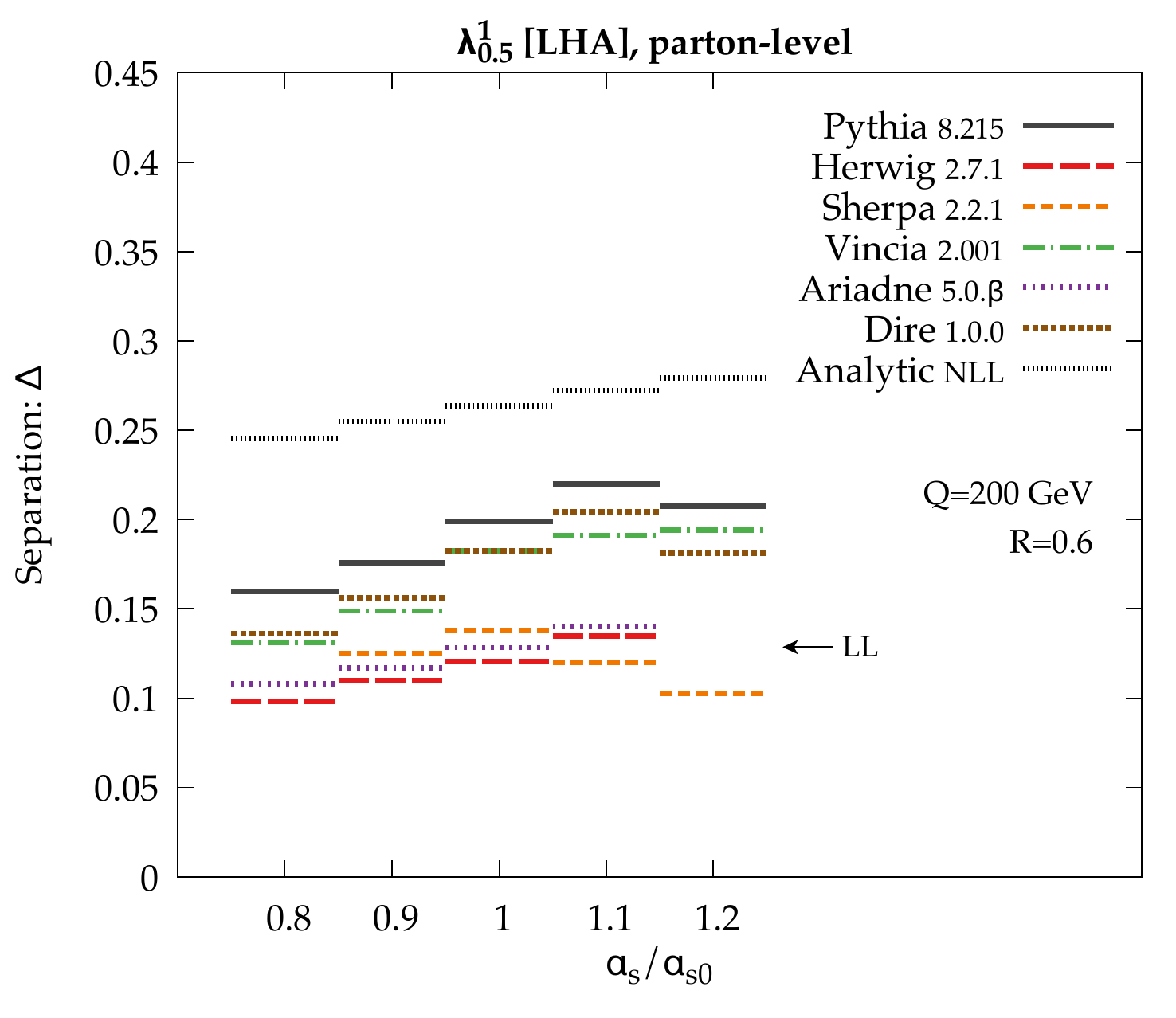}
\label{fig:sweep_as_parton}
}
\caption{Classifier separation $\Delta$ for the LHA, sweeping the collision energy $Q$ (top row), jet radius $R$ (middle row), and coupling constant $\alpha_s/\alpha_{s0}$ (bottom row).  Results are shown at hadron level (left column) and parton level (right column).}
\label{fig:ee_sweep}
\end{figure}

Given the large absolute differences in discrimination power seen above, we next want to check if the parton-shower generators exhibit similar or dissimilar trends as parameters are varied.  We perform three parameter sweeps, using the boldface values below as defaults:
\begin{equation}
\label{eq:ee_sweep_values}
\begin{aligned}
\text{Collision Energy}: \quad Q &= \{ 50, 100, \mathbf{200}, 400, 800\}~\GeV, \\
\text{Jet Radius}: \quad R &= \{ 0.2, 0.4, \mathbf{0.6}, 0.8, 1.0\}, \\
\text{Strong Coupling}: \quad \alpha_s / \alpha_{s0} &= \{0.8,0.9,\mathbf{1.0},1.1,1.2\}, \\
\end{aligned}
\end{equation}
where $\alpha_{s0}$ is the default value of the strong coupling, which is different between the generators (and sometimes different between different aspects of the same generator).

The resulting values of $\Delta$ for the LHA are shown in \Fig{fig:ee_sweep}, at both the hadron level and parton level.   There are number of surprising features in these plots.  Perhaps the most obvious (and seen already in \Fig{fig:summary_all}) is that even for the IRC-safe angularities, the effect of hadronization is rather large, both on the absolute scale of discrimination and the trends.  The main exception to this is \textsc{Herwig}, which does not exhibit as much of a shift from hadronization, though an effect is still present.

The next surprising feature is that the parton-level trends for sweeping $\alpha_s$ do not necessarily correspond to those for sweeping $Q$ and $R$.  According to the perturbative NLL logic in \Ref{Larkoski:2013eya}, quark/gluon discrimination should depend on $\alpha_s$ evaluated at the scale $Q R / 2$, with larger values of $\alpha_s(Q R / 2)$ leading to improved discrimination power.  This is indeed seen in the parton-level curves obtained from the analytic NLL calculation in \Sec{sec:analytic}, and parton-level \textsc{Pythia}, \textsc{Herwig}, \textsc{Vincia}, \textsc{Ariadne}, and \textsc{Dire} also show improved performance with larger $\alpha_s$.  However, larger values of $Q$ and $R$ correspond to smaller values of $\alpha_s$, so the NLL logic would predict that increasing $Q$ or $R$ should lead to worse discrimination power.  Instead, at parton-level, all of the generators show the opposite $Q$ and $R$ trend from the analytic NLL result.

One reason to expect quark/gluon discrimination to improve at higher energies is that the phase space available for shower evolution increases as $Q$ increases.  The scale $\mu$ of the shower splitting is $\mu_0^2 < \mu^2 < Q^2$, where $\mu_0 = \mathcal{O}(\GeV)$ is the shower cutoff scale.  With more range for shower evolution at higher $Q$, there is a greater possibility to see that a quark jet is different from a gluon jet.  Similarly, larger values of $R$ allow for more emissions within a jet, and from scaling symmetry, one expects that parton-level discrimination power should depend on the combination $Q R$.\footnote{At small values of $R$, one has to worry about the flavor purity of a jet sample, since scale evolution can change the leading parton flavor \cite{Dasgupta:2014yra,Dasgupta:2016bnd}.  Similarly, the restriction in \Eq{eq:Erestrict} can impose a non-trivial bias on the jet flavor at small $R$.}  By contrast, the NLL logic says that quark/gluon discrimination should be dominated by the leading emission(s) in a jet, and since $\alpha_s$ is smaller at higher values of $Q R$, those leading emissions are more similar between quarks and gluons.  Given these two different but equally plausible logics, both of which undoubtably play some role in the complete story, this motivates experimental tests of quark/gluon separation as a function of $Q$ and $R$.

For many of the generators, going from parton-level to hadron-level reverses
or flattens the $Q$ and $\alpha_s$ trends, though the $R$ trends are more
stable.  For the NLL results, including the shape function from
\Sec{subsec:shapefuncdef} leads to an overall improvement in discrimination
power and a slight flattening of the $Q$ and $R$ trends, though the
difference between parton-level and hadron-level is not nearly as dramatic as
for the parton showers.  This is further evidence that the boundary between perturbative and nonperturbative physics is ambiguous, and hadron-level comparisons are the most meaningful. 

\subsection{Impact of generator settings}
\label{sec:ee_settings}

Formally, parton-shower generators are only accurate to modified leading-logarithmic (MLL) accuracy, though they include physically important effects like energy/momentum conservation and matrix element corrections that go beyond MLL.  We can assess the impact of these higher-order effects by changing the baseline parameter settings in each parton-shower generator.  We will also explore similar kinds of changes for the NLL analytic calculation.

Because each generator is different, we cannot always make the same changes for each generator.  Similarly, the spread in discrimination power shown below should \emph{not} be seen as representing the intrinsic uncertainties in the shower, since many of these changes we explore are not physically plausible.  The goal of these plots is to demonstrate possible areas where small parameter changes could have a large impact on quark/gluon discrimination.  Ultimately, collider data and higher-order calculations will be essential for understanding the origin of quark/gluon differences.  In all cases, we show both hadron-level and parton-level results, even if a setting is only expected to have an impact at the hadron level.  

\begin{figure}
\centering
\subfloat[]{
\includegraphics[width = 0.45\columnwidth]{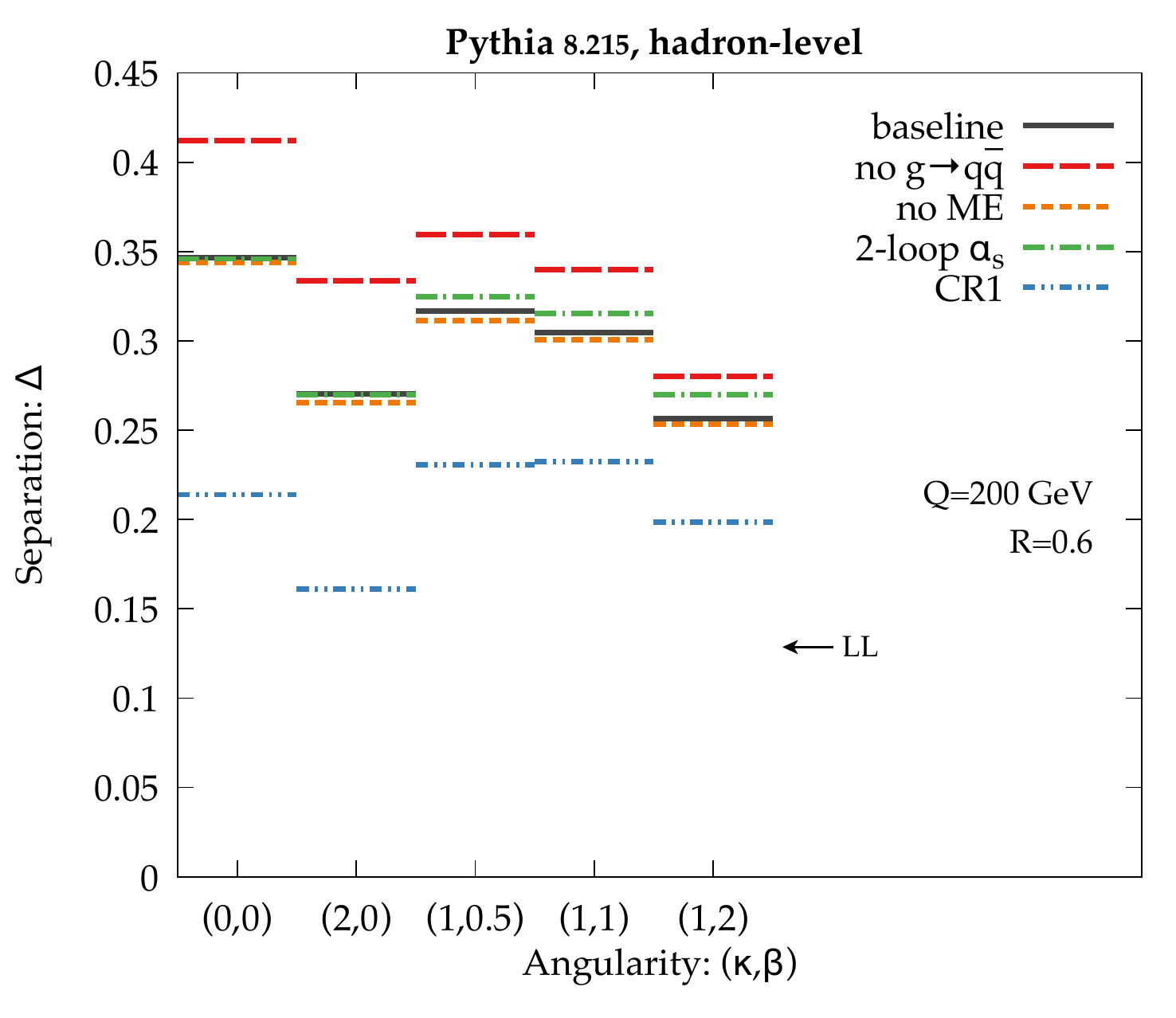}
\label{fig:summary_hadron_pythia}
}
$\qquad$
\subfloat[]{
\includegraphics[width = 0.45\columnwidth]{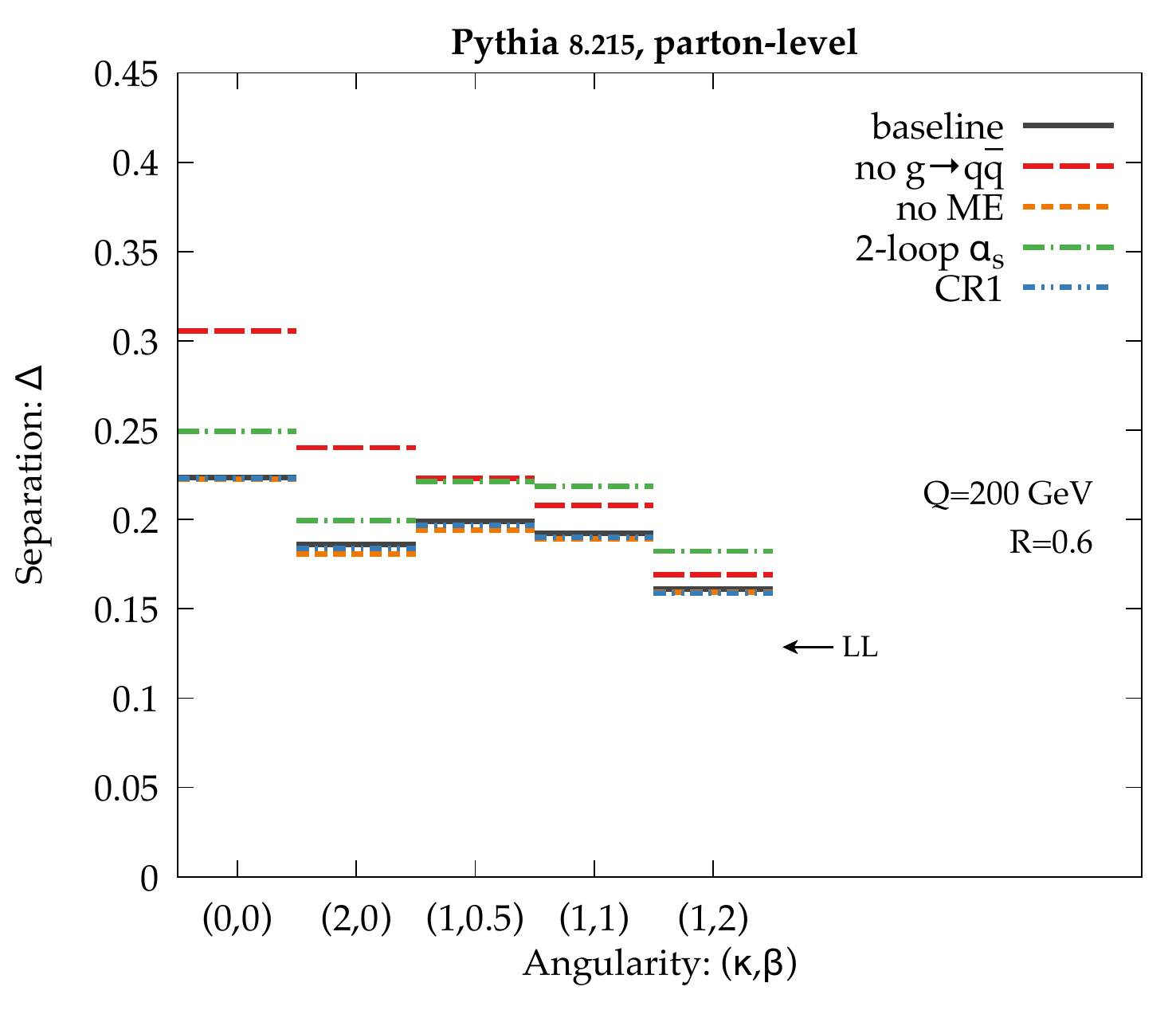}
\label{fig:summary_parton_pythia}
}
\caption{Settings variations for \textsc{Pythia 8.215}.  Shown are (a) hadron-level and (b) parton-level results for the classifier separation $\Delta$ derived from the five benchmark angularities.}
\label{fig:settings_variation_pythia}
\end{figure}

Our $\textsc{Pythia}$ baseline is based on the Monash 2013 tune, with parameters described in \Ref{Skands:2014pea}.  In \Fig{fig:settings_variation_pythia}, we consider the following \textsc{Pythia} variations:
\begin{itemize}
\item \textsc{Pythia: no $g \to q\bar{q}$}.  While the dominant gluon splitting in the parton shower is $g \to gg$, \textsc{Pythia}---and every other shower in this study---also generates the subleading $g \to q \bar{q}$ splittings by default.  This variation turns off $g \to q \bar{q}$, which makes gluon jets look more gluon-like, thereby increasing the separation power.
\item \textsc{Pythia: no ME}.  The first emission in \textsc{Pythia} is improved by applying a matrix element correction \cite{Miu:1998ju}, but this variation turns those corrections off, showing the impact of non-singular terms.  No matrix element correction is available for $h^* \to g g$, though, so the true impact of these corrections might be larger than the relatively small effect seen for this variation.
\item \textsc{Pythia: 2-loop $\alpha_s$}.  The default \textsc{Pythia} setting is to use 1-loop running for $\alpha_s$.  This variation turns on 2-loop running for $\alpha_s$, which has a small (beneficial) effect at parton level which is washed out at hadron level.
\item \textsc{Pythia: CR1}.  Often, one thinks of color reconnection as being primarily important for hadron colliders, but even at a lepton collider, color reconnection will change the Lund strings used for hadronization.  Compared to the baseline, this variation uses an alternative ``\text{SU}(3)''-based color reconnection model~\cite{Christiansen:2015yqa} (i.e.~\texttt{ColourReconnection:mode = 1}).  No attempts were made to retune any of the other hadronization parameters (as would normally be mandated in a tuning context), so this change simply illustrates the effect of switching on this reconnection model with default parameters, leaving all other parameters unchanged.  At parton level, this variation has no effect as expected.  At hadron level, this variation considerably degrades quark/gluon separation compared to the baseline.
\end{itemize}
The most surprising \textsc{Pythia} effect is the large potential impact of the color reconnection model, which is also important for the \textsc{Herwig} generator described next.

\begin{figure}
\centering
\subfloat[]{
\includegraphics[width = 0.45\columnwidth]{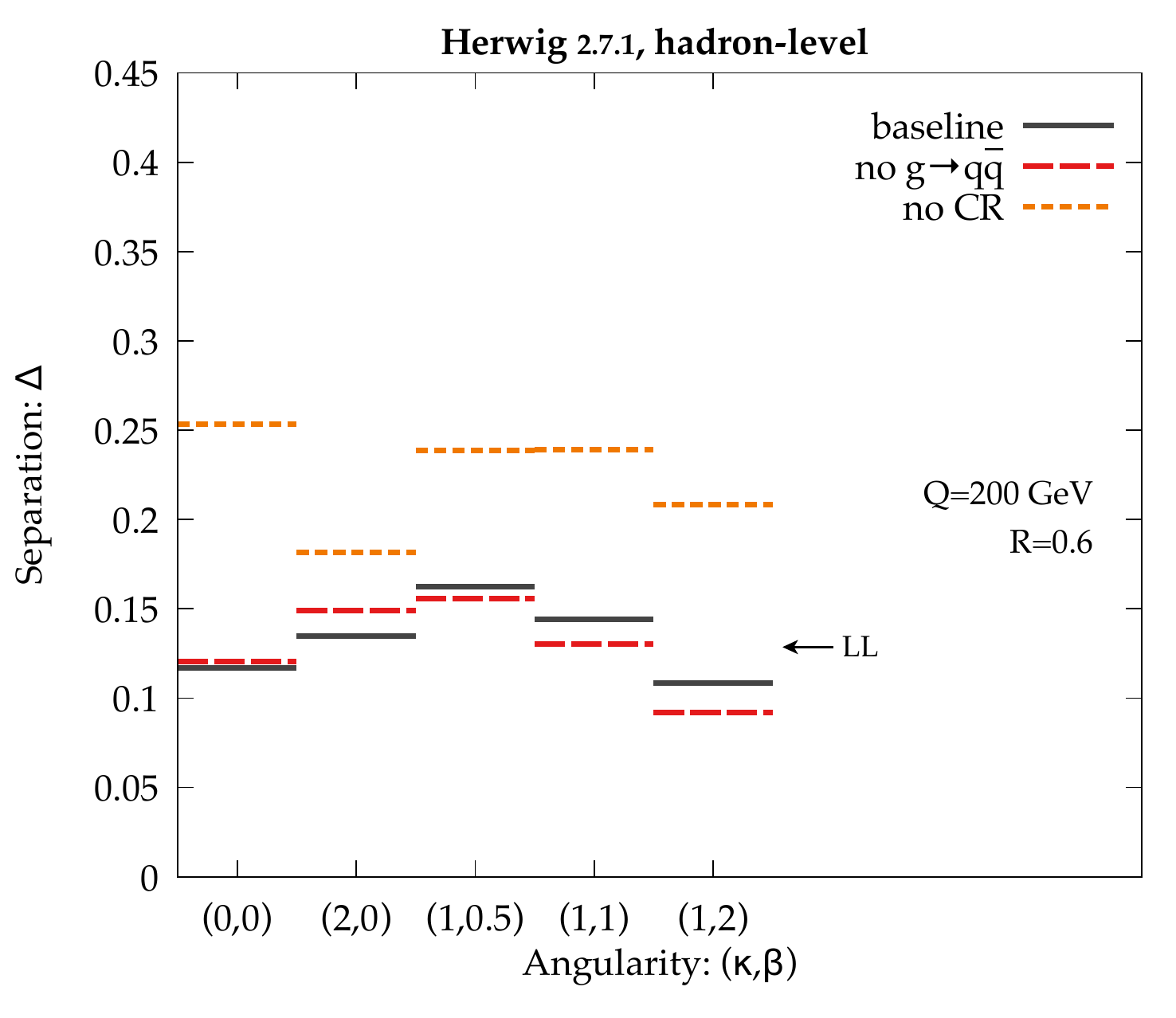}
\label{fig:summary_hadron_herwig}
}
$\qquad$
\subfloat[]{
\includegraphics[width = 0.45\columnwidth]{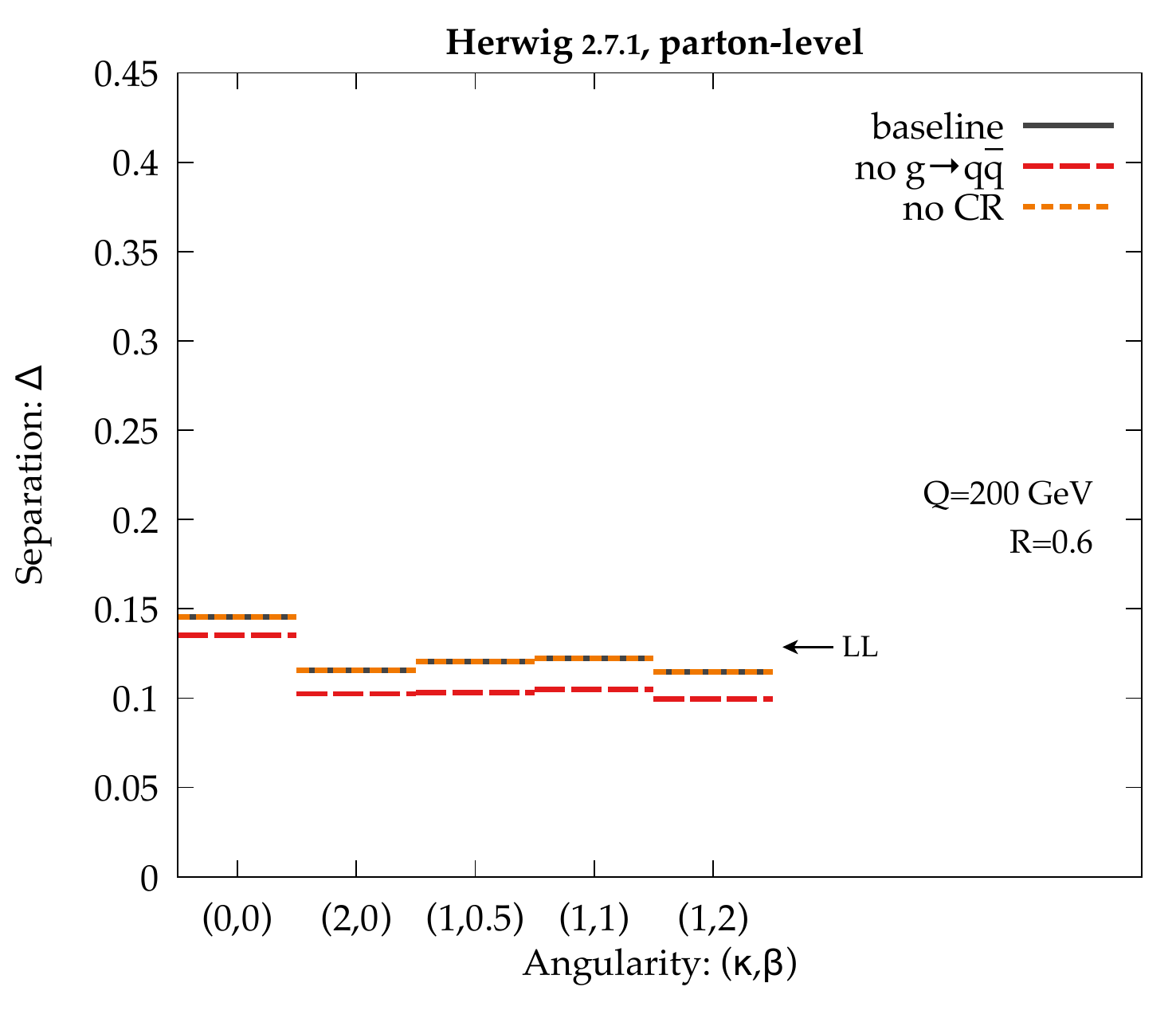}
\label{fig:summary_parton_herwig}
}
\caption{Same as \Fig{fig:settings_variation_pythia}, but for \textsc{Herwig 2.7.1}.}
\label{fig:settings_variation_herwig}
\end{figure}

Our \textsc{Herwig} baseline uses version 2.7.1, with improved modeling of underlying event~\cite{Gieseke:2012ft} and the most recent UE-EE-5-MRST tune~\cite{Seymour:2013qka}, which is able to describe the double-parton scattering cross section~\cite{Bahr:2013gkj} and underlying event data from $\sqrt{s} = 300$~GeV to $\sqrt{s} = 7$~TeV.  In \Fig{fig:settings_variation_herwig}, we consider the following  \textsc{Herwig} variations:
\begin{itemize}
\item \textsc{Herwig: no $g \to q\bar{q}$}.  Turning off $g \to q \bar{q}$ splittings in \textsc{Herwig} has the reverse behavior as seen in \textsc{Pythia}, leading to slightly worse discrimination power, though the effect is modest.
\item \textsc{Herwig: no CR}.  The variation turns off color reconnections in \textsc{Herwig}.  This has no effect at parton level, as expected.  At hadron level, this variation for \textsc{Herwig} gives a rather dramatic improvement in quark/gluon discrimination power.  We think this arises since color reconnection in \textsc{Herwig} allows any color-anticolor pair to reconnect, even if they arose from an initially color octet configuration.  By turning off color reconnection, the gluons look more octet-like, explaining the improvement seen.
\end{itemize}
The importance of color reconnections in \textsc{Herwig} is a big surprise from this study, motivating future detailed studies into which color reconnection models are most realistic when compared to data.  In the future, we also plan to test the default angular-ordered \textsc{Herwig} shower against an alternative dipole shower \cite{Platzer:2011bc}.

\begin{figure}
\centering
\subfloat[]{
\includegraphics[width = 0.45\columnwidth]{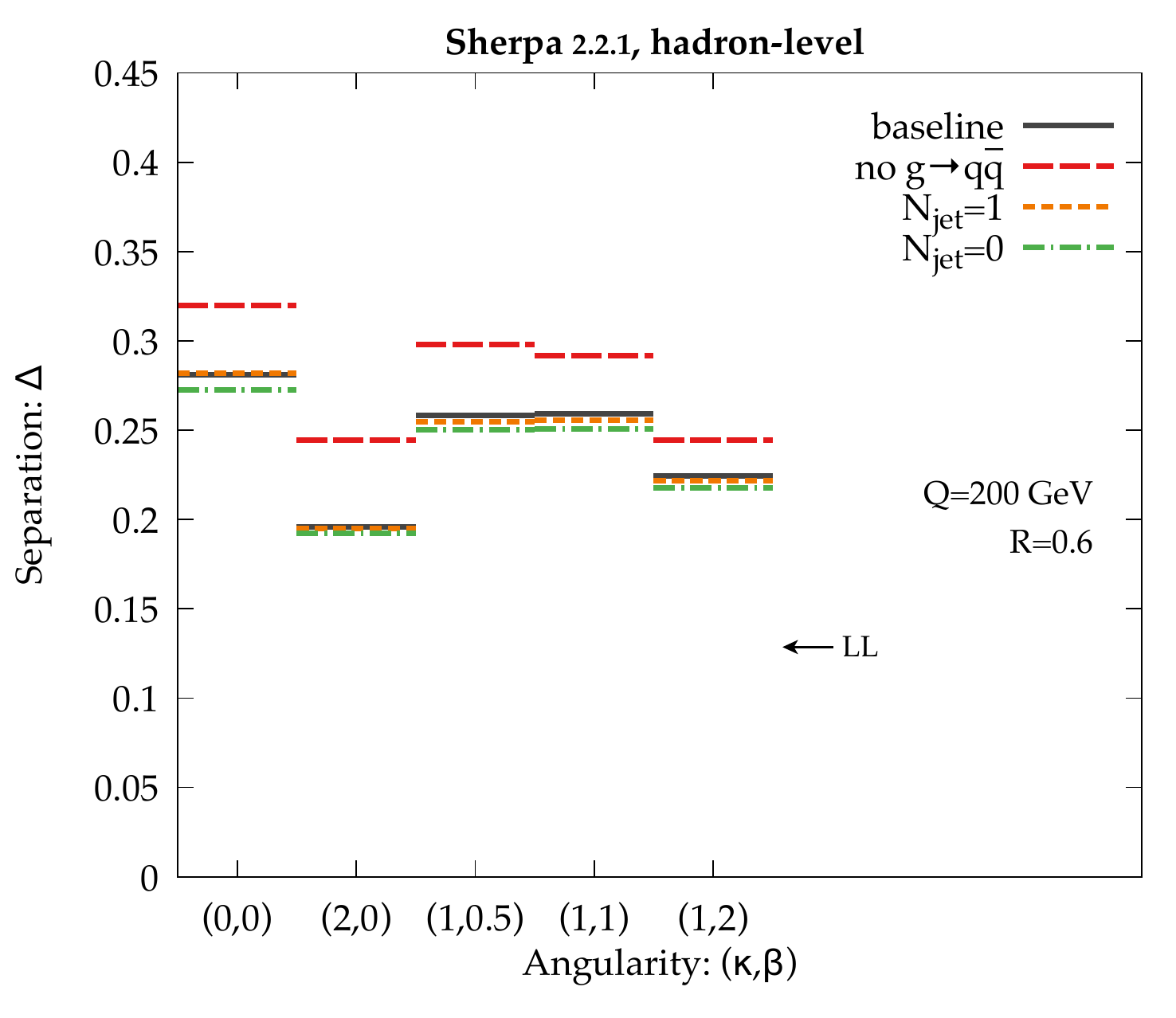}
\label{fig:summary_hadron_sherpa}
}
$\qquad$
\subfloat[]{
\includegraphics[width = 0.45\columnwidth]{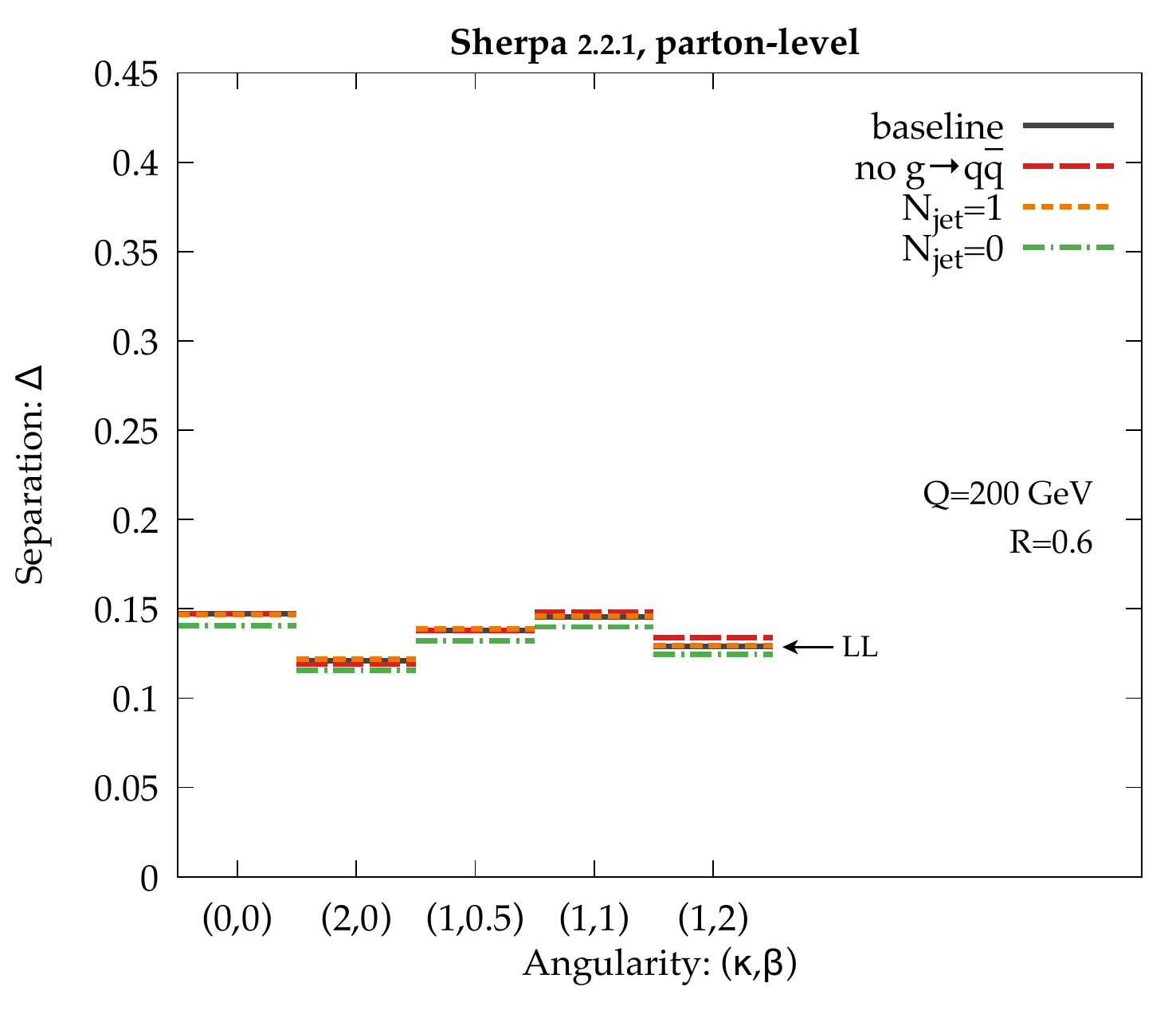}
\label{fig:summary_parton_sherpa}
}
\caption{Same as \Fig{fig:settings_variation_pythia}, but for \textsc{Sherpa 2.2.1}.}
\label{fig:settings_variation_sherpa}
\end{figure}

Our  \textsc{Sherpa} baseline uses matrix element corrections for the first two emissions ($N_\text{jet} = 2$) with CKKW-style matching \cite{Catani:2001cc}.  In \Fig{fig:settings_variation_sherpa}, we consider the following \textsc{Sherpa} variations:
\begin{itemize}
\item \textsc{Sherpa: No $g \to q\bar{q}$}.  Turning off $g \to q \bar{q}$ splittings in \textsc{Sherpa} has a negligible effect at parton level, but it leads to a large jump in discrimination power at hadron level, again due to an interplay between the perturbative shower and nonperturbative hadronization.
\item \textsc{Sherpa: $N_\mathrm{jet} = 1$}.  This variation only performs CKKW matching for the first emission, leading to negligible changes in the discrimination performance.
\item \textsc{Sherpa: $N_\mathrm{jet} = 0$}.  Turning off all matrix element corrections in \textsc{Sherpa} slightly decreases the predicted quark/gluon discrimination power, in agreement with the behavior of \textsc{Pythia}.
\end{itemize}
Within \textsc{Sherpa}, matrix element corrections appear to have a very small effect at parton level.  The large changes seen at hadron level from turning off $g \to q \bar{q}$ splittings motivates further investigations into the shower/hadronization interface.

\begin{figure}
\centering
\subfloat[]{
\includegraphics[width = 0.45\columnwidth]{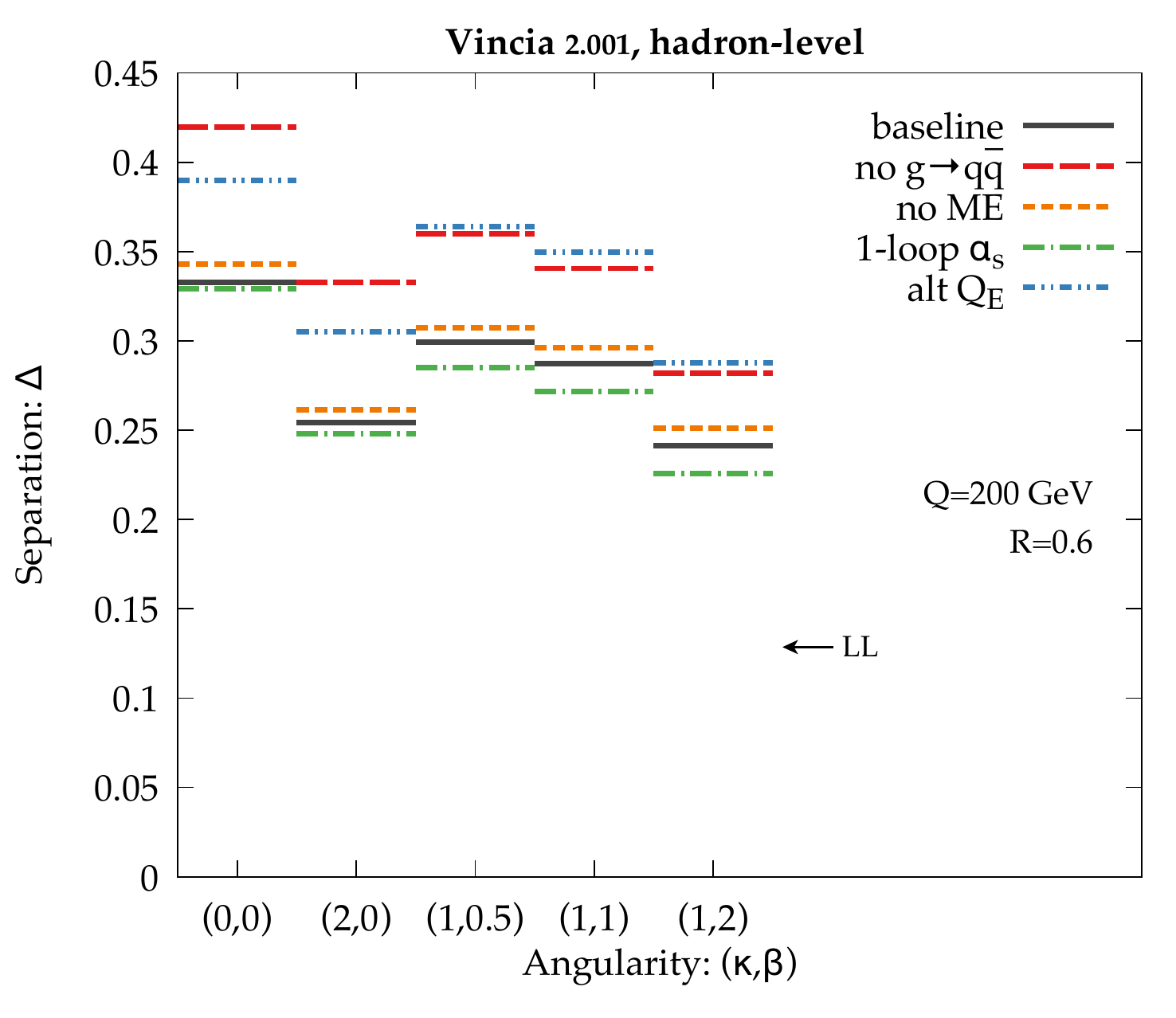}
\label{fig:summary_hadron_vincia}
}
$\qquad$
\subfloat[]{
\includegraphics[width = 0.45\columnwidth]{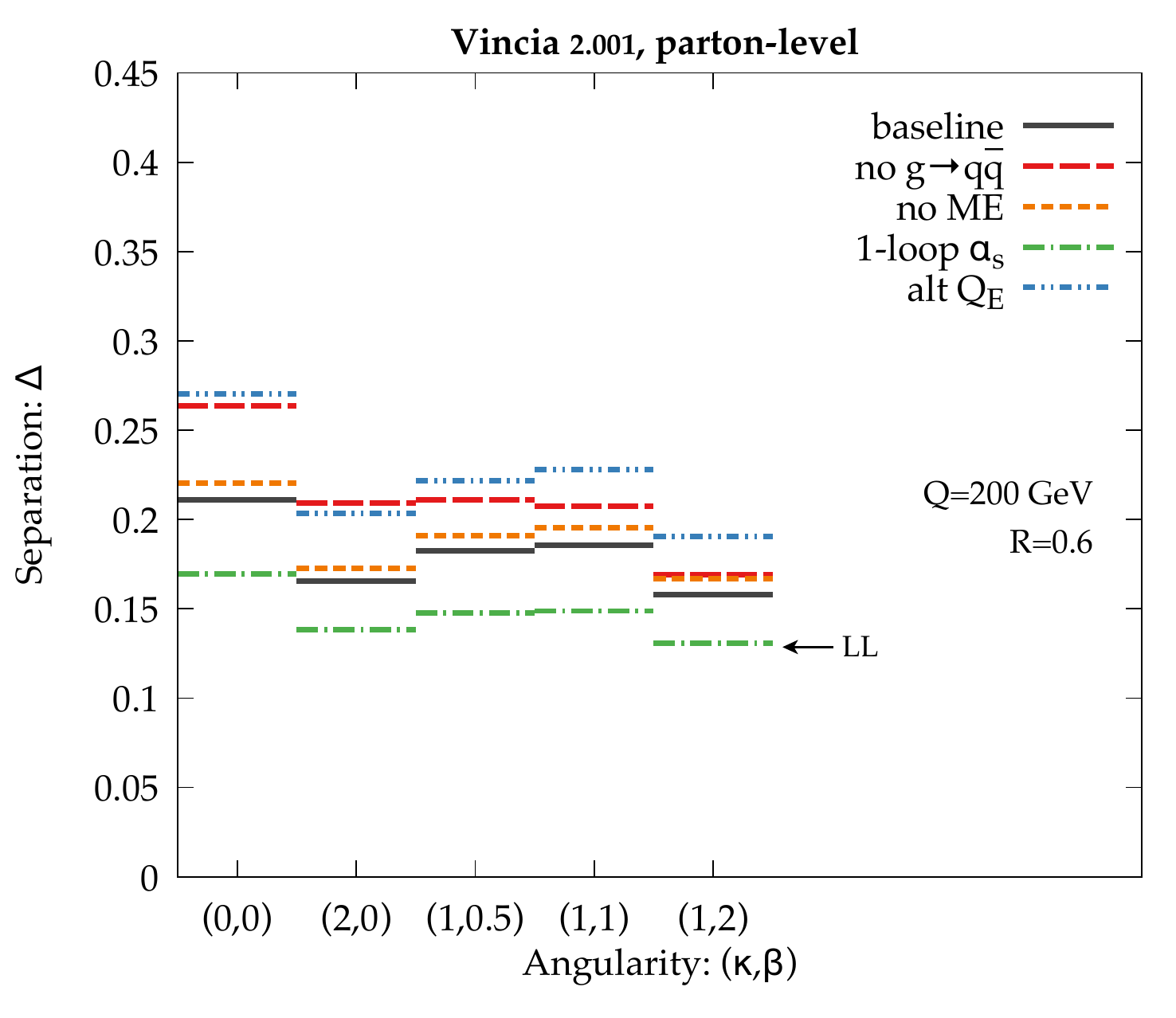}
\label{fig:summary_parton_vincia}
}
\caption{Same as \Fig{fig:settings_variation_pythia}, but for \textsc{Vincia 2.001}.}
\label{fig:settings_variation_vincia}
\end{figure}

Our \textsc{Vincia} baseline uses the default setup for version
2.001~\cite{Fischer:2016vfv}, which includes ``smooth ordering'' and LO matrix-element corrections \cite{Giele:2011cb} up
to ${\cal O}(\alpha_s^3)$ for both $e^+ e^- \to q \bar{q}$ and $e^+ e^- \to gg$.
The coupling $\alpha_s$ is evaluated with 2-loop running defined by $\alpha_s(M_Z)=0.118$
(reinterpreted according to the CMW scheme~\cite{Catani:1990rr}) with 
$\mu_R = 0.6 p_\perp$ as the renormalization scale for gluon emissions
and $\mu_R = 0.5 m_{q\bar{q}}$ for $g\to q\bar{q}$ branchings.
In \Fig{fig:settings_variation_vincia}, we consider the following \textsc{Vincia} variations: 
\begin{itemize}
\item \textsc{Vincia:  no $g \to q\bar{q}$}.  This variation turns off $g \to q \bar{q}$, leading to the expected increase in separation power as seen in \textsc{Pythia}.
\item \textsc{Vincia: no ME}.  By default, each $2 \to 3$ antenna in \textsc{Vincia} has an associated matrix element correction factor.  Since the antennae are already rather close to the true matrix elements, turning off these matrix elements has a modest effect on quark/gluon discrimination power.
\item \textsc{Vincia: 1-loop $\alpha_s$}.  This variation switches
  from 2-loop to 1-loop $\alpha_s$ running, yielding a
  parton-level difference which goes in the same direction as the
  equivalent \textsc{Pythia} variation (note the baseline in
  \textsc{Pythia} is 1-loop running) and a modest hadron-level
  difference, again in agreement with the observation for \textsc{Pythia}. 
\item \textsc{Vincia: alt $Q_E$}.  By default, \textsc{Vincia} uses a
  transverse-momentum scale (the same as in \textsc{Ariadne}) as the
  evolution variable for gluon emissions. This variation
  instead uses a virtuality-like quantity. This changes the Sudakov
  factors to slightly enhance wide-angle emissions over collinear
  ones (see e.g.~\cite{Fischer:2014bja}).  The resulting increase in separation
  power is mainly due to increased activity in the $H\to gg$ shower. 
\end{itemize}
Since \textsc{Vincia} and \textsc{Pythia} share the same hadronisation
model and both have dipole-style showers, it is not surprising that
they exhibit similar behaviors as parameters are changed.
The biggest surprise is the significant change observed when using an alternative shower
evolution variable (``\textsc{alt $Q_E$}''), which persists at hadron
level.
Although this variation is theoretically disfavored (the
default $p_\perp$ evolution variable has been shown to reproduce the logarithmic structure of the
  $q\bar{q}\to qg\bar{q}$ antenna function to second order in
  $\alpha_s$~\cite{Hartgring:2013jma}), formal control of the
  ambiguity would depend on one-loop corrections.  It would therefore be
  interesting to determine the extent to which multi-leg NLO
  merging techniques (such as UNLOPS~\cite{Lonnblad:2012ix}) would
  reduce it, and/or whether second-order corrections to the shower
  kernels are required (for which only a proof of concept currently
  exists~\cite{Li:2016yez}). 

Our \textsc{Deductor} baseline uses leading color plus (LC+) showering, which includes some subleading color structures. We find that switching from LC+ to LC showering at parton level has a negligible impact on quark/gluon discrimination power.  When \textsc{Deductor} interfaces with the default tune of \textsc{Pythia 8.212} for hadronization, only leading color is used in the showering, such that partons with their LC color information can be directly passed to the Lund string model.  No \textsc{Deductor} variations are shown here, though it would be interesting to study the effect of $g \to q \bar{q}$ splitting in future work.

\begin{figure}
\centering
\subfloat[]{
\includegraphics[width = 0.45\columnwidth]{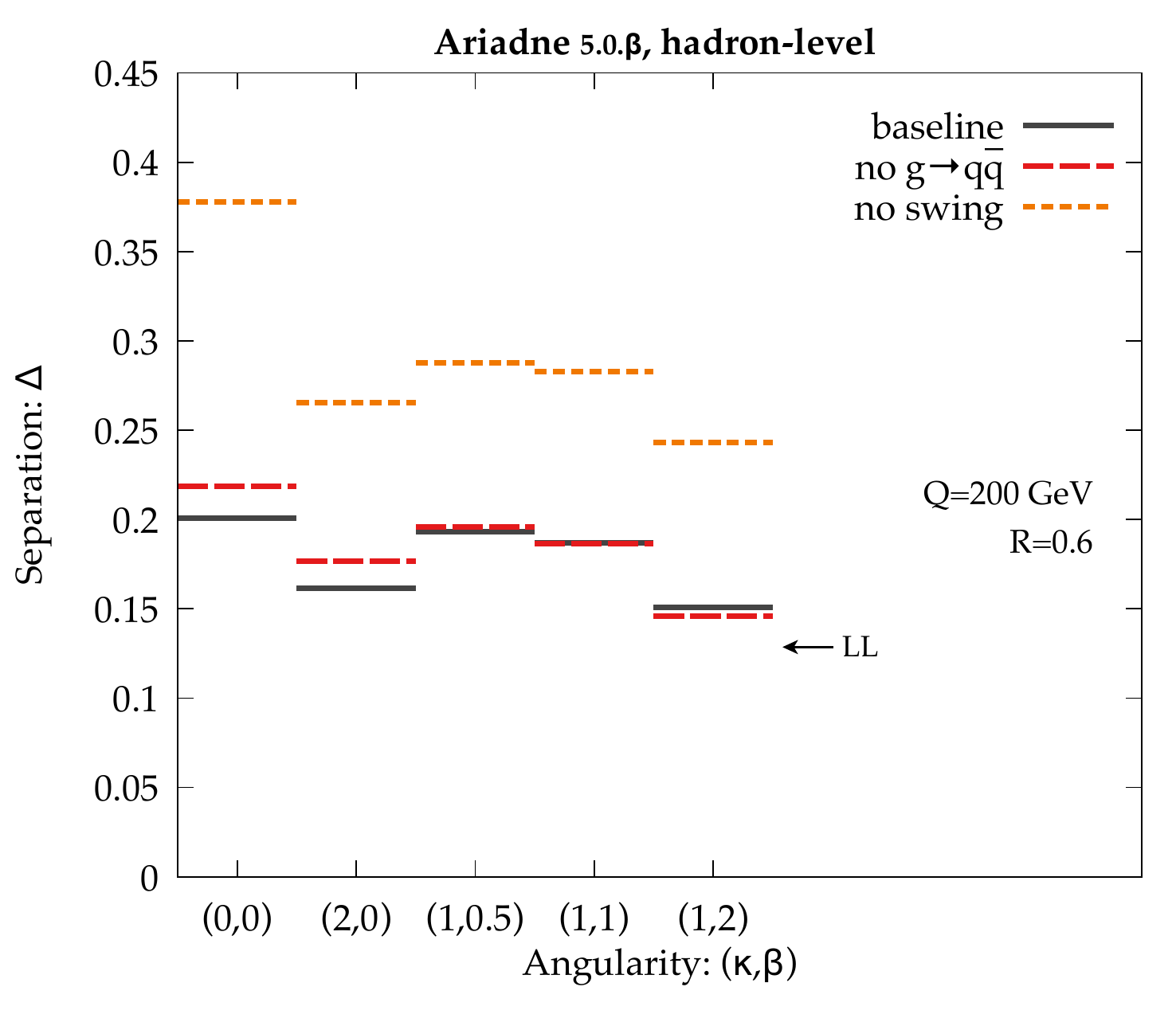}
\label{fig:summary_hadron_ariadne}
}
$\qquad$
\subfloat[]{
\includegraphics[width = 0.45\columnwidth]{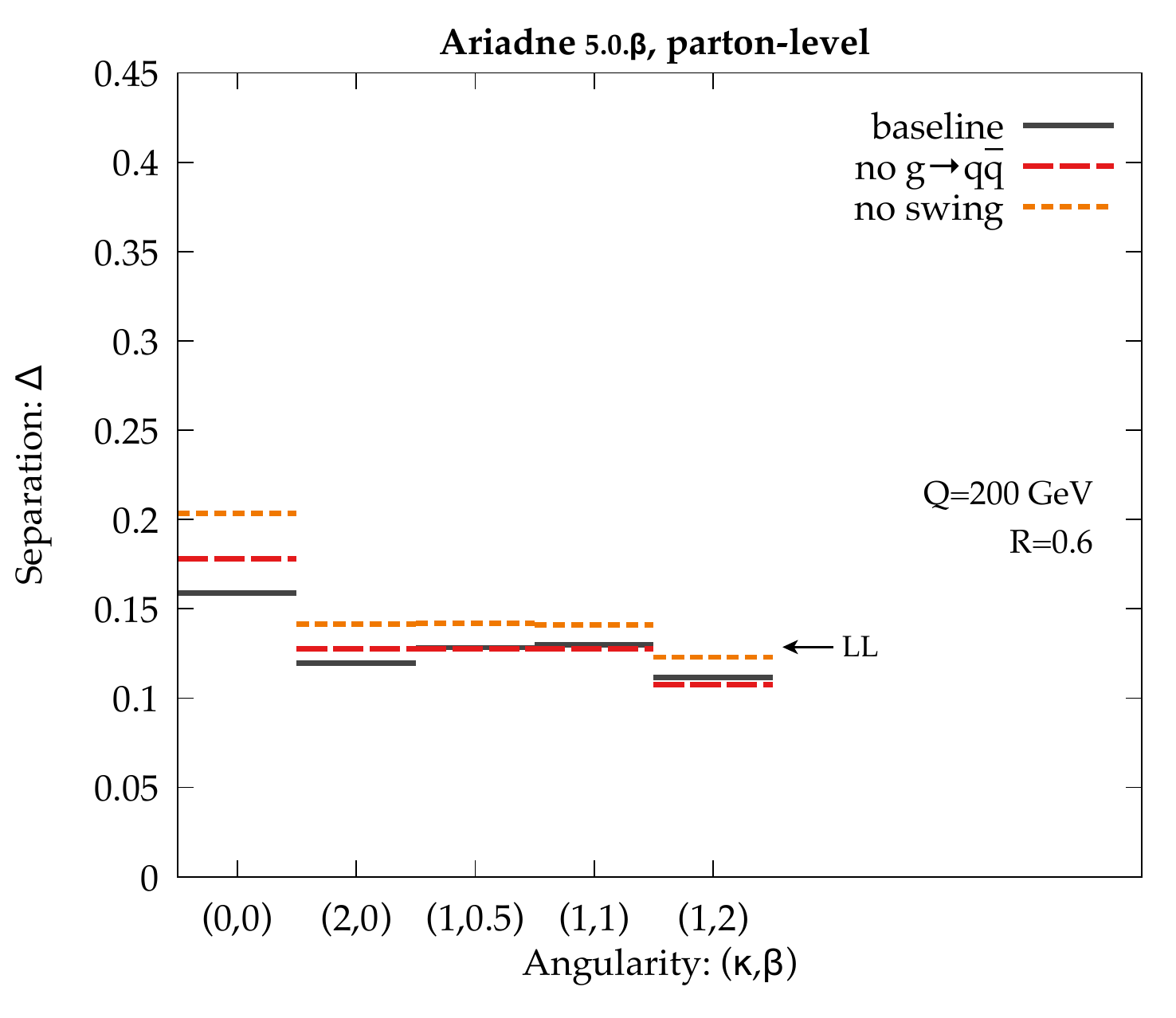}
\label{fig:summary_parton_adriadne}
}
\caption{Same as \Fig{fig:settings_variation_pythia}, but for \textsc{Ariadne 5.0.$\beta$}.}
\label{fig:settings_variation_ariadne}
\end{figure}

Our \textsc{Ariadne} baseline is based on a beta release of version 5.  In \Fig{fig:settings_variation_ariadne}, we consider the following \textsc{Ariadne} variation:
\begin{itemize}
\item \textsc{Ariadne:  no $g \to q\bar{q}$}.  This variation turns off $g \to q \bar{q}$, leading to modest improvement in separation power, similar in magnitude to \textsc{Herwig} though in the opposite direction.
\item \textsc{Ariadne: no swing}.  Swing refers to color reconnections performed during the perturbative cascade, where dipoles in the same color state are allowed to reconnect in a way which prefers low-mass dipoles  \cite{Flensburg:2011kk,Bierlich:2014xba}.  Turning off swing has an effect already at parton level, which is amplified at hadron level, leading to improved quark/gluon separation.
\end{itemize}
Like for \textsc{Pythia} and \textsc{Herwig}, color reconnections play a surprisingly important role in \textsc{Ariadne}.

\begin{figure}
\centering
\subfloat[]{
\includegraphics[width = 0.45\columnwidth]{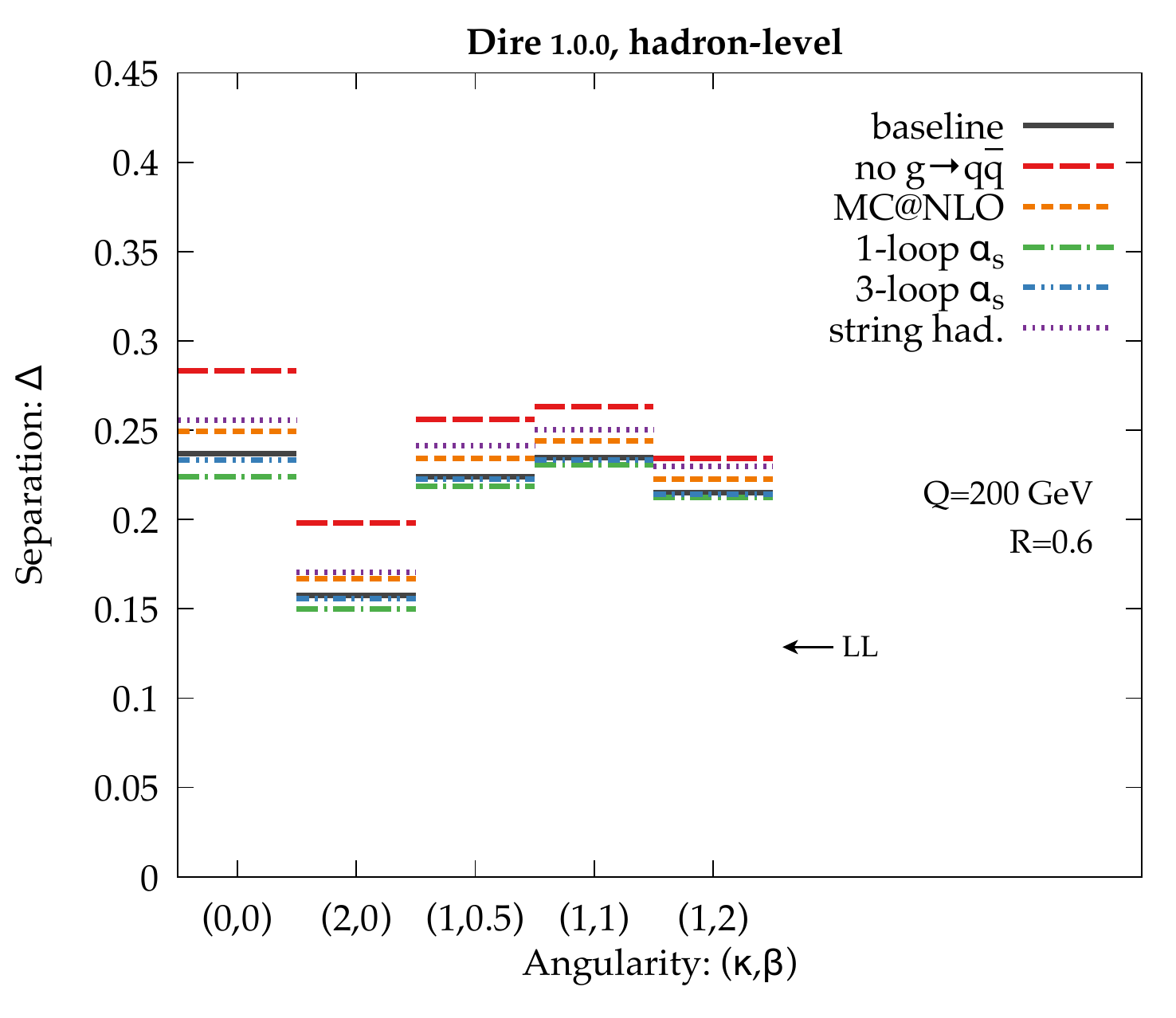}
\label{fig:summary_hadron_dire}
}
$\qquad$
\subfloat[]{
\includegraphics[width = 0.45\columnwidth]{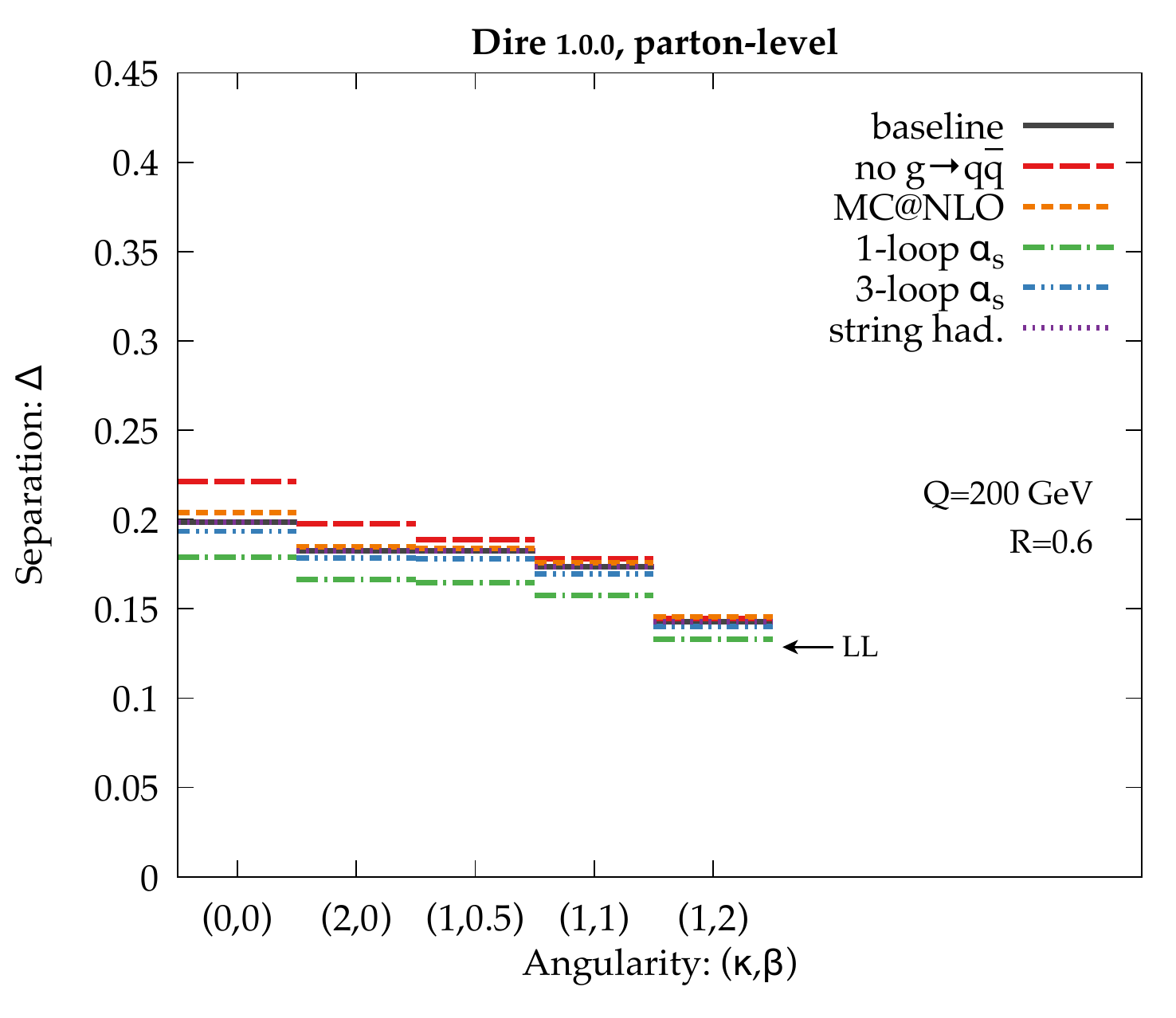}
\label{fig:summary_parton_dire}
}
\caption{Same as \Fig{fig:settings_variation_pythia}, but for \textsc{Dire 1.0.0}.}
\label{fig:settings_variation_dire}
\end{figure}

Our \textsc{Dire} baseline is based on the initial release, interfaced with \textsc{Sherpa} for cluster hadronization.  In \Fig{fig:settings_variation_dire}, we consider the following \textsc{Dire} variations:
\begin{itemize}
\item \textsc{Dire:  no $g \to q\bar{q}$}.  This variation turns off $g \to q \bar{q}$, yielding an improvement in separation power at both the parton level and hadron level, intermediate between \textsc{Ariadne} and \textsc{Vincia}.
\item \textsc{Dire: MC@NLO}.  This variation uses MC@NLO \cite{Frixione:2002ik} as implemented in \textsc{Sherpa} to provide a matrix element correction.  The discrimination power slightly improves at both parton and hadron level, though not that much, since the \textsc{Dire} shower already is very close to capturing the matrix element for the first emission.
\item \textsc{Dire: 1-loop $\alpha_s$}.  The default within \textsc{Dire} is to perform 2-loop $\alpha_s$ running.  This variation uses just 1-loop running, with a slight degradation of discrimination power.
\item \textsc{Dire: 3-loop $\alpha_s$}.  Using 3-loop running also degrades performance, but by a very small amount.
\item \textsc{Dire: string had}.  This variation uses \textsc{Pythia} for Lund string fragmentation, which only has an effect at hadron level.  This leads to a modest improvement in discrimination power, suggesting that long-range color connections can play an important role in quark/gluon discrimination.  Note that the shower cutoff scale is the same for cluster and string fragmentation in \textsc{Dire}.
\end{itemize}
Of the generators we tested, \textsc{Dire} is the only one that interfaces with two different hadronization routines, motivating further studies into the differences between cluster and string fragmentation.

\begin{figure}
\centering
\subfloat[]{
\includegraphics[width = 0.45\columnwidth]{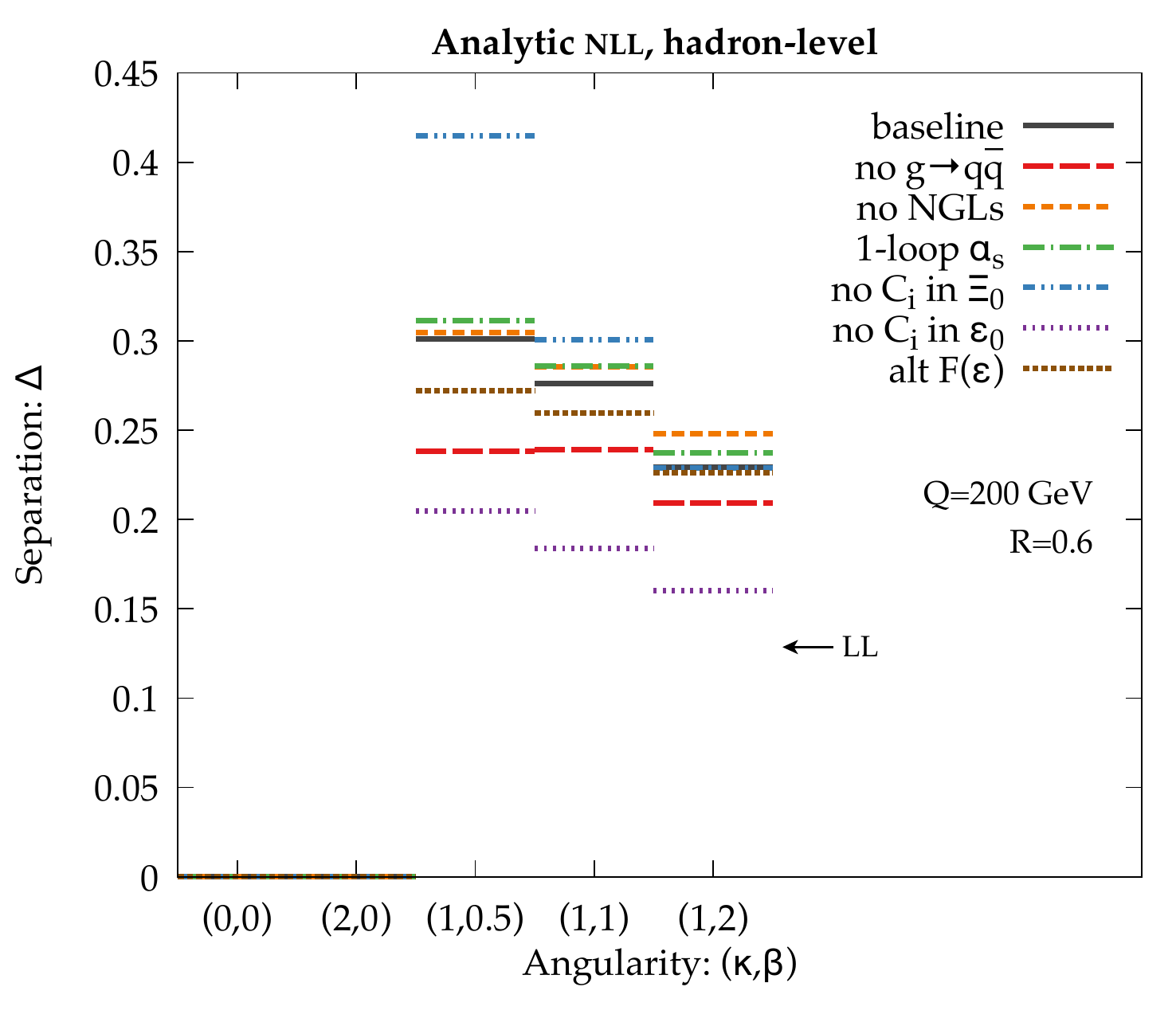}
\label{fig:summary_hadron_analytic}
}
$\qquad$
\subfloat[]{
\includegraphics[width = 0.45\columnwidth]{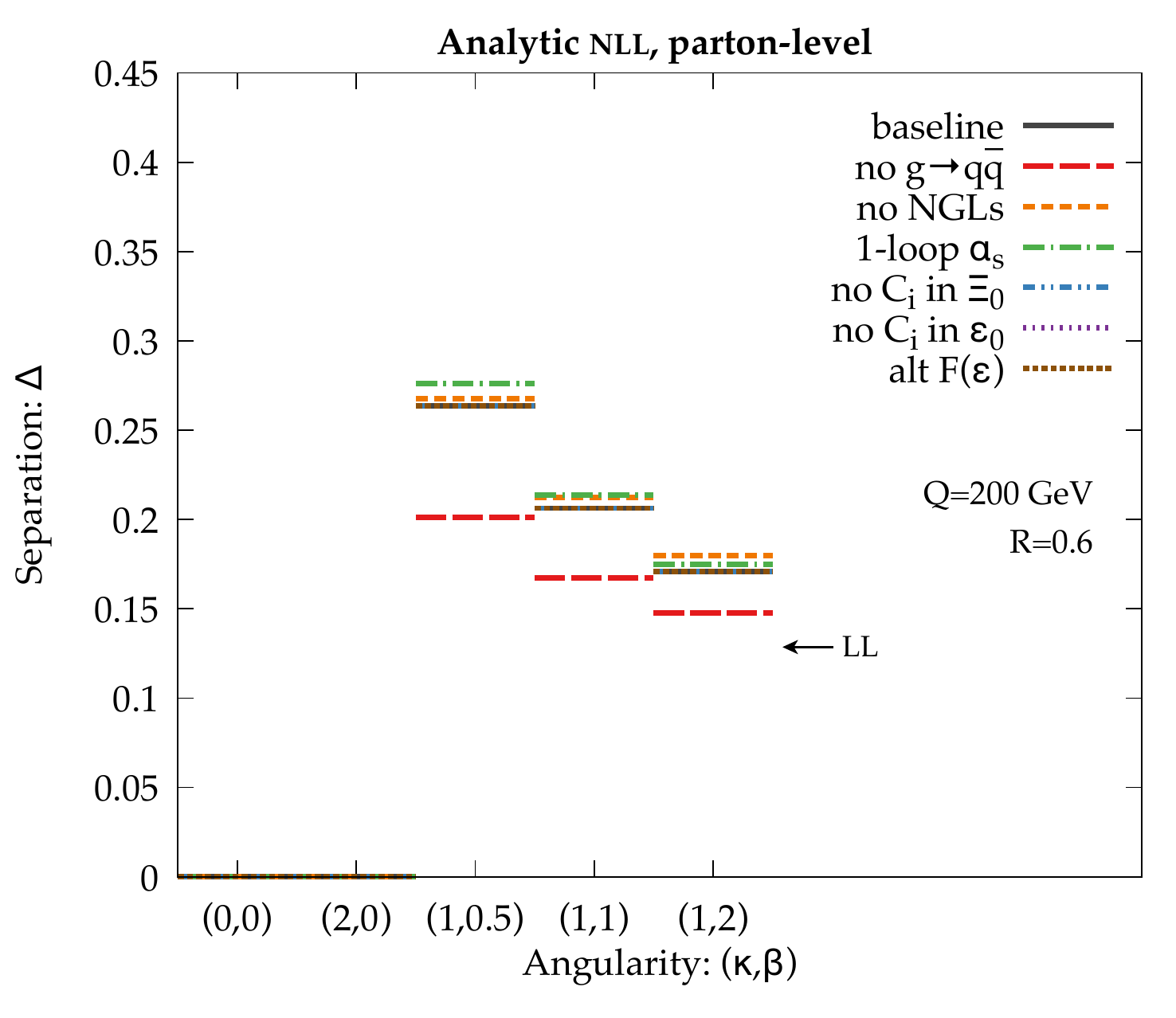}
\label{fig:summary_parton_analytic}
}
\caption{Same as \Fig{fig:settings_variation_pythia}, but for the analytic NLL calculation from \Sec{sec:analytic}.}
\label{fig:settings_variation_analytic}
\end{figure}

Finally in \Fig{fig:settings_variation_analytic}, we consider the analytic NLL calculation from \Sec{sec:analytic}.  Here, we can only study the IRC-safe angularities with $\kappa = 1$.  
\begin{itemize}
\item \textsc{Analytic NLL: no $g \to q\bar{q}$}.  To turn off gluon splitting to quarks, we set $n_f = 0$ in \Eq{eq:reducedsplitting}, without adjusting the running of $\alpha_s$.  This effectively decreases the number of emissions from gluons, making them look more quark-like.  The resulting decrease in discrimination power is the opposite of the behavior seen in the parton-shower generators (except \textsc{Herwig}), suggesting that at higher perturbative orders, the effect of $g \to q\bar{q}$ will go beyond just changing the reduced splitting functions.
\item \textsc{Analytic NLL: no NGLs}.  Here, we set $f_{\rm NGL} = 0$ in \Eq{eq:NLLcaesar}.  Since nonglobal logarithms obey Casimir scaling in the $N_c \to \infty$ limit, this is expected to have a mild impact on quark/gluon separation power, which is indeed the case. 
\item \textsc{Analytic NLL: 1-loop $\alpha_s$}.  The analytic NLL calculation uses 2-loop $\alpha_s$ running by default.  This option uses only 1-loop running, which has a relatively small (beneficial) impact.
\item \textsc{Analytic NLL: no $C_i$ in $\Xi_0$}.  The default choice for the average nonperturbative shift $\epsilon_0$ assumes Casimir scaling as in \Eq{eq:OmegaXiValues}.    This option uses instead the shift in \Eq{eq:OmegaXiValuesHalfAlt}, which only has Casimir scaling for $\Omega_0$ and not for $\Xi_0$.  This makes a dramatic impact for $\beta < 1$ at hadron-level, since $\Xi_0$ dominantly controls the impact of nonperturbative collinear emissions.  Specifically, the default $\epsilon_0$ scales like $C_i^\beta$ for $\beta \ll 1$ whereas this option has linear scaling with $C_i$, leading to increased discrimination power.
\item \textsc{Analytic NLL: no $C_i$ in $\epsilon_0$}.  This option uses the nonperturbative shift in \Eq{eq:OmegaXiValuesAlt}, which is the same for quarks and gluons.  As expected, this reduces the difference between quark and gluon jets at hadron-level, leading to a large reduction in discrimination power.
\item \textsc{Analytic NLL: alt $F(\epsilon)$}.  Here, we change the functional form of the shape function in \Eq{eq:Foptions} from $F(\epsilon)$ to $F_{\rm alt}(\epsilon)$, keeping the same value of $\epsilon_0$.  Since $F_{\rm alt}$ has a larger high-side tail, there is more overlap of the quark and gluon distributions, reducing somewhat the discrimination power.
\end{itemize}
The key lesson from this analytic study is that the form of the nonperturbative shape function has a large effect on quark/gluon discrimination power, especially the assumed dependence of $\epsilon_0$ on the Casimir factor.  So while higher-order perturbative calculations of quark/gluon radiation patterns are essential, quantitative control over nonperturbative physics will be required to make robust statements about the predicted discrimination power.

\section{Quark/gluon tagging at the LHC}
\label{sec:pp}

It is clear from our $e^+e^-$ study that quark/gluon radiation patterns face considerable theoretical uncertainties, as seen from the differing behaviors of parton-shower generators and from the importance of the shape function in the analytic calculation.  This is true even accounting only for final-state physics effects, so additional initial-state complications can only increase the uncertainties faced in $pp$ collisions at the LHC.  Beyond just the application to quark/gluon tagging, this is an important challenge for any analysis that uses jets.  For example, a proper experimental determination of jet energy scale corrections requires robust parton-shower tools that correctly model effects like out-of-cone radiation.

Eventually, one would like to perform improved analytic calculations to address these radiation pattern uncertainties.  In the near term, though, measurements from the LHC will be essential for improving the parton-shower modeling of jets.  In this section, we perform an example LHC analysis that highlights the kind of information one can gain about quark/gluon radiation patterns, despite the additional complications faced by hadronic collisions.

\subsection{Defining enriched samples}

\begin{figure}
\centering
\subfloat{
\includegraphics[width = 0.45\columnwidth]{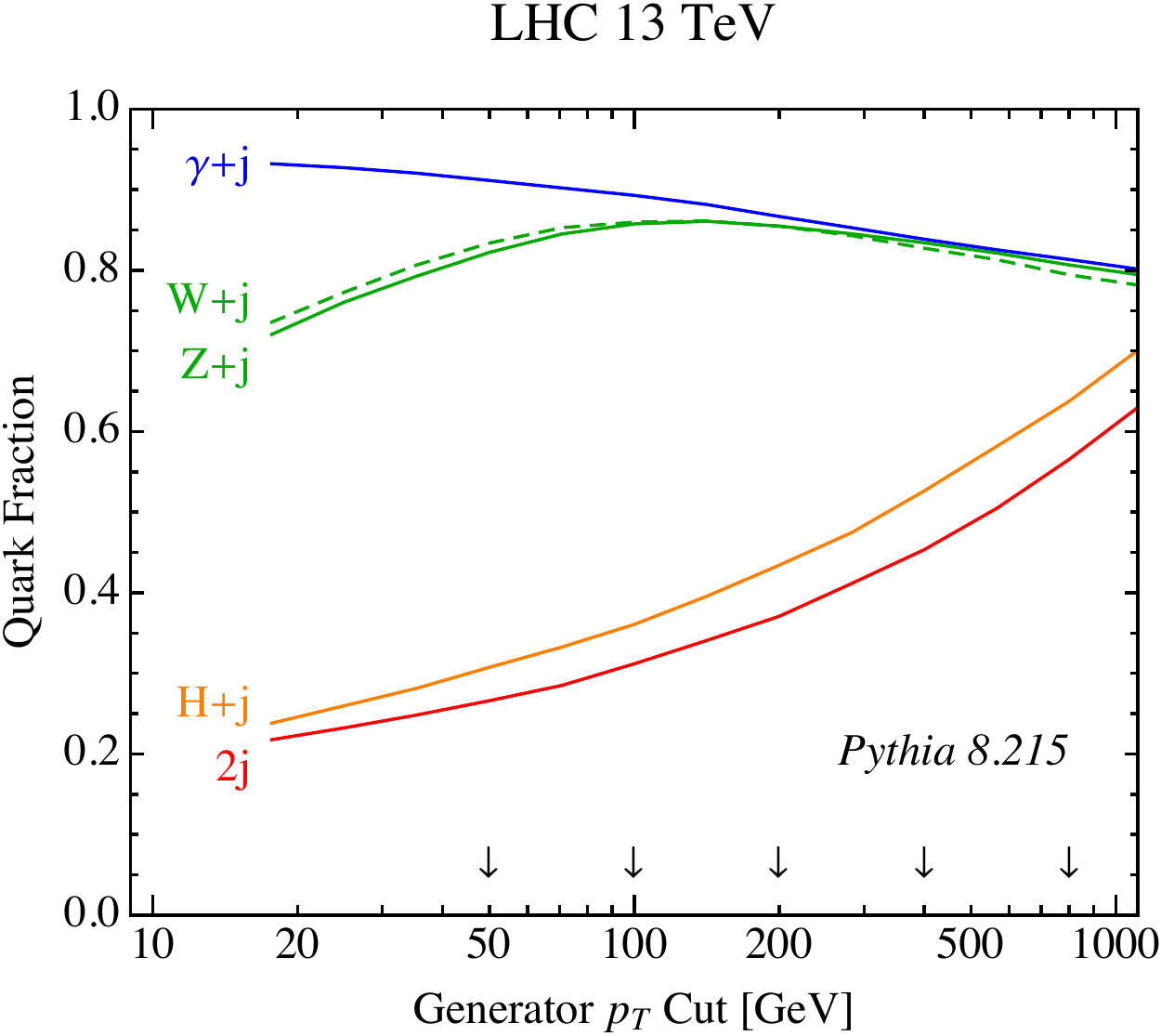}
}
\caption{Quark fraction of jets at parton level, as defined by the Born-level parton flavor.  The arrows indicate the $p_T^{\rm min}$ values used for the study in \Sec{subsec:pp_para}.}
\label{fig:parton_level_qg_composition}
\end{figure}

As discussed in \Sec{sec:def}, there is no way to isolate pure samples of quark or gluon jets at the LHC, but one can isolate quark/gluon-enriched samples, as defined by the flavor label of the jet in the corresponding Born-level partonic process.  As shown in \Fig{fig:parton_level_qg_composition}, the Born-level jet in $W/Z/\gamma + \text{jet}$ is more than 70\% quark enriched over the entire jet $p_T$ range of interest.  For jets softer than around 200 GeV, the Born-level jet in dijets or $H+\text{jet}$ is more than 60\% gluon enriched, with that fraction decreasing as the jet $p_T$ increases.  More sophisticated enrichment procedures are described in \Ref{Gallicchio:2011xc}.

In principle, one could try to ``diagonalize" some combination of vector boson plus jet and dijet samples in order to define separate quark or gluon samples (see e.g.~\cite{Aad:2014gea}).  In the spirit of \Sec{sec:def}, though, we think it is more beneficial for the LHC experiments to perform process-specific measurements without trying to directly determine their quark and gluon composition.  Instead of quark/gluon separation, here we ask the more well-defined question of whether one can tell ``the jet in $Z$ plus jets" (quark-enriched) apart from ``the jet in dijets" (gluon-enriched).\footnote{See, however, \Ref{Dery:2017fap} for a machine-learning approach to handle mixed quark/gluon samples.}  In a similar spirit, one could test for differences within a single jet sample, such as comparing central jets versus forward jets in dijet production.  This process-based strategy can help sidestep the known process dependance of defining quarks and gluons at the LHC, where color correlations have an important impact on observed jet radiation patterns.

For this study, we study proton-proton collisions at the 13 TeV LHC.  We consider four different hadron-level generators---\textsc{Pythia 8.215} \cite{Sjostrand:2014zea}, \textsc{Herwig 2.7.1} \cite{Bahr:2008pv,Bellm:2013hwb}, \textsc{Sherpa 2.2.1} \cite{Gleisberg:2008ta}, and \textsc{Vincia 2.001} \cite{Fischer:2016vfv}---using $Z \to \mu^+ \mu^-$ plus jets as our quark-enriched sample and dijets as our gluon-enriched sample.  All of these generators are used with their default settings, including underlying event modeling and hadronization.  We set $R = 0.4$ as the default jet radius, with jets defined by the anti-$k_t$ algorithm, in keeping with current jet studies at the LHC, exploring other values in \Sec{subsec:pp_para}.  Hadrons with rapidity $|y| < 2.5$ are used for jet clustering, and the resulting jets are restricted to have $|y_{\rm jet}| < 1.5$.  We apply a minimum $p_T$ cut with default value $p^{\min}_T = 100~\GeV$, similar in spirit to the $Q/2$ value used in the $e^+e^-$ study, though the precise meaning of $p_T^{\rm min}$ differs between the two samples. 

The specific analysis routines used for this $pp$ study are available from \url{https://github.com/gsoyez/lh2015-qg}.  For the $Z$ plus jets analysis (with \textsc{Rivet} routine \verb|MC_LHQG_Zjet.cc|), the selection criteria for the reconstructed $Z$ boson and jet are:
\be
\text{$pp \to Z + j$ (``quark-enriched'')} : \quad p^{Z}_T  > p^\text{min}_T, \qquad \frac{p^\text{jet}_T}{p^{Z}_T} > 0.8, \qquad |y_{\rm jet} - y_Z| < 1.0.
\ee
In addition, we apply a $p_T > 5~\GeV$ cut on each muon.  For the dijet analysis (with \textsc{Rivet} routine \verb|MC_LHQG_dijet.cc|), our selection is based on the two hardest jets (labeled $1$ and $2$), both of which are used for analysis if they satisfy:
\be
\text{$pp \to 2j$ (``gluon-enriched'')} : \quad \frac{p_{T,1} + p_{T,2}}{2} > p^\text{min}_T, \qquad \frac{p_{T,2}}{p_{T,1}} > 0.8, \qquad |y_1 - y_2| < 1.0.
\ee 
We study the same five benchmark angularities from
\Eq{eq:benchmarkang}, but we also test the impact of soft radiation
removal using mMDT grooming ($\mu = 1$ and $z_{\rm cut} = 0.1$), with
the grooming condition given in \Eq{eq:mMDTcriteria}.
Prior to both the computation of $\lambda^{\kappa}_{\beta}$ and the application of the
mMDT procedure, the jet constituents are reclustered with the
C/A algorithm, using the winner-take-all recombination scheme.

\subsection{Baseline analysis}

\begin{figure}
\centering
\subfloat[]{
\includegraphics[width = 0.45\columnwidth]{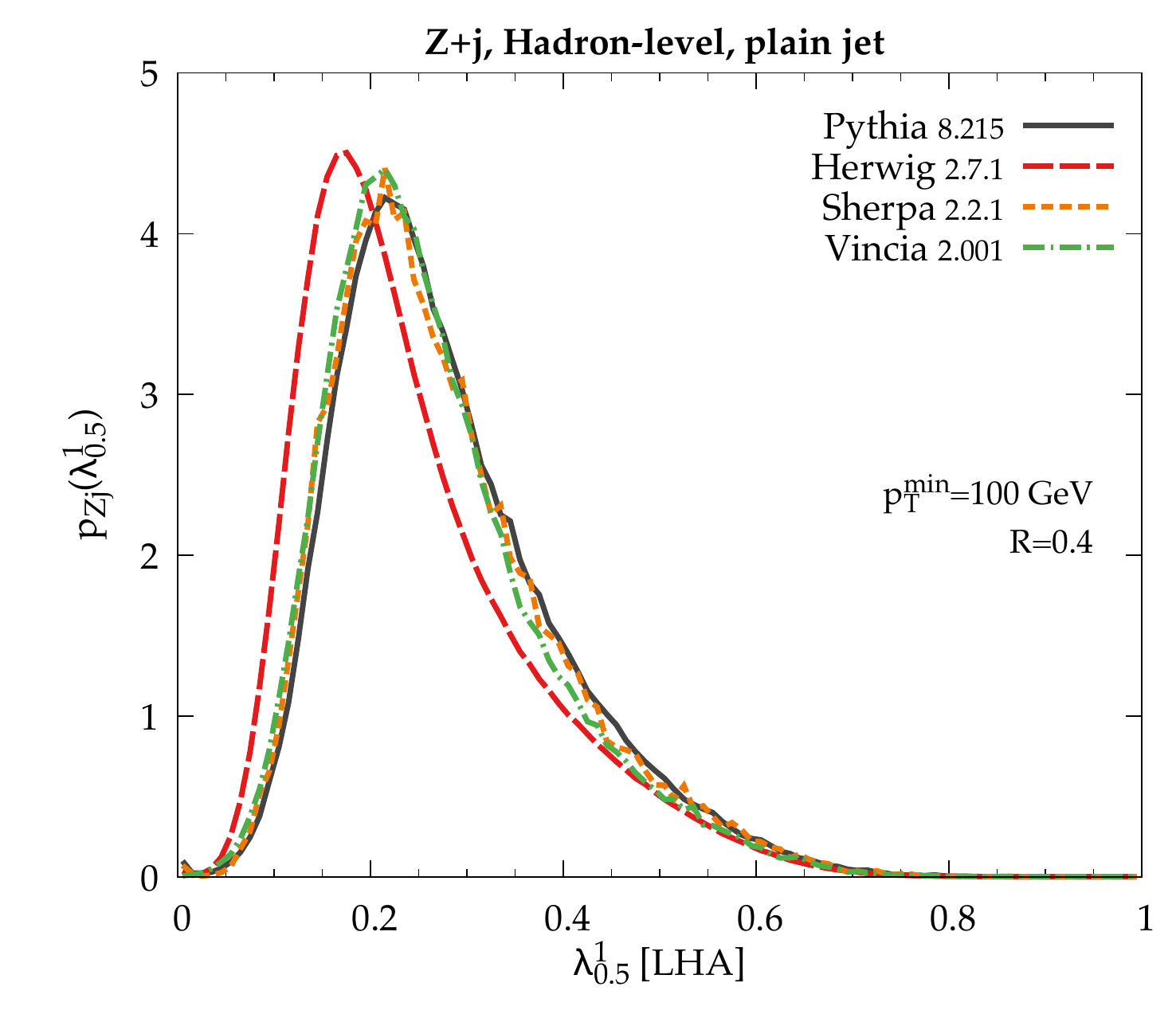}
\label{fig:LHA_hadron_pp_quark}
}
$\qquad$
\subfloat[]{
\includegraphics[width = 0.45\columnwidth]{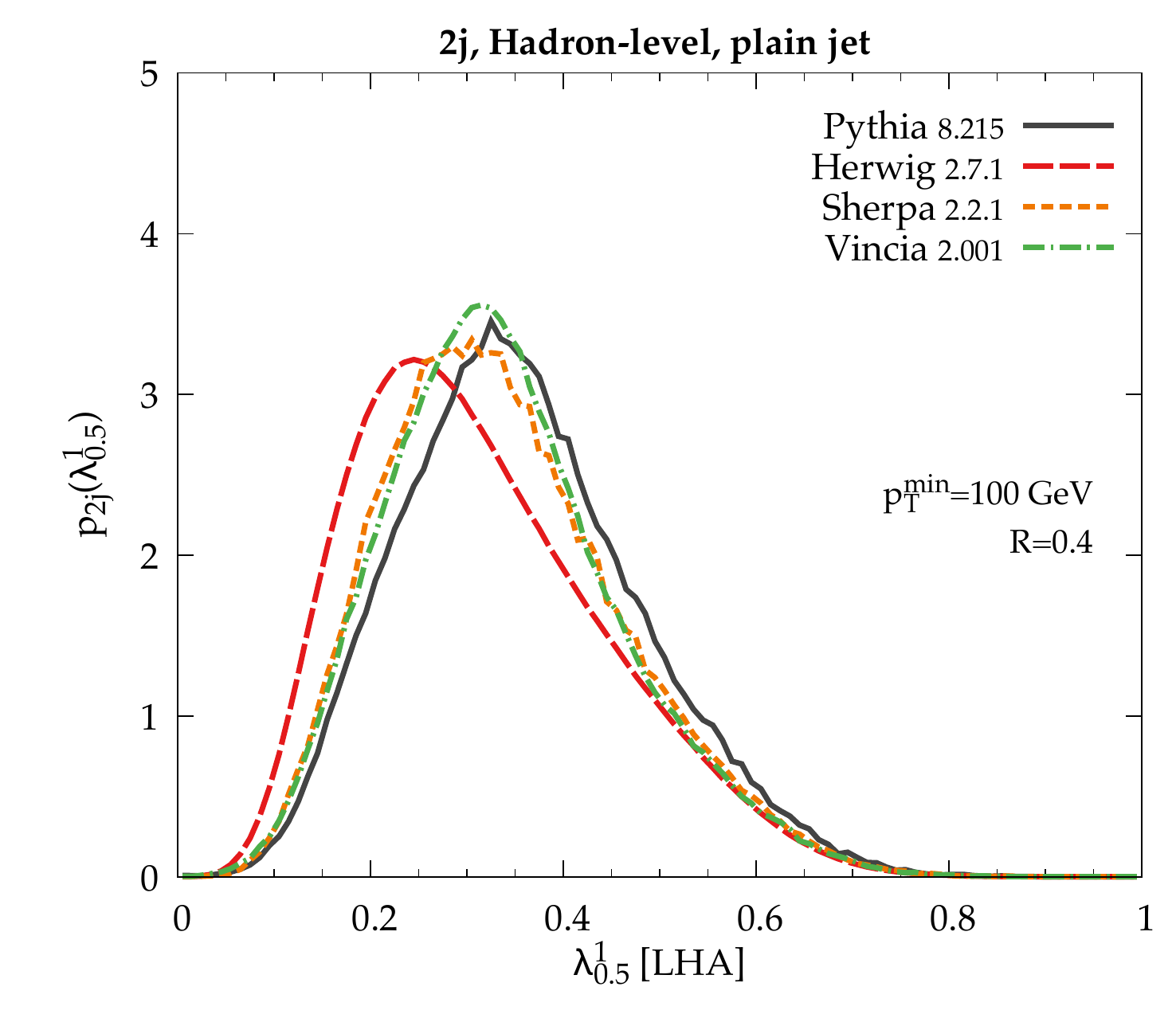}
\label{fig:LHA_hadron_pp_gluon}
}

\subfloat[]{
\includegraphics[width = 0.65\columnwidth]{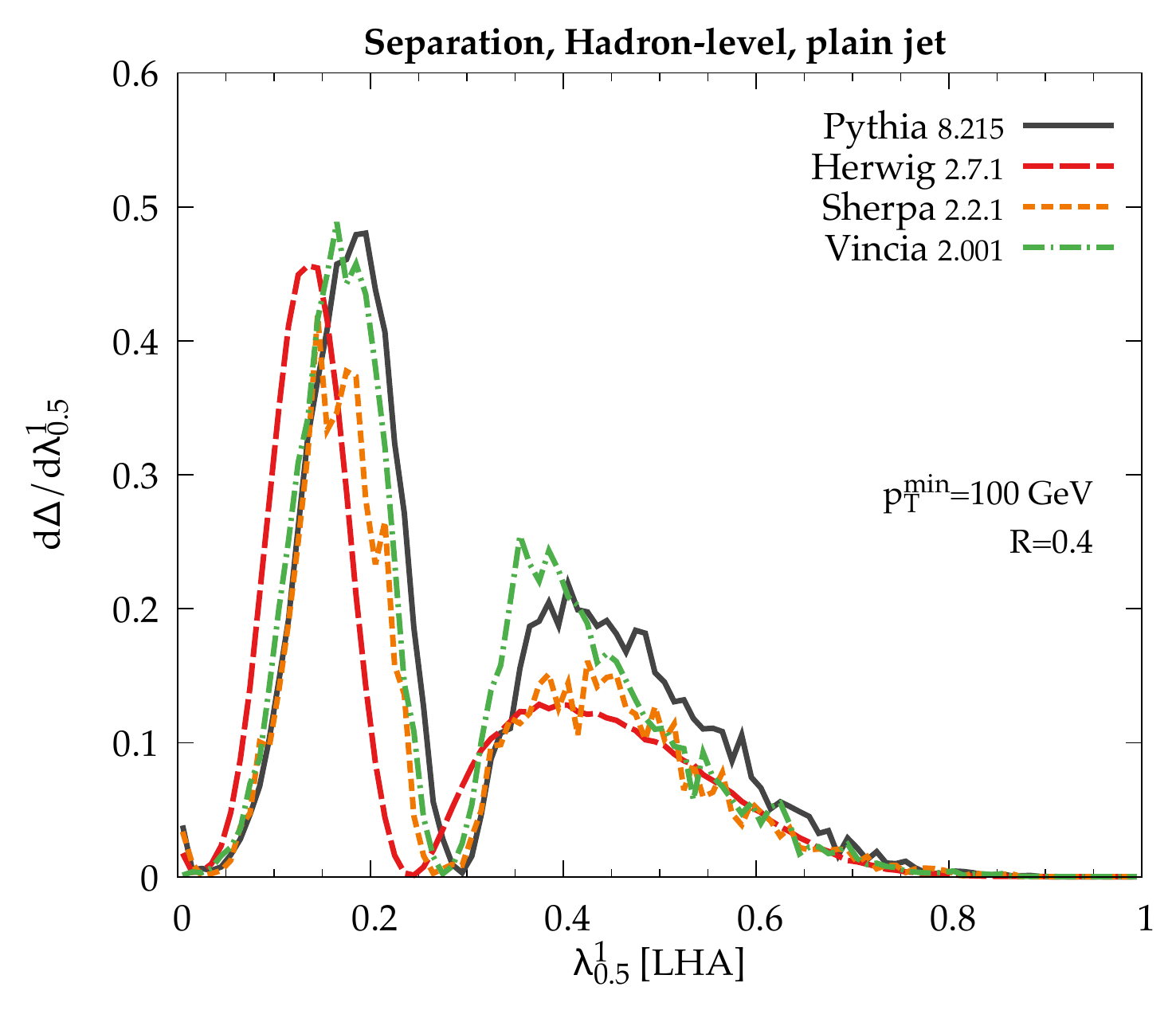}
\label{fig:LHA_hadron_pp_separation}
}
\caption{Distributions of the LHA at the LHC for (a) the $pp \to Z+j$
  (``quark-enriched'') sample, (b) the $pp \to 2j$
  (``gluon-enriched'') sample, and (c) the classifier separation
  integrand in \Eq{eq:deltaintegrand}.  Four parton-shower
  generators---\textsc{Pythia 8.215}, \textsc{Herwig 2.7.1},
  \textsc{Sherpa 2.2.1}, and \textsc{Vincia 2.001}---are run at their
  baseline settings with $p_T^{\rm min} = 100~\GeV$ and jet radius
  $R= 0.4$.  Note that the plotted range is different from
  \Fig{fig:LHA_hadron}.}
\label{fig:LHA_hadron_pp}
\end{figure}

\begin{figure}
\centering
\subfloat[]{
\includegraphics[width = 0.45\columnwidth]{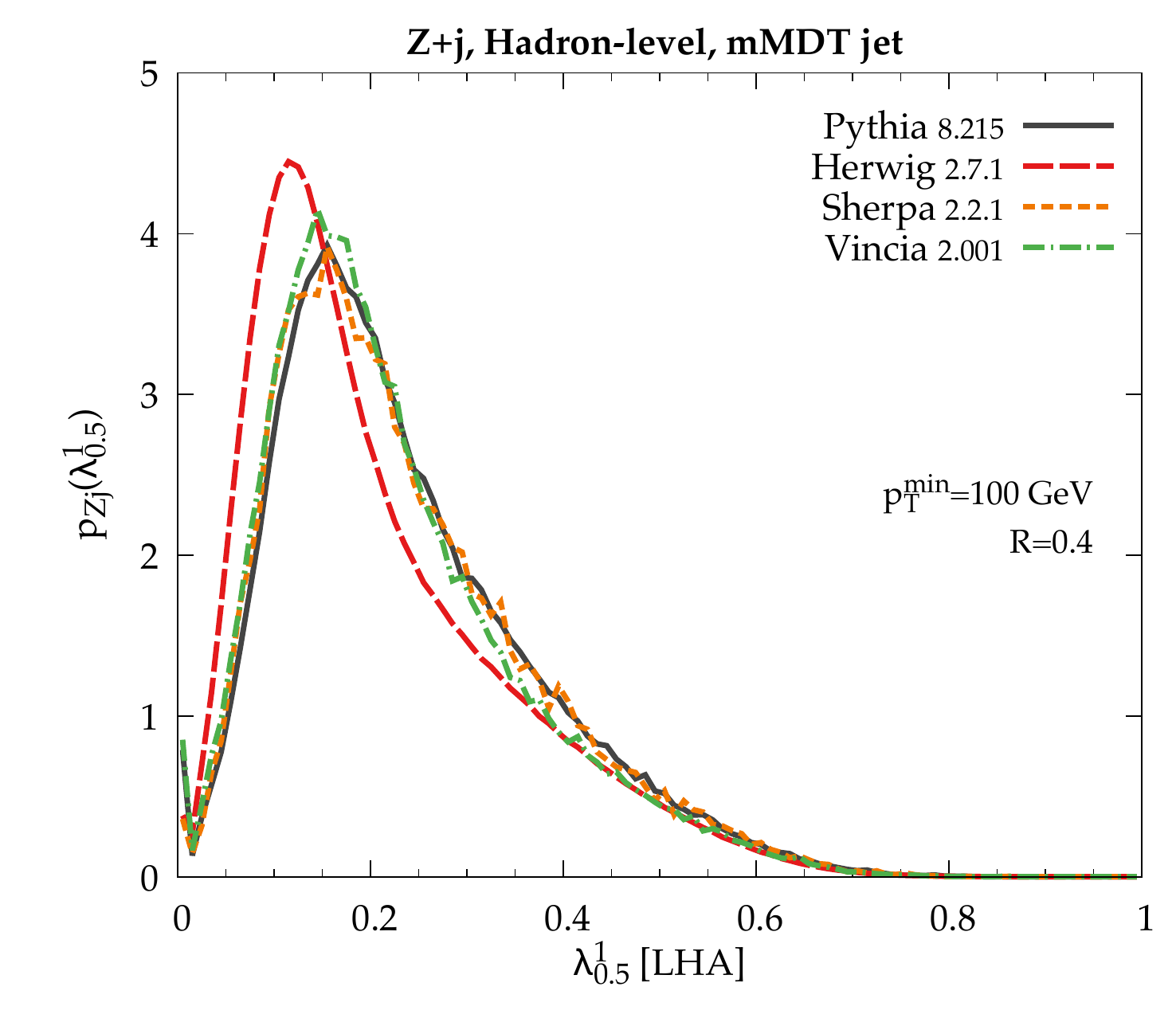}
\label{fig:LHA_hadron_pp_mmdt_quark}
}
$\qquad$
\subfloat[]{
\includegraphics[width = 0.45\columnwidth]{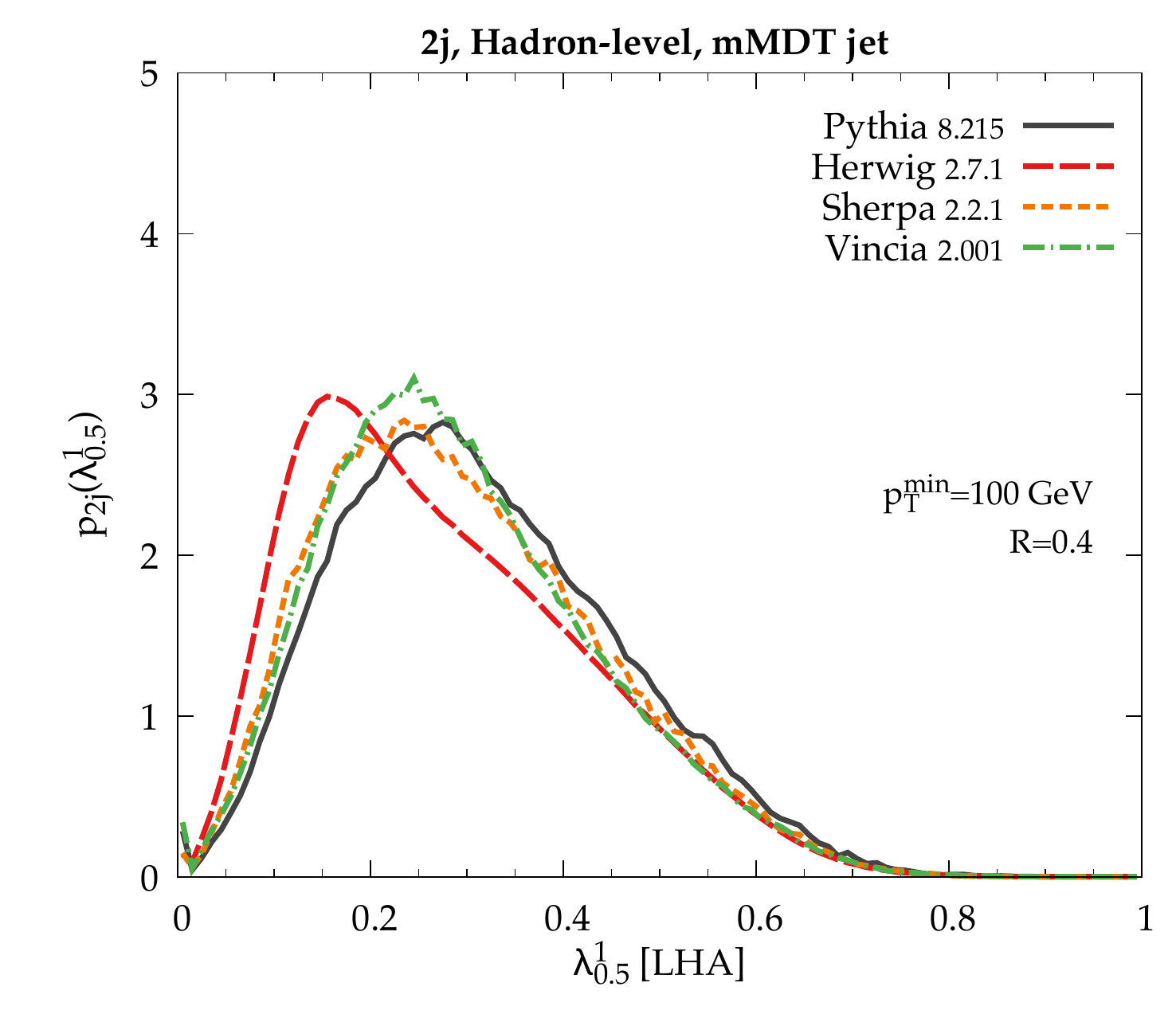}
\label{fig:LHA_hadron_pp_mmdt_gluon}
}

\subfloat[]{
\includegraphics[width = 0.65\columnwidth]{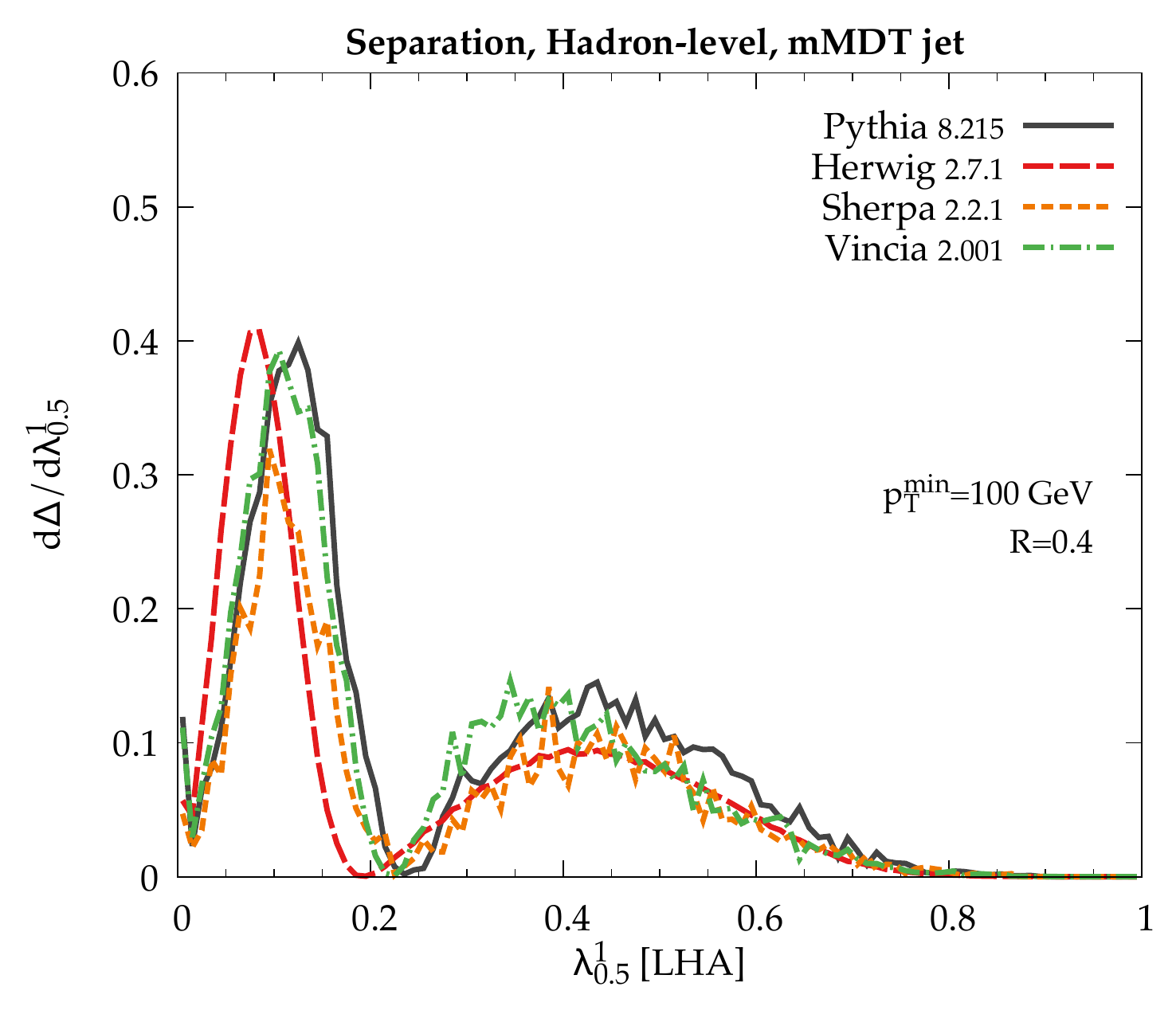}
\label{fig:LHA_hadron_pp_mmdt_separation}
}
\caption{Same as \Fig{fig:LHA_hadron_pp} but after mMDT jet grooming with $\mu = 1$ and $z_{\rm cut} = 0.1$.}
\label{fig:LHA_hadron_pp_mmdt}
\end{figure}

\begin{figure}
\centering
\subfloat[]{
\label{fig:summary_hadron_pp_all}
\includegraphics[width = 0.45\columnwidth]{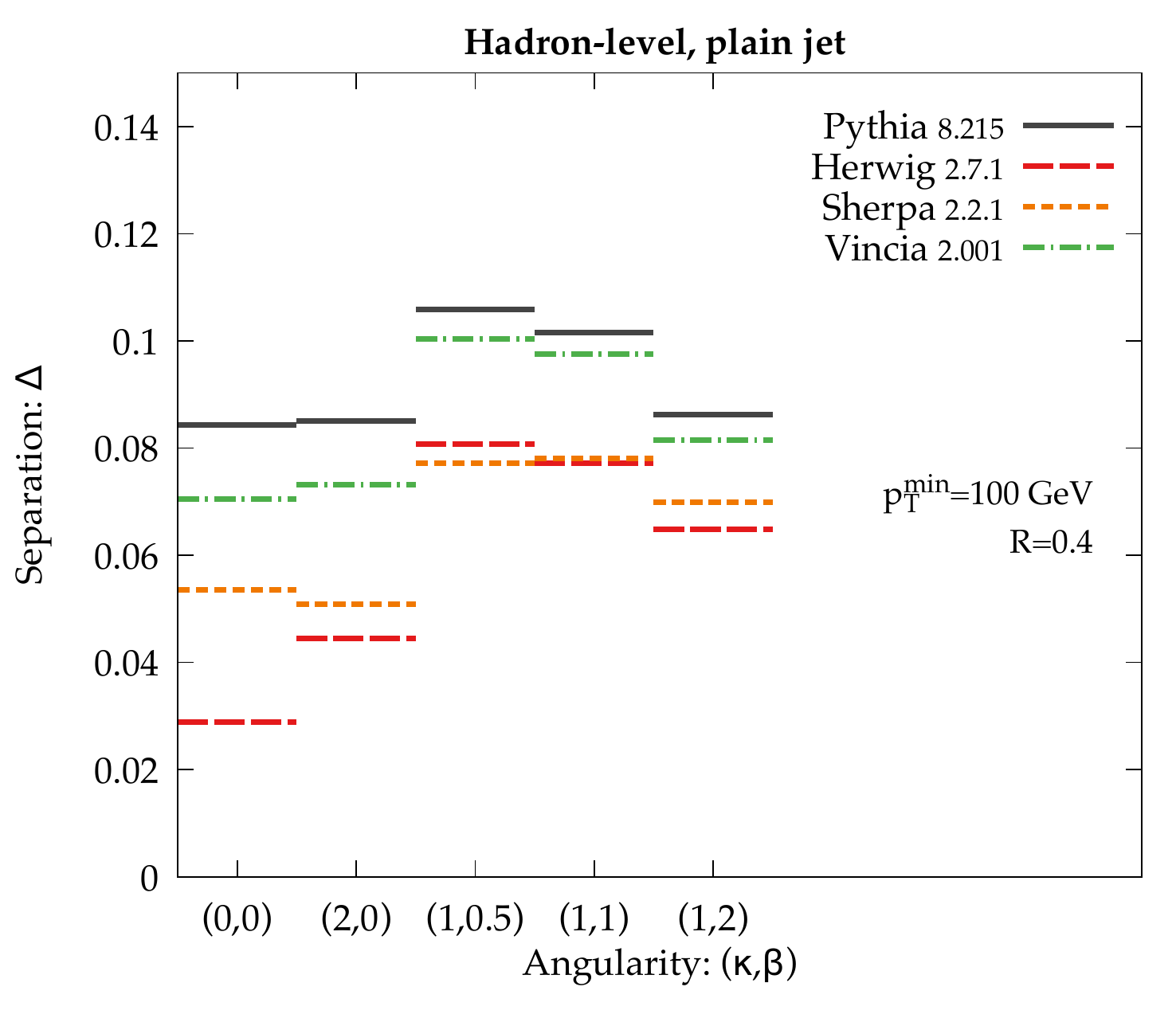}
}
$\qquad$
\subfloat[]{
\label{fig:summary_hadron_pp_mmdt_all}
\includegraphics[width = 0.45\columnwidth]{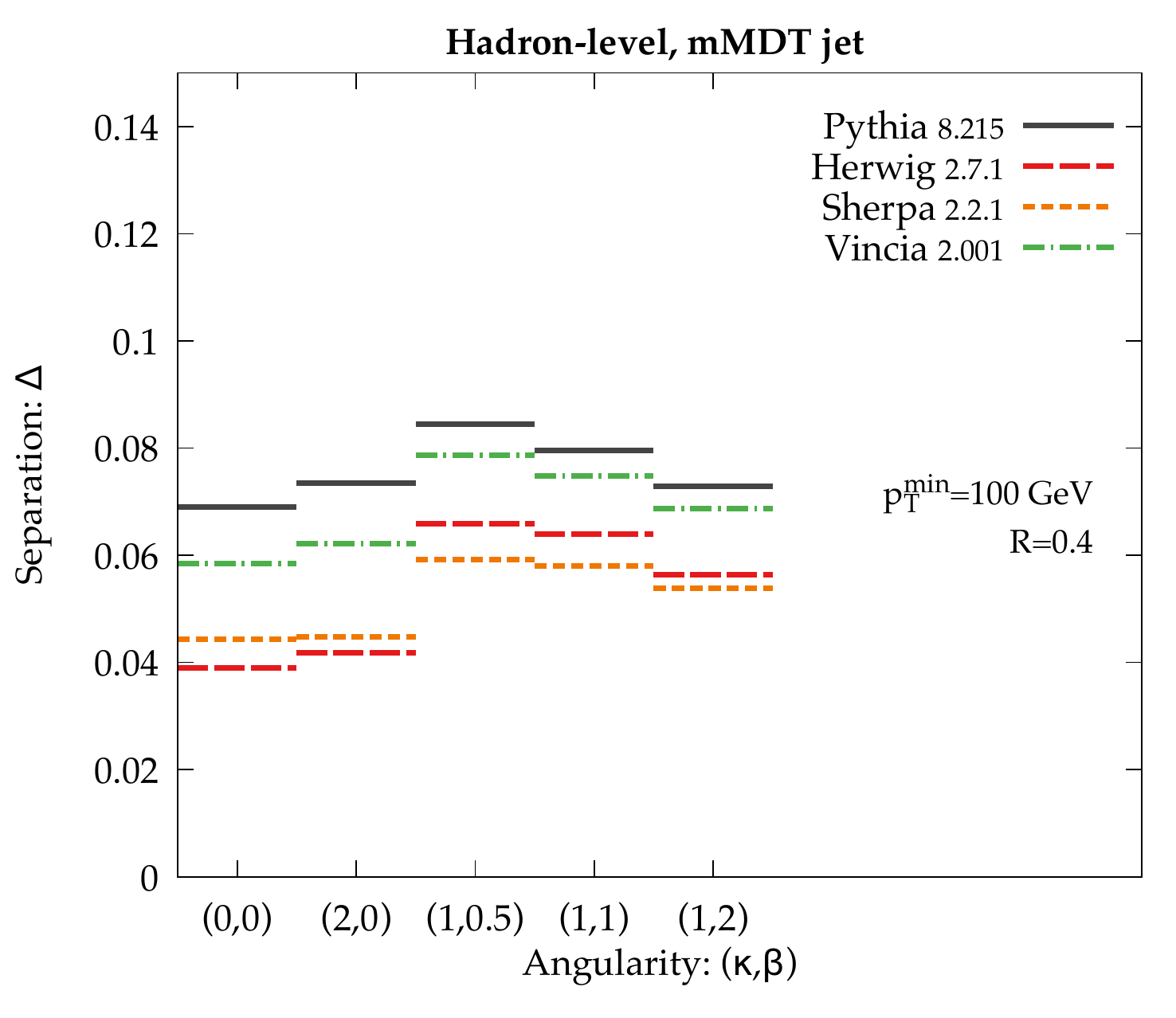}
}
\caption{Classifier separation $\Delta$ for the five benchmark
  angularities in our LHC study, determined from the various
  generators (a) using all jet constituents and (b) after mMDT
  grooming.  Note that the plotted range is different from
  \Fig{fig:summary_all}, reflecting the decreased discrimination
  power in the realistic $pp$ case compared to the idealized $e^+e^-$
  study with pure quark/gluon samples.}
\label{fig:summary_pp_all}
\end{figure}

In \Fig{fig:LHA_hadron_pp}, we show LHA distributions for the quark-enriched and gluon-enriched samples, using the default values $p_T^{\rm min} = 100~\GeV$ and $R = 0.4$.  For the quark-enriched $Z$ plus jets sample, all of the generators except \textsc{Herwig} yield similar distributions, as expected from the $e^+e^-$ study where the various generators broadly agreed on quark radiation patterns.  For the gluon-enriched dijet sample, the difference between generators grows noticeably, yielding disagreements that are even larger than the $e^+e^-$ study.  This difference is apparent also in the $\text{d} \Delta / \text{d} \lambda$ distribution, where \textsc{Pythia} and \textsc{Vincia} predict substantially larger separation power than \textsc{Herwig}, again in agreement with the $e^+e^-$ study.  Results from \textsc{Sherpa} appear to be intermediate between these extremes, though the integrated $\Delta$ value turns out to be similar to \textsc{Herwig} (see \Fig{fig:summary_hadron_pp_all}).  Already from these raw distributions, we see that LHC jet shape measurements would help constrain parton-shower uncertainties, especially for gluon-enriched jets.

We next turn to the impact of jet grooming.  Often jet grooming is described as a strategy to mitigate jet contamination from pileup, underlying event, and initial-state radiation \cite{Butterworth:2008iy,Ellis:2009su,Ellis:2009me,Krohn:2009th}.  Even at the level of final-state radiation, though, grooming modifies the observed jet radiation patterns in ways that are interesting from the quark/gluon discrimination perspective \cite{Dasgupta:2013ihk,Larkoski:2014wba}.  The impact of grooming is shown for the LHA after mMDT in \Fig{fig:LHA_hadron_pp_mmdt}.   In general, grooming pushes jet shapes to smaller values, since the effect of grooming is to remove soft peripheral radiation from a jet.  If the parton showers differed primarily in their treatment of wide-angle soft radiation, then one would expect grooming to bring the distributions into closer agreement.  Instead, we see that the generator differences persist even after grooming, suggesting that the parton showers differ already in their treatment of collinear radiation, despite using the same underlying collinear splitting kernels.  This motivates LHC measurements of groomed jet shapes to better understand the description of collinear physics.

In \Fig{fig:summary_pp_all}, we plot the classifier separation $\Delta$ for all five benchmark angularities, with and without jet grooming.  Compared to the $e^+e^-$ study, the overall degree of discrimination power is reduced, as expected because the $Z+j$ and dijet processes do not yield pure quark/gluon samples.  The spread between the generators is fairly large, with the expected trend that \textsc{Pythia} is more optimistic about quark/gluon separation than \textsc{Herwig}.  We see that \textsc{Vincia} has somewhat smaller predicted separation power than \textsc{Pythia}.   Though their raw distributions differ, the discrimination power in \textsc{Sherpa} is comparable to \textsc{Herwig}, with the ordering roughly flipped for unsafe versus safe observables.

One surprising outcome of this study is the relatively modest impact of grooming on discrimination power.  From the calculations in \Refs{Dasgupta:2013ihk,Larkoski:2014wba}, one generically expects quark/gluon discrimination power to degrade after jet grooming, since the soft radiation that is being removed carries information about the color structure of the jet.  This predicted degradation, however, is only seen modestly here, possibly because soft correlations with the initial state already blurred the distributions in the ungroomed case.   One advantage of working with groomed samples is that jet grooming reduces the process dependence in quark/gluon radiation patterns \cite{Frye:2016okc,Frye:2016aiz}.   In this way, groomed angularities should yield a more robust theoretical definition for quark and gluon jets, with only a small performance penalty.

\subsection{Parameter dependence}
\label{subsec:pp_para}

To test parameter dependence, we now consider five different minimum $p_T$ values and five different jet radii, with the boldface values corresponding to the defaults used above:\footnote{As a technical note, in order to test all values of $p^\text{min}_T$ in a single Monte Carlo run, we generate $p_T$-weighted events in each generator.}
\be
\label{eq:pp_pTRsweep}
\begin{aligned}
\text{Minimum $p_T$}: p^\text{min}_T &= \{ 50, \mathbf{100}, 200, 400, 800\}~\GeV, \\
\text{Jet Radius}: R &= \{ 0.2, \mathbf{0.4}, 0.6, 0.8, 1.0\}. \\
\end{aligned}
\ee
These $p_T$ values are effectively twice those used for the $e^+ e^-$ study in \Eq{eq:ee_sweep_values} (where $E_{\rm jet} \simeq Q/2$), and one should keep in mind from \Fig{fig:parton_level_qg_composition} that the degree of quark/gluon enrichment changes as a function of $p^\text{min}_T$.

\begin{figure}[t]
\centering
\subfloat[]{
\includegraphics[width = 0.45\columnwidth]{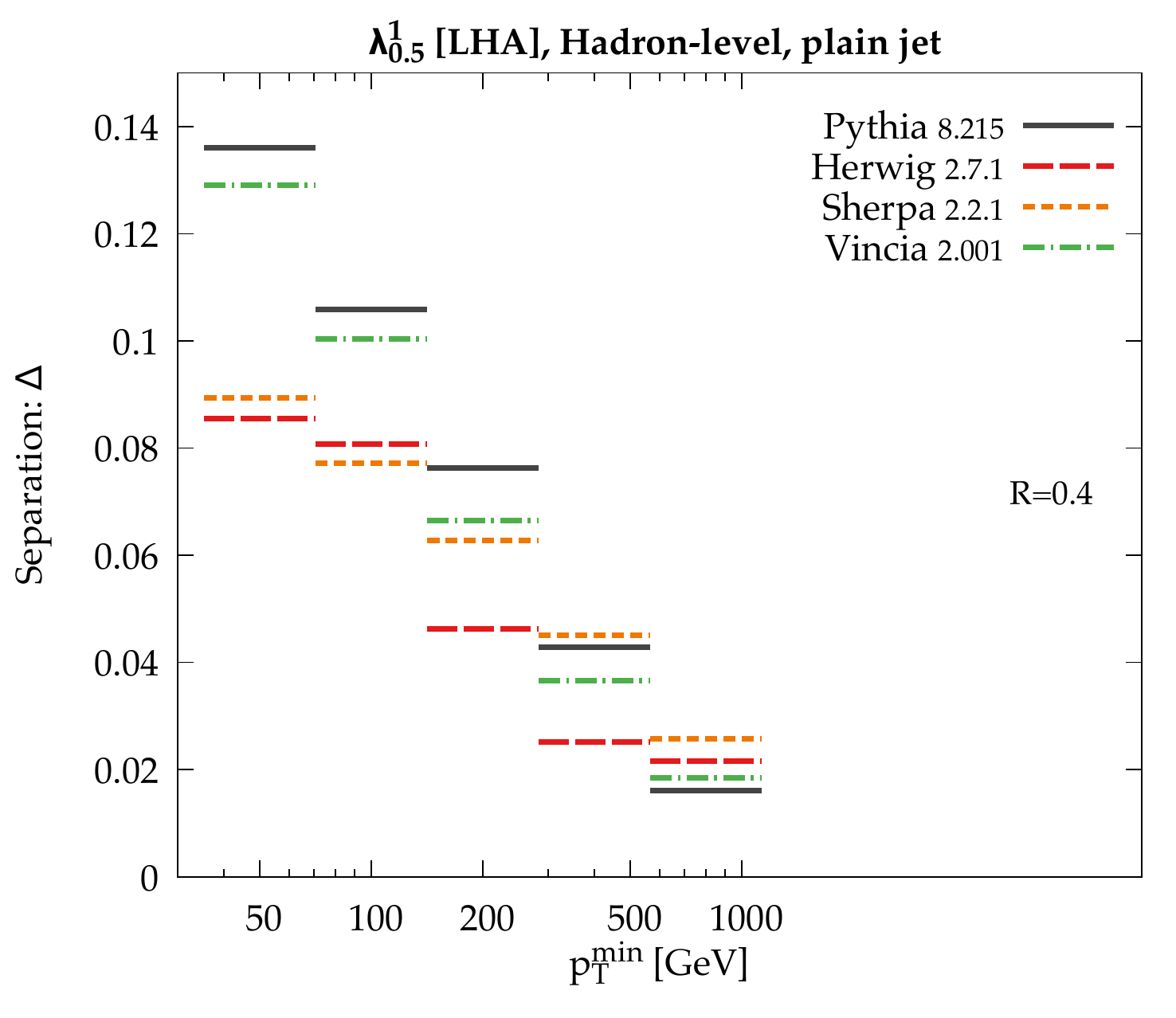}
\label{fig:sweep_Q_hadron_pp}
}
$\qquad$
\subfloat[]{
\includegraphics[width = 0.45\columnwidth]{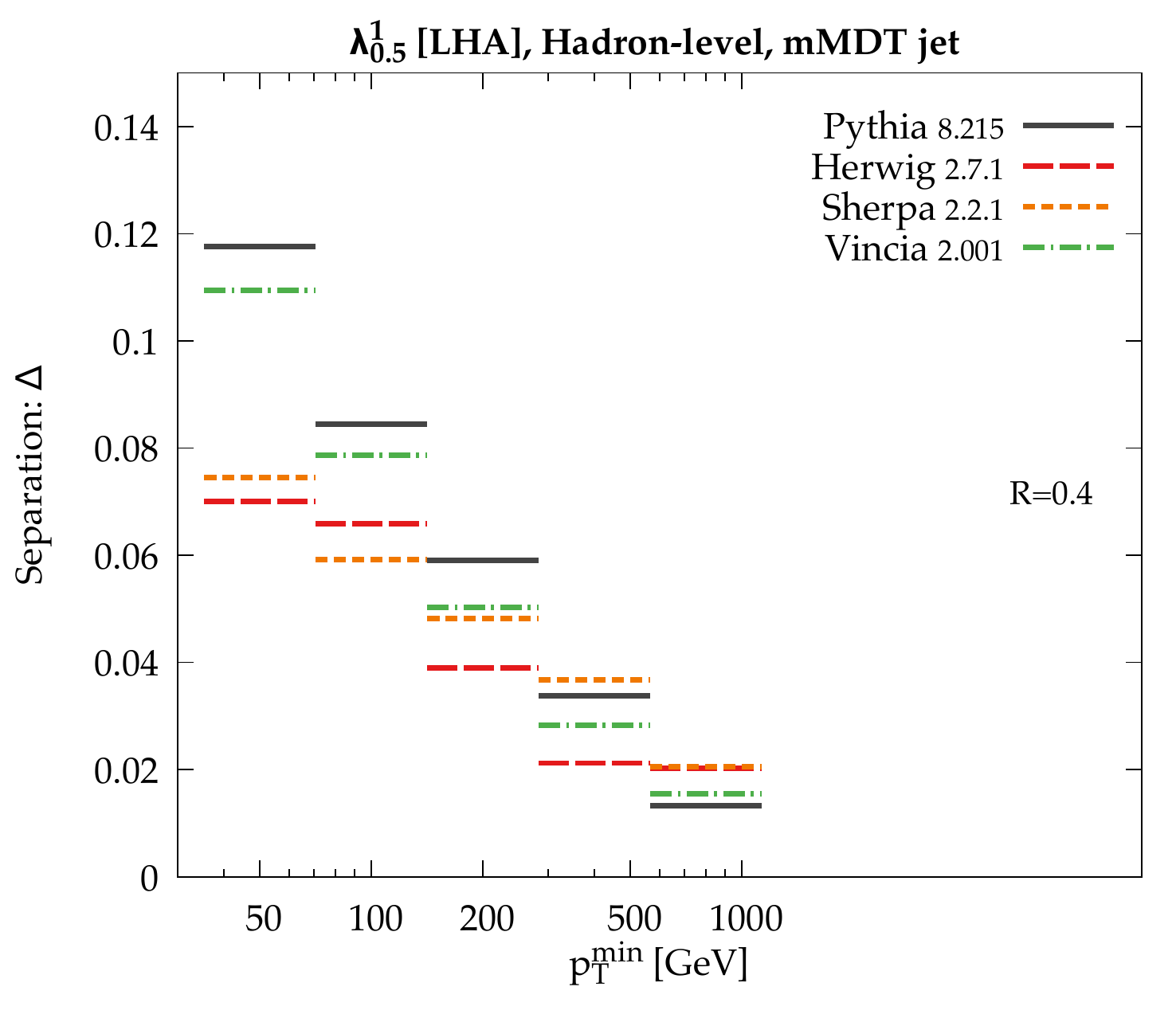}
\label{fig:sweep_Q_hadron_pp_mmdt}
}

\subfloat[]{
\includegraphics[width = 0.45\columnwidth]{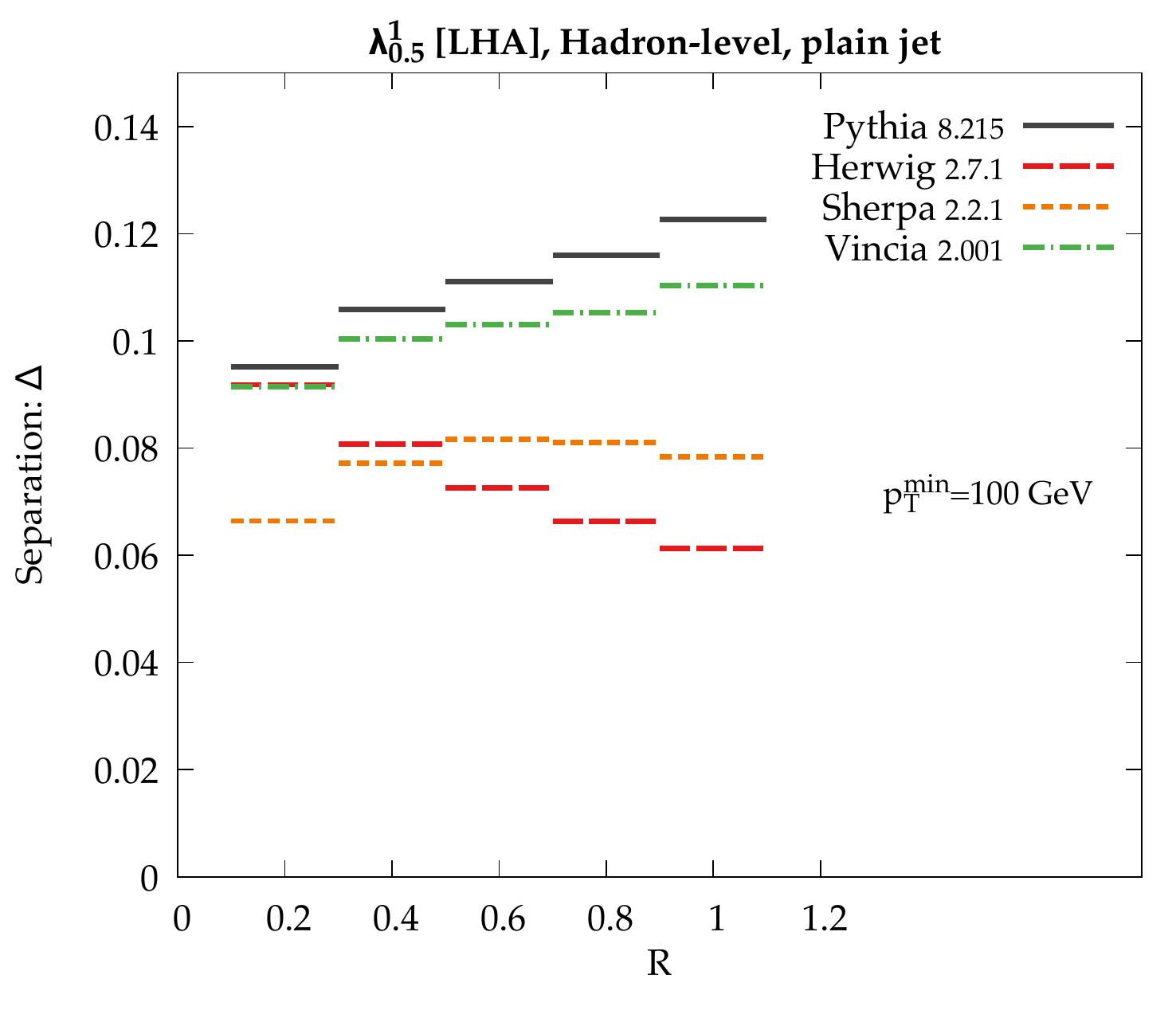}
\label{fig:sweep_R_hadron_pp}
}
$\qquad$
\subfloat[]{
\includegraphics[width = 0.45\columnwidth]{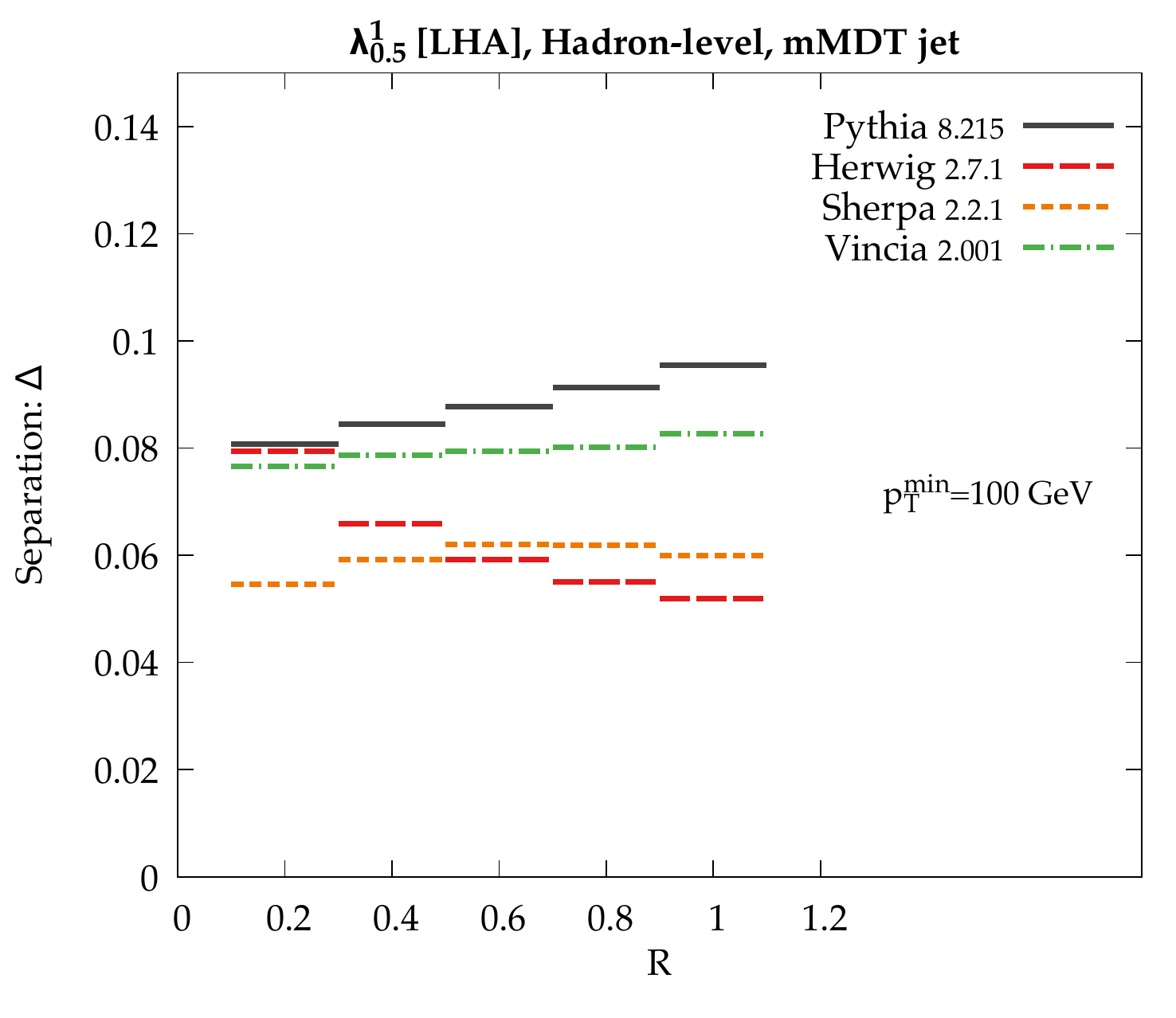}
\label{fig:sweep_R_parton_pp_mmdt}
}
\caption{Classifier separation $\Delta$ for the LHA in our LHC study, sweeping $p_T^{\rm min}$ (top row) and jet radius $R$ (bottom row).  Results are shown with all jet constituents (left column) and after mMDT grooming (right column).}
\label{fig:pp_sweep}
\end{figure}

The results of sweeping $p^\text{min}_T$ and $R$ are shown in \Fig{fig:pp_sweep}.  In general, increasing $p^\text{min}_T$ leads to a degradation of separation power, though this is due in large part to the change in sample composition shown in \Fig{fig:parton_level_qg_composition}.  In the context of these mixed samples, it is difficult to disentangle the impact of reduced quark/gluon enrichment at high $p_T$ with actual trends in discrimination power (cf.~\Fig{fig:sweep_Q_hadron} from the $e^+e^-$ study).  That said, the trends are sufficiently different between generators that the relative differences cannot be ascribed to sample composition alone.

With respect to changing the jet radius $R$, the discrimination trends are noticeably different between the generators.  In \textsc{Pythia} and \textsc{Vincia}, the discrimination power rises with increasing jet radius, whereas in \textsc{Herwig}, the discrimination power degrades with larger $R$; \textsc{Sherpa} has relatively little $R$ dependence.  The trends are rather similar before and after jet grooming, again pointing to differences between the generators in collinear physics and not just soft physics.  We conclude that varying $p_T^{\rm min}$ and $R$ provides important information about quark/gluon radiation patterns that cannot captured by focusing on a single kinematic regime.  We therefore encourage LHC measurements of jet shapes at multiple energy scales with multiple jet radii.  

\section{Summary and recommendations}
\label{sec:conclude}

By measuring the substructure of jets, one can gain valuable information about the relative quark/gluon composition of a jet sample.  The challenge we have identified in this study is that the precise radiation patterns of quark and gluon jets are poorly understood, in the sense that parton-shower generators give rather different predictions for absolute quark/gluon discrimination power as well as relative trends as a function of the jet kinematics.  From our analytic NLL studies including nonglobal logarithms and shape functions, we see that both perturbative and nonperturbative physics play an important role in determining jet shape distributions.  That said, analytic calculations are not yet at the level of accuracy where they could directly guide the tuning of event generators.  Therefore, LHC measurements are the best near-term strategy to constrain quark/gluon radiation patterns and enable quark/gluon discrimination to become a robust experimental tool.

Our five benchmark angularities probe both the perturbative and nonperturbative structure of jets, so we think they would be a good starting point for a more comprehensive quark/gluon jet shape analysis at the LHC.  In this spirit, we are encouraged by the track multiplicity study of \Ref{Aad:2016oit}, though for parton-shower tuning is it is important to have measurements not only of jet shape averages but also of the full jet shape probability distributions.  In terms of specific measurements that should be highest priority for ATLAS and CMS, our study has not revealed a silver bullet.  Rather, all of the observables studied in this paper show similar levels of disagreement between generators, so a systematic LHC study of even one observable is likely to offer crucial new information.

What does seem to be essential is to make LHC measurements at multiple
jet $p_T$ scales with multiple jet radii $R$ in multiple different
quark/gluon-enriched samples.  Unfolded distributions would be the
most useful for constraining parton-shower uncertainties, but even
detector-level measurements compared to detector-simulated parton
showers could help spot troubling trends.  For the IRC-safe
angularities in particular, studying the $\beta$ dependence would help
separate information about collinear and soft radiation patterns,
especially given the fact that the $\beta$ trends seen in the
parton-shower generators here disagree with those seen in
\Ref{Aad:2014gea}.  In addition, measurements of both groomed and
ungroomed jet shapes could help disentangle collinear versus soft
effects.

If possible, it would be interesting to study the LHA ($\beta = 1/2$) on archival LEP data, since this angularity probes the core of jets in a new way, distinct from broadening-like ($\beta = 1$) or thrust-like ($\beta = 2$) observables.  Among the IRC-safe angularities studied here, the LHA has the best predicted discrimination power, making it (and other $0<\beta < 1$ angularities) a well-motivated target for future lepton collider measurements.  Similarly, it would be worthwhile to improve our analytic understanding of the LHA.  From \Fig{fig:LHA_hadron_separation}, we see that the LHA has discrimination power both at small values of $\lambda_{0.5}^1$ (where nonperturbative corrections play an important role) as well as at larger values of $\lambda_{0.5}^1$ (where fixed-order corrections are important).  Therefore, one must go beyond an NLL understanding to accurately describe the quark/gluon performance of the LHA.

The key lesson to parton-shower authors is that, contrary to some standard lore, existing LEP measurements used for tuning \emph{do not} constrain all of the relevant aspects of the final state parton shower.  While we have extensive information about quark-jet radiation patterns from LEP event shapes, gluon-jet radiation patterns are largely unconstrained.  This has important implications for parton-shower tuning strategies, since LHC data can and should be used to adjust final-state shower parameters.  For example, the ATLAS A14 tune of \textsc{Pythia} has a 10\% lower value of $\alpha_s$ in the final-state shower compared to the Monash tune, which yields better agreement with charged-particle multiplicity distributions \cite{Aad:2016oit}.  However, A14 has not been tested on LEP event shapes, suggesting that a global tuning strategy is needed.   In addition, it is worth mentioning that similar quark/gluon studies have been carried out in deep inelastic electron-proton scattering \cite{Chekanov:2004kz}, which offer an intermediate step between $pp$ and $e^+ e^-$ collisions, and this $ep$ data could also be valuable for parton-shower tuning.

Interestingly, there are LEP measurements that do constrain gluon
radiation patterns, as recently summarized by K.~Hamacher in~\Ref{Anderle:2017qwx}.
Unfortunately, these are not currently implemented in \textsc{Rivet}~\cite{Buckley:2010ar} and, 
to our knowledge, are not used in any present-day parton-shower
tuning strategy.
The cleanest LEP studies focused on $Z\to b\bar{b}g$
events, i.e.\ 3-jet events with two heavy-flavor tagged
jets~\cite{Alexander:1991ce,Acton:1993jm}.
By applying appropriate event-selection cuts, these studies identified ``symmetric
events'' where the gluon was relatively isolated~\cite{Buskulic:1995sw,Abreu:1995hp,OPAL:1995ab,Abreu:1998ve};
this strategy was extended to more general 3-jet topologies using Lorentz-invariant $p_\perp$ scales~\cite{Abreu:1999af}.
In the rare case that the two tagged jets appeared in the same hemisphere of an event, the opposite hemisphere could be used to define an inclusive
sample of gluon-like jets~\cite{Alexander:1996qr,Ackerstaff:1997xg,Abbiendi:1999pi,Abbiendi:2003ri}.
With a relatively pure gluon-jet sample, one could then study various aspects of gluon-jet fragmentation, including hadron multiplicity, single-hadron energy fractions, and $y$-splitting scales.
It should be noted, however, that in at least some of the above analyses, the corrections to hadron level made use of Monte Carlo truth information
to correct not only for photon initial-state radiation and detector effects but also for impurities in the gluon-jet selection.
We therefore encourage efforts to determine the extent to which \textsc{Rivet} implementations of these measurements are practicable, and to begin that
process if the corrections are deemed to be sufficiently model independent.
This would enable a broader suite of LEP measurements to be included in the next round of parton-shower tunes and in global comparisons
such as
\href{http://mcplots.cern.ch}{\mbox{\textsc{MCplots}}}~\cite{Karneyeu:2013aha}.

In a similar spirit, a future high-luminosity lepton collider would allow for measurements of the above processes with a high precision (see
e.g.~\cite{Anderle:2017qwx}).  At sufficiently high collision energies, one can also measure other interesting processes, such as associated Higgs production with the Higgs boson decaying to bottom quarks or gluons.  Such measurements would provide an invaluable source of data in the context of quark/gluon discrimination and, more generically, for parton-shower tuning.

Based on this study, we have identified three aspects of the final-state parton shower that deserve closer scrutiny.
\begin{itemize}
\item \textit{Gluon splitting to quarks}.  Some of the largest differences between generators came from turning on and off the $g \to q \overline{q}$ splitting process.  While \textsc{Pythia}, \textsc{Sherpa}, \textsc{Vincia}, \textsc{Ariadne} and \textsc{Dire} suggest that (unphysically) turning off $g \to q \overline{q}$ would improve quark/gluon separation, \textsc{Herwig} (and the analytic calculation from \Sec{sec:analytic}) suggests the opposite conclusion.  Beyond quark/gluon discrimination, it would be helpful to identify other contexts where $g \to q \overline{q}$ might play an important role (see e.g.~\cite{Ilten:2017rbd}).
\item \textit{Color reconnection in the final state}.  Color reconnection is often thought of as an issue mainly at hadron colliders, but we have seen that it can have an impact in $e^+ e^-$ collisions as well.  This is particularly the case with the default color reconnection model in \textsc{Herwig}, since it allows the reconnection of color/anticolor lines even if they originally came from an octet configuration.  We also saw large changes from ``\textsc{Pythia: CR1}'' and  ``\textsc{Ariadne: no swing}'', suggesting that one should revisit color reconnection physics when tuning generators to LEP data.
\item \textit{Reconsidering $\alpha_s$ defaults}:  In the context of parton-shower tuning, the value of $\alpha_s$ used internally within a code need not match the world average value, since higher-order effects not captured by the shower can often be mimicked by adjusting $\alpha_s$.  That said, one has to be careful whether a value of $\alpha_s$ tuned for one process is really appropriate for another.  For example, \textsc{Pythia} uses a relatively large value of $\alpha_s$ in its final-state shower, which allows it to match LEP event shape data.  The same value of $\alpha_s$, though, probably also leads to too much radiation within gluon jets.
\end{itemize}
Finally, we want to emphasize that despite the uncertainties currently present in parton-shower generators~\cite{Bellm:2016rhh,Bellm:2016voq,Mrenna:2016sih,Bothmann:2016nao}, 
parton showers in particular (and QCD resummation techniques more generally) will be essential for understanding quark/gluon discrimination.  Fixed-order QCD calculations cannot reliably probe the very soft and very collinear structure of jets, which is precisely where valuable information about quark/gluon radiation patterns reside.  Given the ubiquity and value of parton-shower generators, improving the understanding of quark/gluon discrimination will assist every jet study at the LHC.

\begin{acknowledgments}

We thank Marat Freytsis and Davison Soper for contributing to the original Les Houches study \cite{Badger:2016bpw} that formed the basis for this work.  We particularly thank Davison Soper for contributing plots based on Deductor.
We thank L'\'{E}cole de Physique des Houches and the organizers of the 2015 Les Houches workshop on ``Physics at TeV Colliders'' for a stimulating environment while this work was being initiated.
We benefitted from helpful discussions and feedback from Samuel Bein, Andy Buckley, Jon Butterworth, Mario Campanelli, Peter Loch, Ben Nachman, Zoltan Nagy, Chris Pollard, Stefan Prestel, Salvatore Rappoccio, Gavin Salam, Alexander Schmidt, Frank Tackmann, and Wouter Waalewijn.
The work of SH is supported by the U.S. Department of Energy (DOE) under contract DE-AC02-76SF00515.
DK acknowledges support from the Science Faculty Research Council, University of Witwatersrand.
The work of LL, SP, and AS is supported in part by the MCnetITN FP7 Marie Curie Initial Training Network, contract PITN-GA-2012-315877.
LL is also supported by the Swedish Research Council (contracts 621-2012-2283 and 621-2013-4287).
SP acknowledges support from a FP7 Marie Curie Intra European Fellowship under Grant Agreement PIEF-GA-2013-628739.
AS acknowledges support from COST Action CA15213 THOR and the National Science Centre, Poland Grant No.~2016/23/D/ST2/02605.
PS is the recipient of an Australian Research Council Future Fellowship, FT130100744.
The work of GS is supported in part by the Paris-Saclay IDEX under the
IDEOPTIMALJE grant, by the French Agence Nationale de la Recherche,
under grant ANR-15-CE31-0016, and by the ERC Advanced Grant Higgs@LHC
(No.\ 321133).
The work of JT is supported by the DOE under grant contract numbers DE-SC-00012567 and DE-SC-00015476.

\end{acknowledgments}

\bibliographystyle{jhep}
\bibliography{quarkgluon_bib}

\end{document}